\documentclass[aps,prl,twocolumn,superscriptaddress,showpacs,footnotebib]{revtex4-2}
\usepackage{amsmath}
\usepackage{bm}
\usepackage{color}
\usepackage{graphicx}
\usepackage{epstopdf}
\usepackage{epsfig}
\usepackage{amsfonts}
\usepackage[naturalnames]{hyperref}
\usepackage{hypcap}
\usepackage{verbatim}
\usepackage{tabularx}
\usepackage{bbm}
\usepackage{esvect}
\usepackage{subfigure}
\usepackage{slashed}

\def\bea{\begin{eqnarray}}
\def\eea{\end{eqnarray}}
\def\nn{\nonumber}
\def\ba{\begin{array}}
\def\ea{\end{array}}
\def\Tr{\text{Tr}}
\def\nn{\nonumber}

\def\Tr{\text{Tr}}

\hypersetup{colorlinks=true, citecolor=blue, urlcolor=blue, linkcolor=blue}
\begin{document}
\title{New critical states induced by measurement}
\author{Xinyu Sun}
\affiliation{Institute for Advanced Study, Tsinghua University, Beijing 100084, China}
\author{Hong Yao}
\email{yaohong@tsinghua.edu.cn}
\affiliation{Institute for Advanced Study, Tsinghua University, Beijing 100084, China}
\author{Shao-Kai Jian}
\email{sjian@tulane.edu}
\affiliation{Department of Physics and Engineering Physics, Tulane University, New Orleans, Louisiana, 70118, USA}

\date{\today}

\begin{abstract}
Finding new critical states of matter is an important subject in modern many-body physics.
Here we study the effect of measurement and postselection on the critical ground state of a Luttinger liquid theory and show that it can lead to qualitatively new critical states.
Depending on the Luttinger parameter $K$, the effect of measurement is irrelevant (relevant) at $K>1$ ($K<1$).
We reveal that this causes an entanglement transition between two phases, one with logarithmic entanglement entropy for a subregion ($K>1$), and the other with algebraic entanglement entropy ($K<1$).
At the critical point $K=1$, the measurement is marginal, and we find new critical states whose entanglement entropy exhibits a logarithmic behavior with a continuous effective central charge as a function of measurement strength.
We also performed numerical density matrix renormalization group and fermionic Gaussian state simulations to support our results.
We further discuss promising and feasible routes to experimentally realize new critical states in our work.
\end{abstract}

\maketitle

{\it Introduction.---}Critical state underlies various interesting physics in phase transition, hydrodynamics, and even quantum gravity according to AdS/CFT correspondence.
While conformal field theory (CFT) describes a huge class of critical states, it is of impact to find new critical states. 
Recently, studies on quantum trajectory with local measurement reveal a critical point separating two quantum phases with distinct entanglement structures~\cite{li2019measurement, skinner2019measurement, chan2019unitary, gullans2020dynamical, jian2021measurement}.
It is natural to study the effect of measurement in CFT.
The effect of local projective measurement in CFT is described by boundary CFT~\cite{cardy1989boundary, affleck1991universal, cardy2004boundary, rajabpour2015post, rajabpour2016entanglement}, where (a region of) the critical state is projected onto a Cardy state~\cite{cardy2004boundary}. 
Apart from projective measurements in boundary CFT, less is known about general measurement. 

\begin{figure}[t]
    \centering
\subfigure[]{
\includegraphics[width=0.22\textwidth]{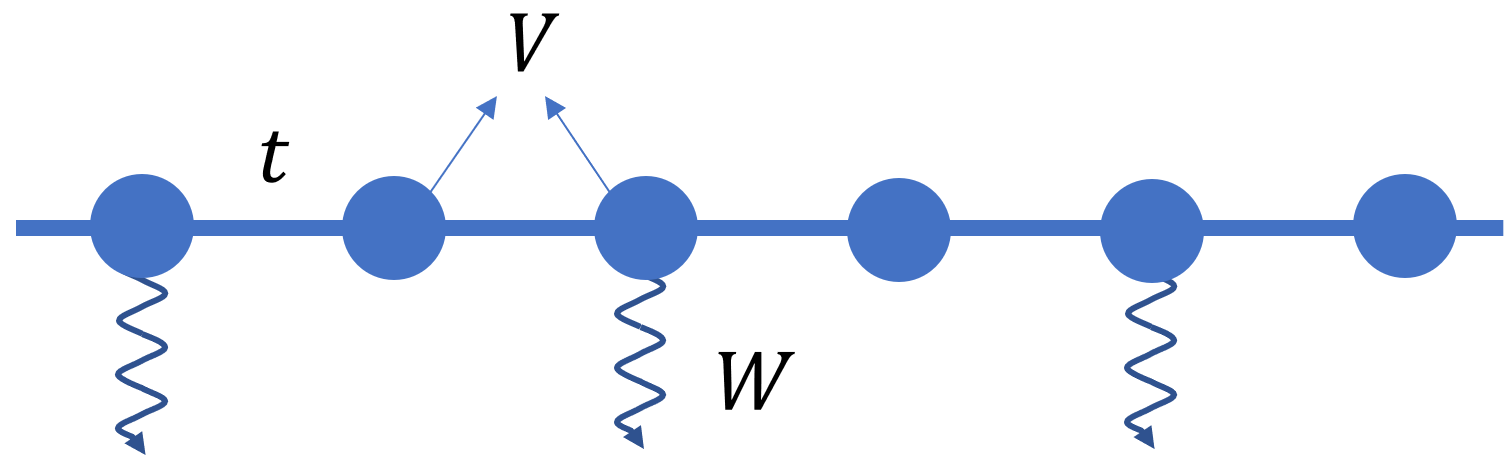}}
\subfigure[]{
\includegraphics[width=0.23\textwidth]{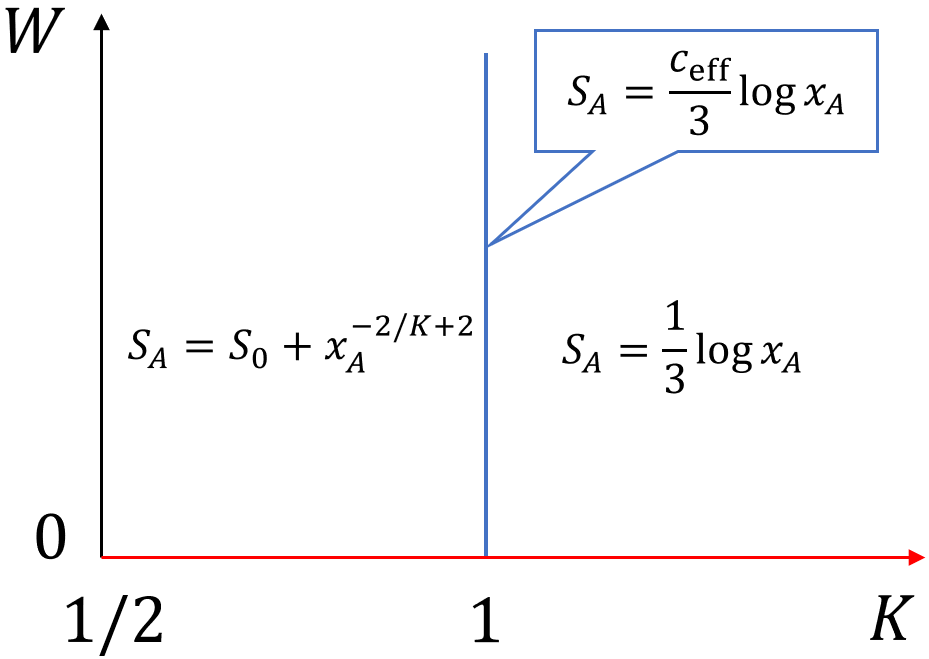}
}
    \caption{(a) A schematic plot of the spinless fermion chain.
    $t$ and $V$ denote the hopping and the nearest-neighbor interaction, respectively, in~(\ref{eq:fermion_Hamiltonian}),
    $W$ is the measurement strength in~(\ref{eq:measurement}).
    (b) Phase diagram of the Luttinger liquid after weak measurement.
    $K$ and $W$ denote the Luttinger parameter and the measurement strength, respectively.
    For $K>1$, the measurement is irrelevant, and the entanglement entropy of a subregion $A$ with length $x_A$ satisfies a log-law with central charge $c=1$.
    For $K<1$, the measurement is relevant, and changes the entanglement entropy to an area law with a subleading algebraic correction.
    At $K=1$, there is a continuous critical line, at which the measurement is marginal.
    The entanglement entropy satisfies a log-law with an effective central charge $c_\text{eff}$ given in~\eqref{eq:effective_central_charge}.
    We use the red line to indicate the non-measurement case $W=0$, where the entanglement entropy is given by $S_A = 1/3 \log x_A$.
    }
    \label{fig:summary}
\end{figure}

In this paper, we are interested in the effect of weak measurement on the ground state of a Luttinger liquid.
Different from the constant monitoring in free fermion systems studied previously~\cite{chen2020emergent,jian2022criticality,alberton2021entanglement,buchhold2021effective,zhang2021emergent,minoguchi2022continuous}, we consider performing weak measurement to the critical ground state of the Luttinger liquid without time evolution.
Ref.~\cite{garratt2022measurements} reports a transition as a function of Luttinger parameter between two phases with algebraic correlation functions of distinct power laws after weak measurement, nevertheless, the entanglement property after weak measurement is left unanswered.
Since measurement can change entanglement radically, the scaling of entanglement entropy in these two phases is not immediately obvious.

We study the entanglement properties of the Luttinger liquid theory after weak measurement and postselection.
A schematic representation of the model and the phase diagram are shown in Fig.~\ref{fig:summary}(a) and (b). 
We find that for $K > 1$ the measurement is irrelevant and the central charge remains $c = 1$. 
For $K < 1$, the measurement becomes relevant, and the state exhibits an area law with a subleading algebraic entanglement entropy. 
The algebraic power-law is obtained in the dual theory by taking advantage of a strong-weak duality.  
We further perform a density matrix renormalization group (DMRG) calculation of an equivalent XXZ model to support our findings.

At the critical point $K = 1$, the measurement is marginal.
We identify a critical line for different measurement strength, on which the entanglement entropy exhibits a logarithmic behavior with an effective central charge continuously changing with the measurement strength. 
This is calculated by mapping the model into two decoupled transverse field Ising chains at low energies.
We further perform a fermionic Gaussian state simulation to verify our prediction.

We also discuss two feasible experimental realizations of our model. 
The first one implements the postselection via ancillas~\cite{garratt2022measurements}. 
We propose tuning the filling factor can enhance the success probability of postselection exponentially. 
The second realization uses variational quantum algorithm~\cite{mcclean2016theory} to implement the postselection.

{\it Model.---} We consider spinless fermions in a 1D chain with the Hamiltonian
\bea\label{eq:fermion_Hamiltonian}
    H = -t \sum_i (c_i^\dag c_{i+1} + h.c.) + V \sum_i (n_i - \frac12 ) (n_{i+1} - \frac12 ),~~~
\eea
where $c_i^\dag$ ($c_i$) denotes the fermion creation (annihilation) operator at site $i$, and $n_i = c_i^\dag c_i$ is the density operator.
$t$ and $V$ denote the hopping and the interaction between nearest neighbor sites, respectively.
We define $\Delta = V/t$, and set $t = 1$ without loss of generality (i.e., energy is measured in unit of $t$).
It is well known~\cite{gogolin2004bosonization} that for $|\Delta| < 1 $ the ground state is described by a free compact boson with the Luttinger parameter [see \eqref{eq:action} below] $K = \frac\pi{2(\pi - \arccos \Delta)}$.

We now consider a weak measurement to the ground state of such a Luttinger liquid. 
For the model we considered, there is one qubit (given by the occupation number of a spinless fermion) at each site.
The measurement at site $i$ is described by the following Kraus operator 
$ \{ e^{- W P_i},  \sqrt{1-e^{-2 W}} P_i \}$, where
$W \ge 0$ is the measurement strength, and $P_{2i-1} = 1-n_{2i-1}$, $P_{2i} = n_{2i}$.
With $P_i^2=P_i$, one can check that $(e^{- W P_i})^2 + (\sqrt{1-e^{-2 W}} P_i)^2 = 1$ as expected for Kraus operators.
Here $P_i$ is related to the measurement of the fermion density at site $i$, and the different definitions on the even and odd sites are because for half-filling with $2k_F=\pi$ the periodicity of measurement should match Fermion momentum $k_F$ in the low-energy theory to achieve the transition.
A physical implementation of this Kraus operator is given in Ref.~\cite{garratt2022measurements}.
Briefly speaking, we can couple each site to an ancilla.
After proper time evolution and measurement on ancilla, it will effectively induce a non-unitary transformation on each site.
We can post-select one outcome such that the effective transformation is $e^{- W P_i}$.
Because measurement operators at different sites commute, we arrive at the total measurement operator (up to an unimportant constant)
\bea \label{eq:measurement}
M  = \prod_i{e^{- W P_i}} = e^{- W \sum_i (-1)^i n_i},
\eea
and the post-selected density matrix $\rho_m = \frac{M \rho M^\dag}{\Tr[M \rho M^\dag]}$, where $\rho = \lim_{\beta \rightarrow \infty} e^{-\beta H} $ is an unnormalized projection onto the ground state.

We are interested in the entanglement properties of $\rho_m$.
To this end, using the path integral representation (see Supplemental Material for the derivation), the post-selected density matrix is proportional to $\langle \tilde\phi(x) | M \rho M^\dag | \tilde\phi'(x) \rangle = \int_\text{b.c.} D\phi e^{-S}$, with the action
\bea \label{eq:action}
    S = \int d\tau dx \left[ \frac1{2\pi K} [(\partial_\tau \phi)^2 + (\partial_x \phi)^2] + \delta(\tau) v \cos 2\phi \right],~~~
\eea
where $\phi$ is the boson field,
and the boundary condition $\phi(x,0^-) = \tilde\phi(x)$, $\phi(x,0^+) = \tilde\phi'(x)$.
Here, $|\tilde \phi \rangle$ is the state with field configuration given by $\tilde\phi$, while $\phi(x,\tau)$ is the compact boson field. Here
$v \propto W$, and the Delta function takes care of the measurement.
The last term should be considered as a sum of two terms at an infinitesimal positive and negative $\tau$, respectively.

On the other hand, at strong measurement strength $W \gg 1$, the post-selected state is close to a product state. In the following, we will first study the weak measurement case in depth, which leads to qualitatively new critical states, and defer the strong measurement case to the discussion in the end.

{\it Measurement induced transition.---} To characterize the critical state at weak measurement, we are particularly interested in
its entanglement entropy.
It can be calculated via replica trick as
$S_A = - \Tr[\rho_A \log \rho_A ] = \lim_{n \rightarrow 1} \frac1{1-n} \Tr[\rho_A^n]$,
where $\rho_A = \Tr_{\bar A} [\rho_m]$ is the reduced density matrix in the subregion $A$ (here $\bar A$ denotes the complement of $A$).
This amounts to replicating the theory $c_{i} \rightarrow c_{i,a}$, $a = 1,...,n$ is the replica index.
Because the measurement operator is bilinear in the fermionic operator, different replica momenta will decouple after we make a Fourier transform w.r.t. the replica index.
More explicitly, the measurement operator in the replicated theory is $M = e^{-W \sum_{i,a} (-1)^i n_{i,a}} $ (with abuse of notation, we use the same symbol $M$).
The replica Fourier transform is defined by $c_{i,k} = \frac1{\sqrt n} \sum_{a} c_{i,a} e^{i \frac{2\pi k a}n} $, with $k$ the replica momentum.
In the replica momentum basis, $M = \prod_{k= -\frac{n-1}2}^{\frac{n-1}2} M_k$, and $ M_k = e^{-W \sum_{i} (-1)^i n_{i,k}}$ can be straightforwardly bosonized to get $\delta(\tau) v \cos 2\phi_k$.
Combining the quadratic term originated from the pre-measured Hamiltonian, we arrive at the following decoupled action for each replica momentum $k$ (see Supplemental Material for details),
\bea \label{eq:replica_action}
    s_k = \frac1{\pi K} \int \frac{dq}{2\pi} |q| |\phi_k(q)|^2 + v \int dx  \cos 2\phi_k(x).
\eea
Here we have further integrated over the time direction.
$q$ denotes the momentum from Fourier transform $\phi_k(q) = \int dx \phi_k(x) e^{i q x}$, and $\phi_k(x) = \phi_k(x,\tau = 0)$.

The entanglement entropy $S_A$ of compact bosons boils down to the expectation value of the twist operator~\cite{cardy2009entanglement,calabrese2009entanglement} $T_{A} = \prod_{k= -\frac{n-1}2}^{\frac{n-1}2}  T_{A,k}$ with
$T_{A,k} = e^{-i \frac{k}{n} \sqrt{\frac{4}{K}} (\phi_k(x_A) - \phi_k(0))}$,
where we assume that the interval $A = \{ x| x \in (0, x_A)\}$, and take the replica limit.

We first discuss the renormalization group (RG) flow at low energies to determine the phase diagram.
In~(\ref{eq:replica_action}), the momentum $k$ is a dummy index because different replica momenta decouple; therefore, the RG equation for $v$ is the same as that of a single replica.
It is given by~\cite{kane1992transport,altland2010condensed, garratt2022measurements} $\frac{d v}{d l} = (1-K) v$.
On the other hand, $K$ is exactly marginal ($\frac{d K}{dl}=0$) because the first term in (\ref{eq:replica_action}) is non-analytical that does not receive a correction from the RG process.
Heuristically, the measurement term is only present at $\tau = 0$, so it cannot renormalize $K$ that is present in 1+1D.
Combining these two facts, the flow of $v$ is simple: it is relevant (irrelevant) for $K<1$ ($K>1$), and marginal at $K=1$.

For $K>1$, $v$ is irrelevant, so the entanglement entropy of $A$ after measurement reduces to that of a free boson with central charge $c=1$, $S_A = \frac{1}3 \log x_A$.
For $K<1$, $v$ is relevant, we will show that its entanglement structure changes qualitatively after weak measurement. 

To study the entanglement entropy for $K<1$, because $v$ is relevant, it is easier to work with the dual field~\cite{altland2010condensed,gogolin2004bosonization} $\theta$, defined via $[\partial_x \phi(x), \theta(x')] = i \pi \delta(x-x')$.
We give a detailed derivation of the dual field theory in Supplemental Material.
The dual action reads $s_k = \frac{K}{4\pi} \int \frac{dq}{2\pi} |q| |\theta_k(q) |^2 + \gamma \int dx \cos \theta_k(x)$, where $\gamma = 2 e^{-a v - 4 \sqrt{v}}$ ($a$ is a constant, whose expression is given in Supplemental Material), and $k$ denotes the replica momentum.
When $v$ flows to a large number, it means $\gamma$ is small, so we can perform a perturbation calculation.
Thanks to decoupling between different replica momenta, $\langle T_A \rangle = \prod_k \langle T_{A,k} \rangle $, where $\langle \cdot \rangle = \Tr [ \cdot \rho_m^{\otimes n}]$.
At the leading order, we arrive at (see Supplemental Material for detail)
$
    \langle T_{A,k} \rangle = \exp\left[ \gamma^2 f_k(K) x_A^{- \frac2{K}+2} \right]
$.
Finally, taking the replica limit, the entanglement entropy for $K<1$ reads
\bea \label{eq:algebraic_ee}
    S_A = \gamma^2\left( S_0 +   f(K) x_A^{- \frac2{K}+2} \right).
\eea
Here, the expressions for $f_k(K)$ and $f(K)$ are given in Supplemental Material.
The first term is a non-universal constant that accounts for the leading area-law contribution.
Therefore, the entanglement entropy for $K<1$ shows an area law with a subleading power-law behavior.

\begin{figure}
    \centering
\subfigure[]{
    \includegraphics[width=0.23\textwidth]{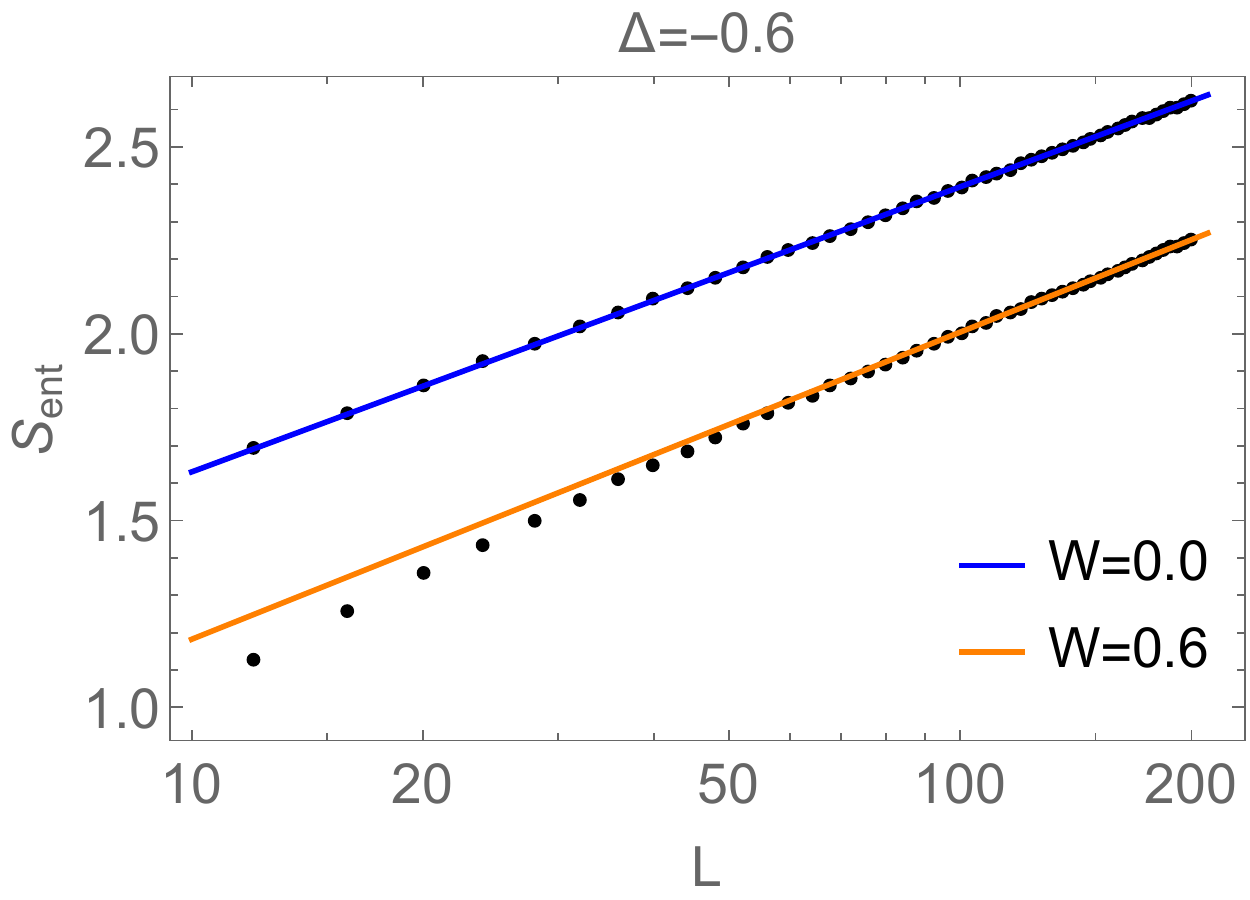}}
\subfigure[]{
    \includegraphics[width=0.23\textwidth]{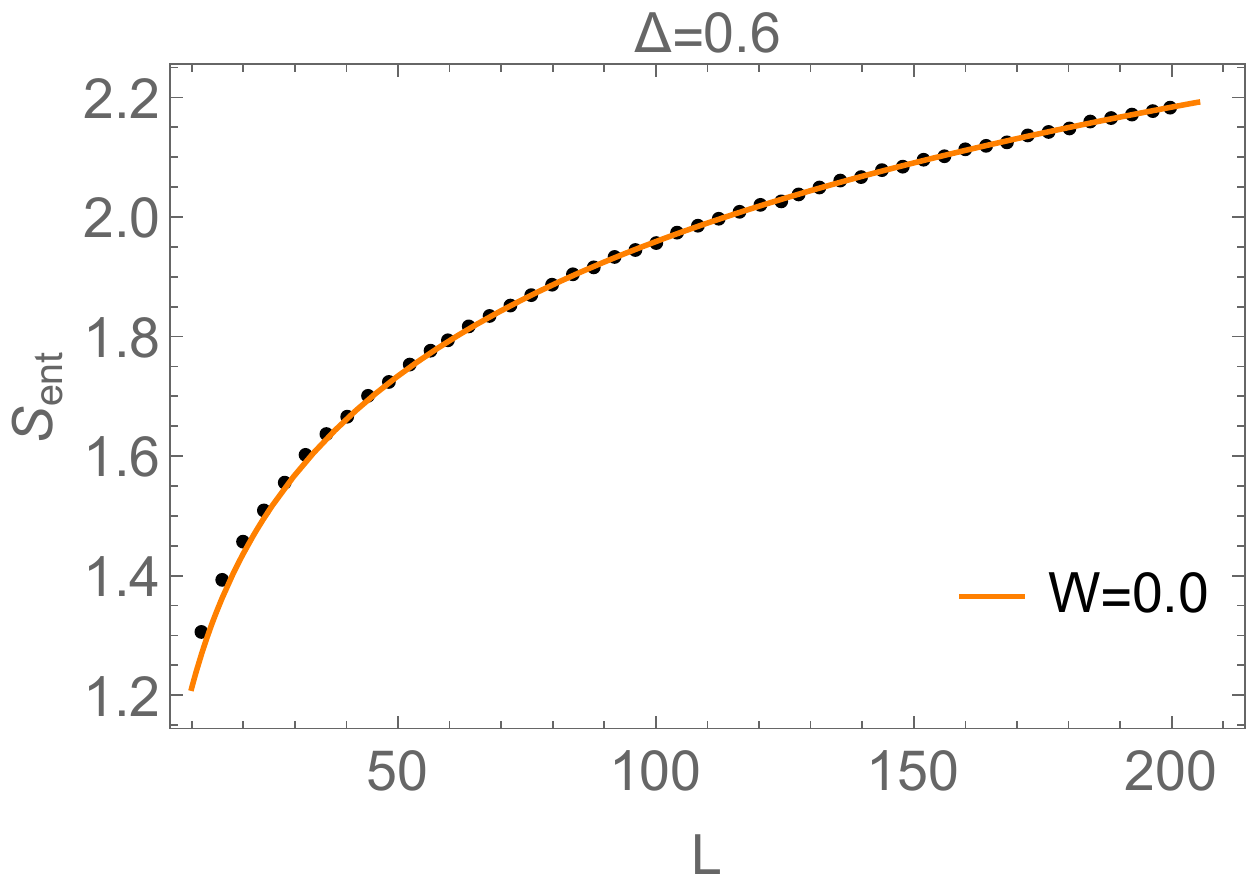}}
\subfigure[]{
    \includegraphics[width=0.23\textwidth]{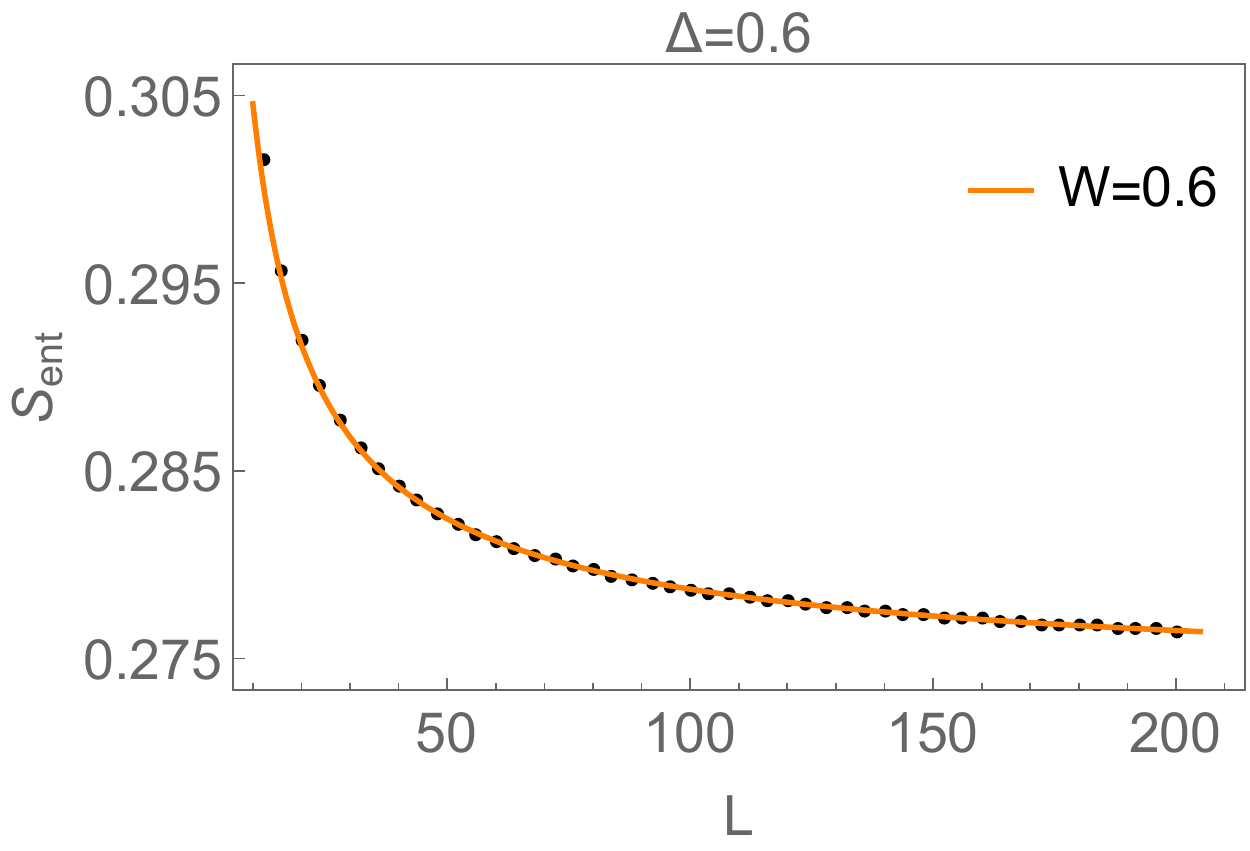}}
\subfigure[]{
    \includegraphics[width=0.23\textwidth]{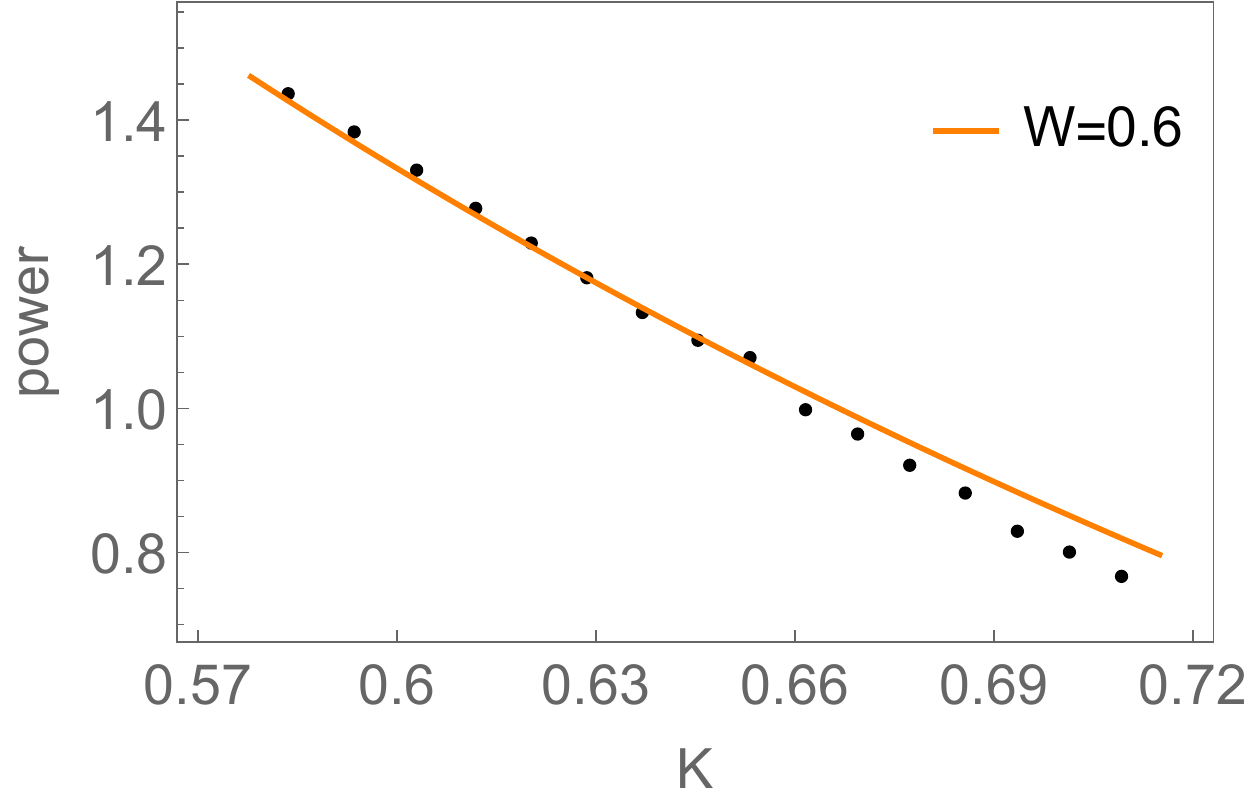}}
    \caption{The half-chain entanglement entropy as a function of different sizes
    at (a) $\Delta = -0.6$, (b,c) $\Delta = 0.6$. 
    The black dots are numerical results and the colored lines are fitting curves.
    (a) The blue (orange) curve is given by measurement strength $W=0$ ($W=0.6$).
    The same slope indicates the effective central charges are the same.
    (b) shows the data at measurement strength $W=0$.
    It shows a logarithmic function with central charge $c=1$.
    (c) shows the data at measurement strength $W=0.6$.
    It shows an algebraic function with power $0.77$.
    (d) The algebraic power as a function of different $K$ for $K<1$.
    The measurement strength is $W=0.6$.
    The black dots show the power fitted by numerical data.
    The orange curve is our prediction $2/K-2$.
    The numerical calculation is obtained with bond dimension $\chi = 100$.}
    \label{fig:noncritical_ee}
\end{figure}

Therefore, we can conclude a measurement-induced entanglement transition (MIPT) between two phases with a logarithmic entanglement and a (subleading) algebraic entanglement, respectively.
To further support our conclusion, we perform DMRG calculation~\cite{hauschild2018efficient} of entanglement entropy for $\rho_m$.
Details of the simulation can be found in Supplemental Material.
For $K>1$, Fig.~\ref{fig:noncritical_ee}(a) shows the entanglement entropy  for $W=0$ (blue) and $W=0.6$ (orange).
The same slope $1/3$ indicates that the central charge of both cases is $c=1$, and the measurement is irrelevant.
For $K<1$, Fig.~\ref{fig:noncritical_ee} shows that the entanglement entropy w.r.t system sizes is a logarithmic function for $W=0$ (b), an algebraic function for $W=0.6$ (c).
It demonstrates that the measurement is relevant for $K<1$ and changes the entanglement entropy from a logarithm law without measurement to an area law with subleading algebraic correction with measurement.
In Fig.~\ref{fig:noncritical_ee} (d), the black dots represent the powers of the entanglement entropy fitted from numerical data for $W>0$ and the orange curve is our prediction $-2/K+2$.
We can see that they are consistent.

{\it Critical point.---} We now discuss the entanglement entropy at the critical point $K=1$, where the measurement parameter $v$ is marginal.
In this case, the interaction $V$ vanishes, and the model reduces to a free fermion theory with measurement~(\ref{eq:measurement}) at $\tau =0$. 
In the continuum limit, the theory is described by a Dirac fermion, $H = \int {\rm d}x  \psi^\dagger (-{\rm i} v_F \partial_x \sigma_z) \psi$,
where $\psi=(\psi_{L}, \psi_{R})^T$ are the left and right movers at $k=\pm \pi/2$ respectively, and the measurement is $M=e^{h}$, where $h =\int{\rm d}x\ \psi^\dagger(W\sigma_x) \psi$.
The measurement~(\ref{eq:measurement}) effective creates a potential that can scatter between the left and right movers, and behaves like a mass term. 
It is easy to see that there is another mass term that can be related to the above one via chiral transformation, $h_1 =\int {\rm d}x \ \psi^\dagger(W \sigma_y) \psi$.
In the following, we will construct a corresponding measurement that can lead to the effective mass term~$h_1$, and moreover, can be mapped to two decoupled transverse field Ising models.
This will enable us to obtain the subsystem entanglement entropy exactly. 
Since in the low-energy limit, these two effective mass terms are related by a local emergent symmetry, the subsystem entanglement entropy from these two measurements will also agree.

First, we map our model to a spin model via the Jordan-Wigner transformation,
$\sigma^{+}_l=c_l^\dagger e^{{\rm i}\pi \sum_{j<l}n_j}$, $\sigma^-_l=e^{-{\rm i}\pi \sum_{j<l}n_j} c_l$, $\sigma^z_l=2c_l^\dagger c_l-1$.
For $K=1$, we arrive at the XX model,
$H_\text{XX} =  -\sum_i (\sigma^x_i \sigma^x_{i+1} + \sigma^y_i \sigma^y_{i+1})$, where $\sigma^\alpha_j, \alpha=x,y,z$ is the Pauli operator at site $i$, and the measurement is $M=e^{- \frac12 W \sum_i (-1)^i \sigma_i^z}$ which is equivalent to \eqref{eq:measurement}.
Now, it is easy to check that the following measurement operator can lead to the effective mass term~$h_1$
\bea \label{eq:measurement_XXYY}
M_1=
e^{\frac{1}{2}W \sum_i (\sigma_{2i}^x \sigma_{2i+1}^x+\sigma_{2i}^y \sigma_{2i+1}^y)}.
\eea

Using Majorana representation~${\gamma}_l^1=c_l^\dagger+c_l, {\gamma}_l^2=(c_l-c_l^\dagger)/{\rm i}$, the XX model and different measurements are decoupled to two quantum Ising chains. 
The Hamiltonian reads 
\begin{equation}
\begin{split}
\label{eq:decoupled XXZ model}
    H_{\rm XX}=\sum_i &\left[({\rm i}\gamma_{2i}^2\gamma_{2i+1}^1+{\rm i}\gamma_{2i+1}^1\gamma_{2i+2}^2)\right.\\
    -&\left.({\rm i}\gamma_{2i}^1\gamma_{2i+1}^2+{\rm i}\gamma_{2i+1}^2\gamma_{2i+2}^1)\right],
\end{split}
\end{equation}
where the first and second line commute  and each of them is equivalent to a transverse field Ising model.
The measurement is $M_1 = e^{\frac{1}2 W \sum_i {\rm i}  \left( \gamma_{2i}^2 \gamma_{2i+1}^1 - \gamma_{2i}^1 \gamma_{2i+1}^2 \right)}$,
where 
each of them effectively corresponds to a weak measurement for the first and second transverse field Ising model~\cite{yang2023entanglement}.
Therefore, the subsystem entanglement entropy for the ground state of the XX model with the measurement~\eqref{eq:measurement_XXYY} is $S_A = \frac{c_\text{eff}}3 \log x_A$ with~\cite{yang2023entanglement,eisler_solution_2010,brehm2015entanglement}
\bea
\label{eq:effective_central_charge}
    c_{\rm eff} &=&-\frac{6}{\pi^2} \Big\{ \left[(1+s)\log(1+s)+(1- s)\log(1-s)\right]\log(s) \nn \\
    && +(1+s){\rm Li}_2(-s)+(1-s){\rm Li}_2(s) \Big\},
\eea
where $s=\frac{1}{\cosh{2W}}$ and ${\rm Li}_2(z)=-\int_0^z {\rm d}x \frac{\ln{(1-x)}}{x}$ is the dilogarithm function.
This also predicts the effective central charge from our original measurement~\eqref{eq:measurement}.

\begin{figure}
    \centering
\subfigure[]{    \includegraphics[width=0.23\textwidth]{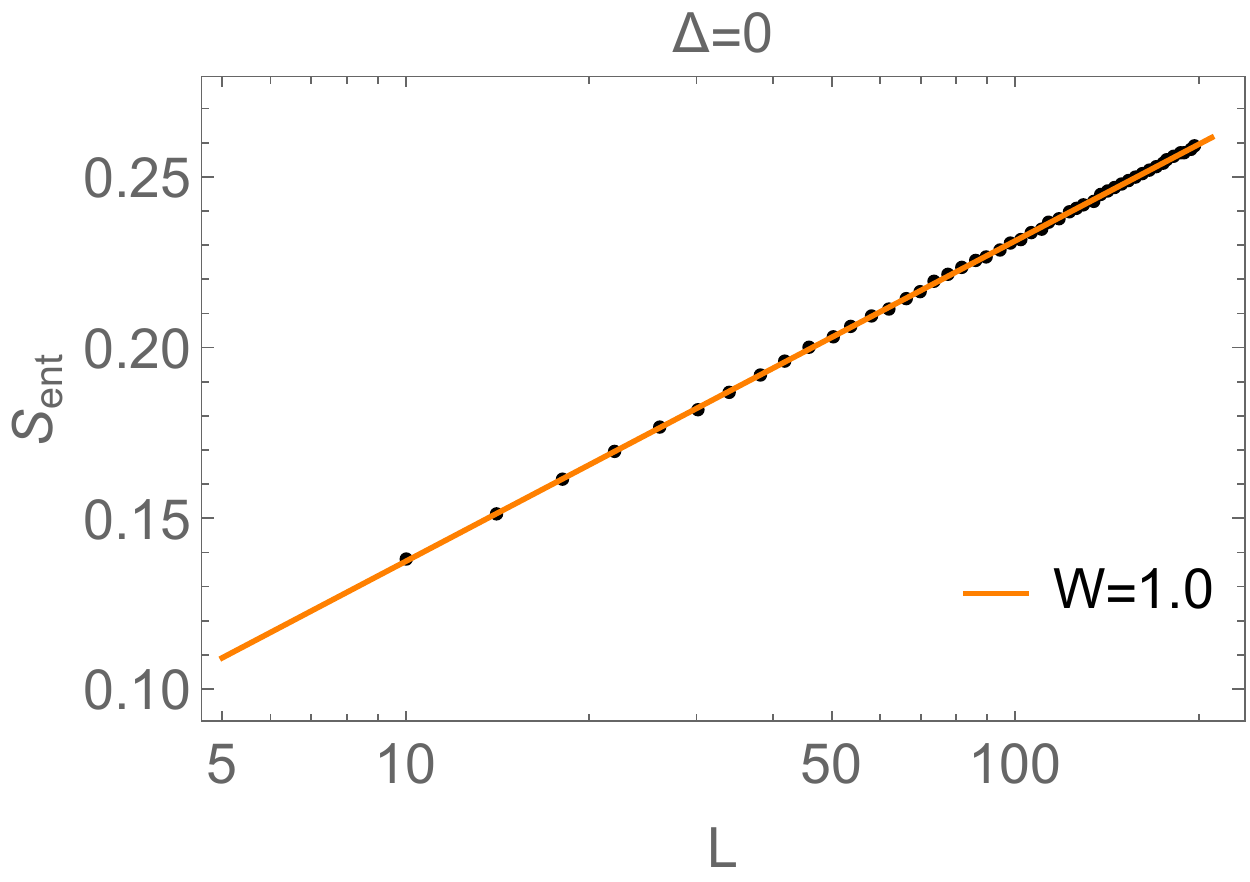}}
\subfigure[]{    \includegraphics[width=0.23\textwidth]{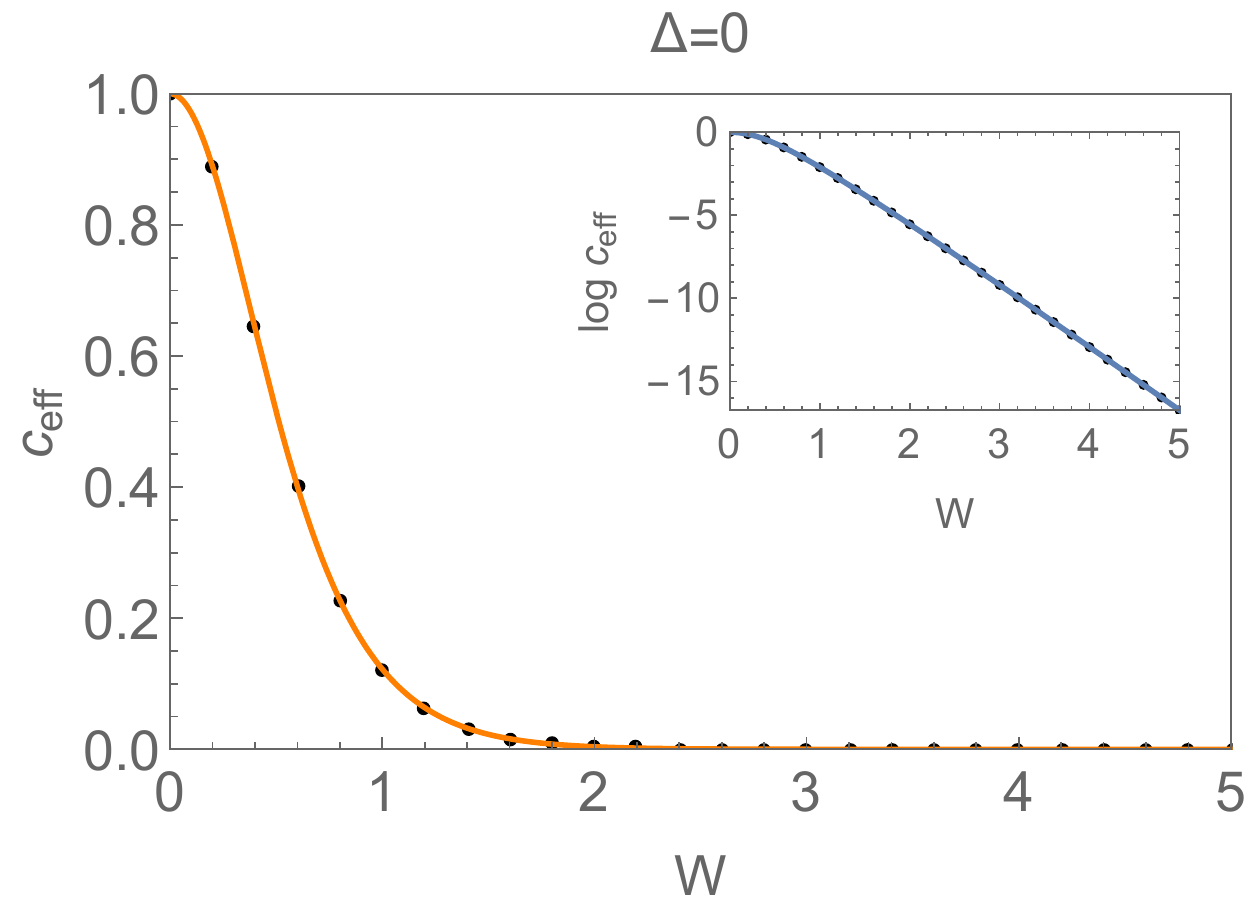}}
    \caption{(a) The half-chain entanglement entropy at the critical point as a function of different system sizes $L$.
    The parameter is chosen to be $t=1$, $V = 0$, $W=1$.
    The black dots represent the numerical data, and the orange line is a fitting with $ 0.04 \log L+0.0435$.
    (b) The effective central charge as a function of different measurement strength $W$.
    The black dots represent effective central charge that is extracted from fitting the numerical data.
    The solid curve is our prediction $c_\text{eff}$.
    In the inner figure, we plot $c_{\rm eff}$-$W$ on a log-linear scale.}
    \label{fig:critical_ee}
\end{figure}

To support our analytical calculation, we perform a Gaussian state simulation~\cite{surace2022fermionic} to calculate half-chain entanglement entropy for different sizes.
Fig.~\ref{fig:critical_ee}(a) shows the half-chain entanglement entropy at critical point, $t=1, V=0, W=1$, for different size $L$.
It exhibits a logarithmic behavior with an effective central charge that deviates from one.
In Fig.~\ref{fig:critical_ee}(b), we show the effective central charge extracted from fitting the numerical data in black dots, and our prediction in a solid curve.

Besides the two effective mass terms that give rise to the same subsystem entanglement entropy, there is one more mass term given by a $p$-wave superconductivity. 
In Nambu space with basis $\Psi=(\psi_R, \psi_L, \psi_R^\dagger,\psi_L^\dagger)^T$ and Pauli matrix $\mu_i$, it is
$h_2=\int_x{\rm d}x\ \Psi^\dagger(-W \sigma_y\mu_x/2)\Psi$.
It is easy to construct a measurement operator that can lead to the effective mass term~$h_2$, $M_2 = e^{ \frac12 W \sum_i \sigma^x_i \sigma^x_{i+1}}$.
We show in the Supplemental Material that all three measurement operators $M$, $M_1$, $M_2$ can induce the same effective central charge given in~\eqref{eq:effective_central_charge}~\footnote{The three measurement operators are equivalent only at $K=1$.}.
We should note that continuous effective central charge has been studied extensively in the context of defect/interface conformal field theory~\cite{eisler_solution_2010,sakai2008entanglement,brehm2015entanglement}.

\begin{figure}
    \centering
    \subfigure[]{    \includegraphics[width=0.22\textwidth]{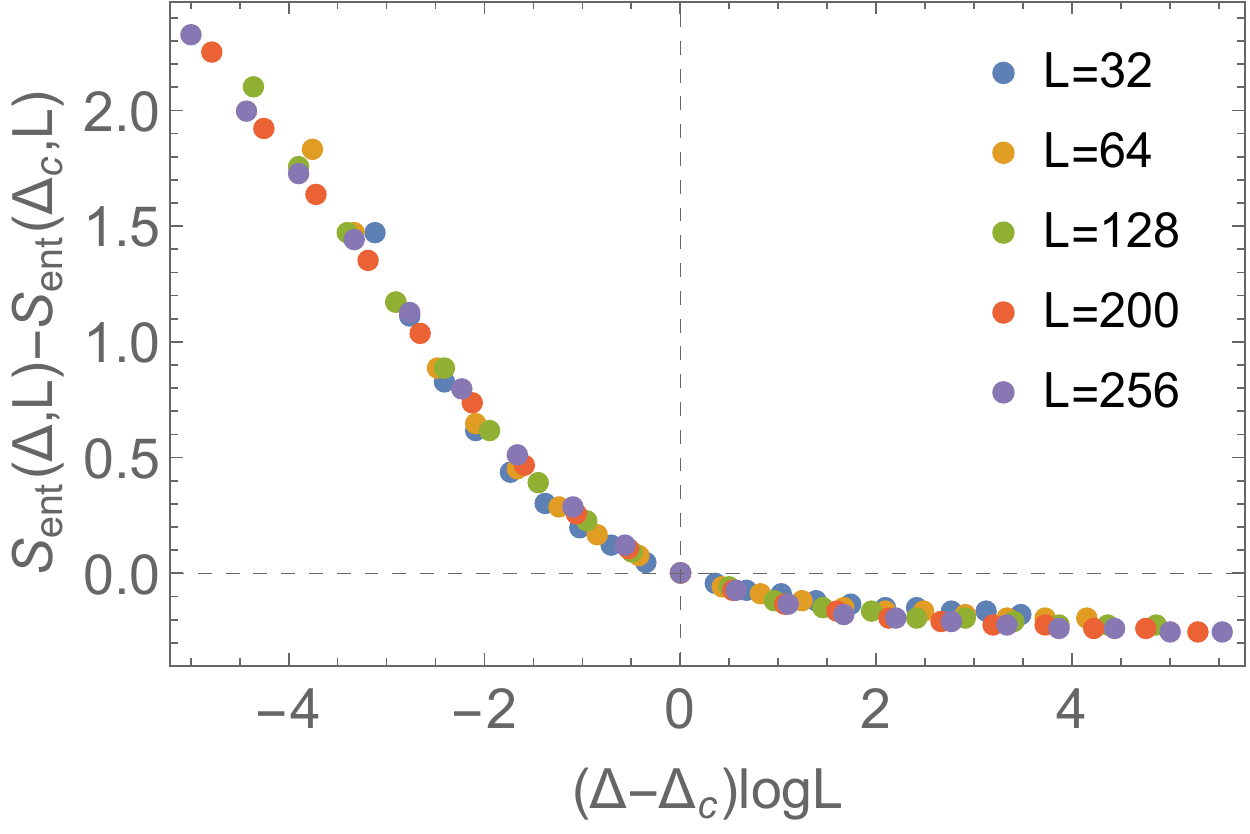}}
    \subfigure[]{    \includegraphics[width=0.23\textwidth]{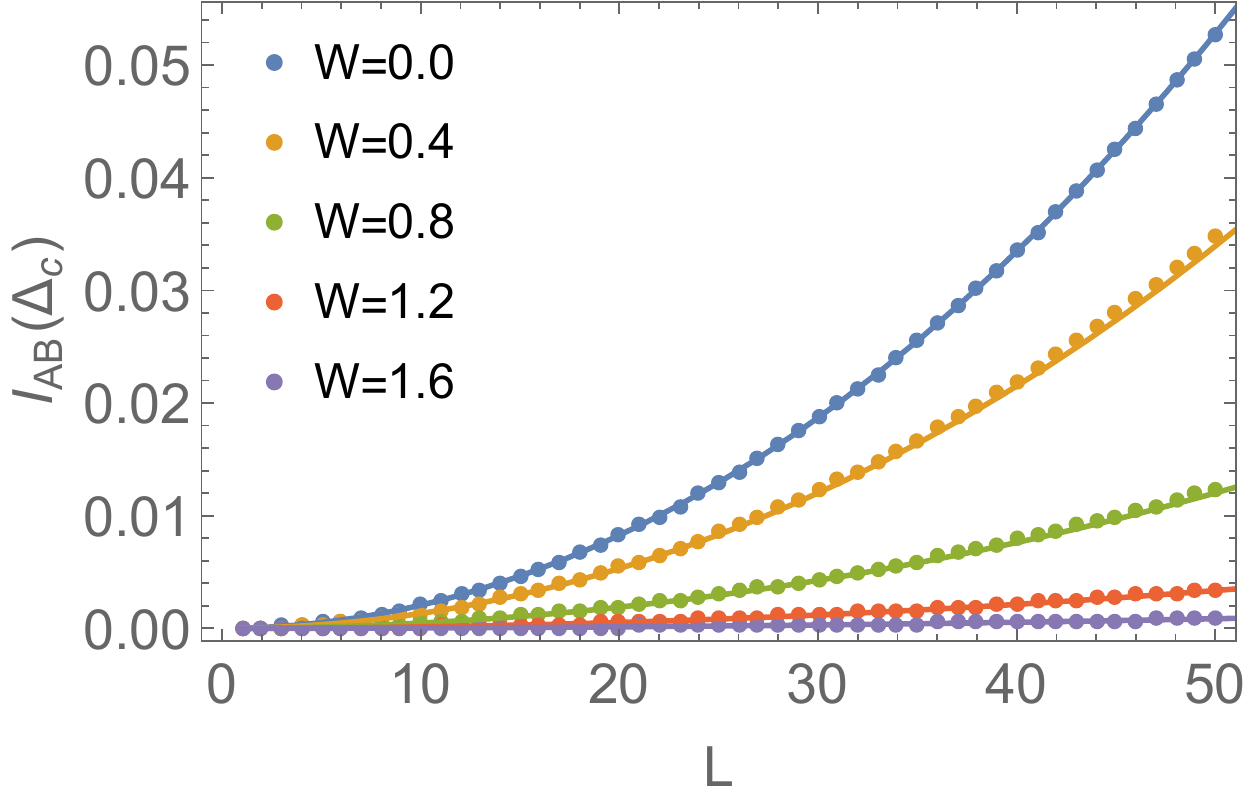}}
    
    \caption{(a) Half-chain entanglement entropy as a function of $\Delta$ for different sizes $L$. 
    The measurement strength is $W=1.0$.
    (b) Mutual information $I_{\rm AB}$ as a function of $L$ at critical point for different measurement strength $W$. The dots (curves) are numerical results (analytical predictions).
    }
    \label{fig:transition}
\end{figure}

{\it Critical exponent.---}The distinct behaviors of entanglement entropy for $\Delta > 0$ and $\Delta < 0$ clearly reveal an entanglement transition at $\Delta_c = 0$.
Unlike conventional transition, the scaling dimension of the tuning parameter is zero at the critical point, thus $1/\nu = 0$.
In Fig.~\ref{fig:transition} (a), we plot half-chain entanglement entropy as a function of $\Delta$ for different sizes $L$.
All data collapse onto a smooth function when the argument is chosen to be $(\Delta-\Delta_c) \log L$, indicating that $1/\nu = 0$.

We also calculate the critical exponent $\eta$ via mutual information.
To this end, we divide the chain of length $L_{\rm tot}$ (assumed to be an even integer) with periodic boundary condition into four parts A, B, C and D, $[0,L], [\frac{L_{\rm tot}}{2},\frac{L_{\rm tot}}{2}+L], [L,\frac{L_{\rm tot}}{2}]$ and $[\frac{L_{\rm tot}}{2}+L,L_{\rm tot}]$.
The mutual information is defined as $I_{\rm AB}=S_{\rm A}+S_{\rm B}-S_{\rm AB}$.
Without measurement, the mutual information of free fermion is $I_{\rm AB}=-\frac{c}{3}\log{[\cos^2{(\pi L/L_{\rm tot})}]}$ where $c=1$ \cite{furukawa2009mutual}.
With finite measurement, we have an effective central charge $c_{\rm eff}$~\eqref{eq:effective_central_charge}. 
Because $c_{\rm eff}$ is related to the scaling dimension of twist operators, 
the mutual information that is given by the correlation function of twist operators \cite{li2019measurement} accordingly changes to $I_{\rm AB}=-\frac{c_{\rm eff}}{3}\log{[\cos^2{(\pi L/L_{\rm tot})}]}$.
In Fig.~\ref{fig:transition} (b), we plot the mutual information $I_{\rm AB}$ for different measurement strength, where the dots (curves) are numerical results (analytical prediction).
For $L/L_{\rm tot}\rightarrow0$, we have $I_{\rm AB}\propto (L/L_\text{tot})^\eta$, $\eta=2$.
Actually, we can fit the exponent $\eta$ and the prefactor that is related to $c_{\rm eff}$ using the numerical data for $L/L_{\rm tot}\rightarrow0$. 
And both are consistent with the results above.

{\it Experimental realization.---}We discuss the feasibility of experimental realization of our work. 
Because our results are valid for a general filling factor of the fermion model~\eqref{eq:fermion_Hamiltonian} with compatible measurement operators that can induce the scattering process between left and right movers at low energies, we can tune the filling factor to increase the success probability of each run of the experiments.
From Ref.~\cite{garratt2022measurements}, it is possible to design a protocol of post-selection whose Born-rule probability at site $j$ is $p_j=1-(1-e^{-2W_j}) n$, where $0 \le n \le 1/2$ is the filling factor~\footnote{Here we only consider filling factor not greater than $1/2$ for simplicity. The case with $n > 1/2$ is equivalent upon a particle hole transformation of both the model and the measurement operator.}, and $W_j$ is measurement strength compatible with the filling factor. 
For an estimate, using $p_j > 1-n$, the probability of one successful experiment is $P=\prod_j p_j > (1-n)^L$,
which means that a lower filling factor will increase the probability exponentially.
In a concrete experimental setting, one can further design a suitable $W_j$ to achieve a better success probability.
In Supplemental Material, we give an example of a filling factor $n=1/4$ that achieves a success probability exponentially greater than the half-filling case in a model with the same length. 
We further confirm that it has the same transition with exponents $1/\nu=0, \eta=2$. 
This tunability in our model will be crucial for NISQ devices~\cite{preskill2018quantum}.

Finally, the measurement and post-selection discussed in our paper can in principle be realized in a quantum computer using imaginary-time evolution.
For instance, variational quantum algorithm can be implemented to prepare the state after imaginary-time evolution~\cite{mcclean2016theory, yuan2019theory}. 
In Supplemental Material, we demonstrate that the interesting physics and transition of our model can indeed be realized using variational quantum algorithm. 
This opens up a great oppotunity for investigation of postselection physics in quantum computers.

{\it Discussion and outlook.---} We have studied of the resulting state after weak measurement.
When the measurement is strong, $W \gg 1$, it is expected to approach a projective measurement at $W \rightarrow \infty$.
In this case, naively the Luttinger liquid theory is not an appropriate starting point, since measurement can insert high energy into the state.
Remarkably, for $K<1$, in the strong-weak duality, the strong measurement strength indicates $\gamma \rightarrow 0$, our formula~(\ref{eq:algebraic_ee}) predicts vanishing entanglement entropy, which is consistent with a product state. 
At $K=1$, our result for the critical point works well for large $W$, as shown in Fig.~\ref{fig:critical_ee}.
Moreover, our result suggests that the critical state at $K=1$ induced by measurement is described by two copies of interface Ising CFT of the continuous Dirichlet type~\cite{oshikawa1997boundary} with $|\sin 2\phi| = \frac1{\cosh(2W)}$.
We leave a detailed study of the correspondence between measurement and boundary/interface CFT to the future work.

We mention a few open questions.
At $\Delta = 1$, the theory has a larger SU(2) symmetry, though the measurement explicitly breaks it.
It would be an interesting future question to investigate the effect of measurement at this special point.
Criticality under measurement in higher dimensions is a natural extension~\cite{lee2023quantum}.
Moreover, it would be interesting to investigate the resulting state after measurement without postselection~\cite{garratt2022measurements, zou2023channeling}. 
Finally, measurement effect is currently under investigation in the context of holographic duality~\cite{takayanagi2011holographic, fujita2011aspects, numasawa2016epr, antonini2022holographic, jian2022holographic, milekhin2022measurement}.
It would also be interesting to develop a holographic description for general measurement.

{\it Acknowledgement.---} We would like to thank Shi-Xin Zhang for helpful discussions. 
We have used the TeNPy package for the DMRG simulation~\cite{hauschild2018efficient}, the  F$\_$utilities package for the fermionic Gaussian state simulation~\cite{surace2022fermionic}, and the TensorCircuit package for the variational quantum algorithm demonstration~\cite{zhang2023tensorcircuit}.
This work is supported in part by 
the MOSTC under Grant
No. 2021YFA1400100 and by NSFC under Grant No. 11825404 (X.S. and H.Y.). S.-K.J is supported by a startup fund at Tulane University.

{\it Note added:} After we finished the first version, we became aware of two independent and related works~\cite{yang2023entanglement,weinstein2023nonlocal}. 
While they studied the transverse field Ising model, we focused on the Luttinger liquid.

\bibliography{references.bib}

\newpage
\onecolumngrid

\setcounter{secnumdepth}{3}
\setcounter{equation}{0}
\setcounter{figure}{0}
\renewcommand{\theequation}{S\arabic{equation}}
\renewcommand{\thefigure}{S\arabic{figure}}
\renewcommand\figurename{Supplementary Figure}
\renewcommand\tablename{Supplementary Table}
\newcommand\Scite[1]{[S\citealp{#1}]}
\makeatletter \renewcommand\@biblabel[1]{[S#1]} \makeatother

\section*{Supplemental Material}

\subsection{Twist operator in 1+1D free fermion theory and compact boson theory}

In this section, we discuss the twist operator in 1+1D free fermion theory and compact boson theory, which is used to calculate the entanglement entropy.
Recall that the entanglement entropy can be evaluated by replica trick, e.g., the entanglement entropy of subregion $A$ reads
\bea
    S_A = - \Tr[\rho_A \log \rho_A ] = \lim_{n \rightarrow 1} \frac1{1-n} \Tr[\rho_A^n],
\eea
where $\rho_A = \Tr_{\bar A}[\rho]$ is the reduced density matrix of $A$.
$\bar A$ denotes the complement of subregion $A$.
In the path integral approach, we make $n$ replicas of the original systems, $a=1,...,n$, and insert twist operators in the subregion $A$ to change the boundary condition.
Then we can evaluate the path integral for $n$ after which, a continuation of $n$ to a real number is taken followed by the replica limit $n\rightarrow 1$.
In the following two subsections, we discuss the twist operators in free fermion theory and compact boson theory, respectively.

\subsubsection{Free fermion theory}
\label{sec:Free Field Twist Operators}

Let $c_{i,a}$, $c_{i,a}^\dag$ be the spinless fermion operator at site $i$ and replica $a$. 
It satisfies $\{c_{i,a}, c_{j,b}^\dag\} = \delta_{ij} \delta_{ab}$
The twist operator for a subregion $A$ is defined as 
\bea
    T^\dag_A c_{i,a} T_A = \begin{cases}
        c_{i,a} \quad & i \in \bar A \\
        c_{i,a+1} \quad & i \in A, a < n \\
        (-1)^{n+1} c_{i,1} \quad & i \in A, a = n
    \end{cases}.
\eea
The sign $(-1)^{n+1}$ is due to the anticommutation of fermion operators~\cite{casini2005entanglement}. 
This sign can be accounted by making the transformation $c_{i,a} \rightarrow (-1)^a c_{i,a}$.
Then the twist operator $T_A$ is similar to a translation in the replica space. 
Therefore, we can make a Fourier transform in the replica space
\bea \label{eq:fourier}
    c_{i,k} = \frac1{\sqrt{n}} \sum_{a=1}^n c_{i,a} e^{-i \frac{2\pi k a}n},
\eea
with $k$ denotes the replica momentum. 
In the replica momentum basis, the action of twist operator is diagonal, namely,
\bea \label{eq:twist_fermion}
    T^\dag_A c_{i,k} T_{A} = 
    \begin{cases}
        e^{i \frac{2\pi k}n} c_{i,k} \quad & i \in A \\
        c_{i,k} \quad & i \in \bar A
    \end{cases}.
\eea
It is then not hard to deduce the twist operator: 
\bea
    T_A = \prod_{k=-(n-1)/2}^{(n-1)/2} T_{A,k}, \quad T_{A,k} = e^{i \frac{2\pi k}n Q_{A,k}}, \quad Q_{A,k} = \sum_{i\in A} c_{i,k}^\dag c_{i,k}.
\eea

In free fermion theory, different replica momenta decouple, and they are described by a same theory. 
In this case, we can omit the replica momentum index in $Q_{A,k}$.
Then taking the continuum limit, we arrive at
\bea
    Q_{A,k} = \int_{x\in A} dx \psi^\dag_{k}(x) \psi_k(x).
\eea
If we omit the dummy replica momentum index in free fermion theory, we arrive at~(13) in the main text.

\subsubsection{Free compact boson theory}

In 1+1D boson field theory, the entanglement entropy of an interval is related to the two-point correlation function of branch-point twist operator~\cite{calabrese2004entanglement,cardy2008form,calabrese2009entanglement}.
In the following, we simply call the branch-point twist operator, denoted as $\mathcal T_n(x,\tau)$, the twist operator. 
Similar to~(\ref{eq:fourier}), we make a Fourier transformation in replica space which diagonalizes the twist operator, 
\bea
     \phi_{k}(x) = \frac1{\sqrt{n}} \sum_{a=1}^n \phi_{a}(x) e^{-i \frac{2\pi k a}n},
\eea
where $\phi_{a}(x)$ is the boson operator at replica $a$.
Denote the twist operator at replica momentum $k$ as $\mathcal T_{nk}$, which satisfies
\bea
    \mathcal T_{nk}(u,0) \phi_{k}(v) = 
    \begin{cases}
    e^{i \frac{2\pi k}{n}} \phi_k(v), \quad & u< v \\
    \phi_{k}(v), \quad & u > v
    \end{cases},
\eea
the twist operator is $\mathcal T_n = \prod_{k} \mathcal T_{nk} $. 
Note that we use a tilde to denote the branch-point twist operator. 
It is a local operator, which should be contrasted with the twist operator $T_A$ defined in~(\ref{eq:twist_fermion}).
But they are related by $T_A = \mathcal T_{nk}(0,0) \mathcal T_{nk}(x_0,0)$ with $A = \{ x | x \in (0, x_0) \}$.
Notice that in the Supplement Material, we use $x_0$ instead of $x_A$ to denote the length of subregion $A$.
We will see this relation more explicitly in the following.

In the free boson theory, different replica momenta decouple.
It is not hard to check that the twist operators are given by
\begin{equation}
\label{twist operators}
    \mathcal{T}_n(u,0) \mathcal{T}^{-1}_n(v,0)
    =\prod_k \mathcal{T}_{nk}(u,0) \mathcal{T}^{-1}_{nk}(v,0)
    =\prod_k e^{-i\frac{2k}{n}(\phi_k(u)-\phi_k(v))},
\end{equation}
where $n$ and $k$ label the number of replica copies and replica momentum, respectively, and $(u, v)$ is the interval of the system we choose to calculate entanglement entropy. 
Here $\mathcal{T}^{-1}$ denotes the anti-twist operator.

We can relate the twist operator in the free fermion theory to the free boson theory by bosonization~\cite{gogolin2004bosonization},
\bea
    Q_{A,k} =  \int_{0}^{x_0} dx \left[- \frac1\pi \nabla\phi_k(x) \right] = \frac1{\pi} \left(\phi_k(x_0) - \phi_{k}(0) \right).
\eea

The interacting spinless fermion model considered in the main text~(1) is
\bea
    H = -t \sum_i (c_i^\dag c_{i+1} + h.c.) + V \sum_i (n_i - \frac12 ) (n_{i+1} - \frac12 ),
\eea
This model is described by the free compact boson theory, the Luttinger liquid theory.
A nontrivial $V \ne 0$ only changes the radius of the compact boson field, or equivalently the Luttinger parameter~\cite{gogolin2004bosonization},
\bea
    K = \frac\pi{2(\pi - \arccos \Delta)}.
\eea
The twist operator should accordingly be modified to be~\cite{calabrese2009entanglement}
\bea \label{eq:rescaled_twist_operators}
    \mathcal{T}_{nk}(u,0) \mathcal{T}^{-1}_{nk}(v,0)
    = e^{-i\frac{2k}{n} \frac1{\sqrt K}(\phi_k(u)-\phi_k(v))},
\eea
which is (8) in the main text.
We should note that (\ref{eq:rescaled_twist_operators}) only works for the entanglement entropy of a single interval. 
For multi disjoint intervals, the compactness of boson field will couple different replica momenta, and more complicated technique is needed~\cite{calabrese2009entanglement}.

\subsection{Effective field theory in 1+0D with measurement}

Without measurement, the bosonized action of Hamiltonian~(1) reads 
\begin{equation}
\label{Luttinger liquid model}
    S[\phi]=\frac{1}{2\pi K}\int {\rm d}x\int_0^\beta {\rm d}\tau \left[ \Dot{\phi}^2+(\nabla \phi)^2 \right].
\end{equation}
where th Luttinger parameter is related to the interaction through~(2)~\cite{gogolin2004bosonization}.
For the measurement operator $\hat{M}=e^{-W\sum_i(-1)^in_i}=e^{-W\int {\rm d}x \cos{(2k_F x)}n(x)}$, where each site is $x=n\pi/2k_F, n\in \mathbb{Z}$.
Using the bosonization of fermion density operator~\cite{garratt2022measurements}
\begin{equation}
    n(x)=-\frac{1}{\pi}\nabla\phi(x)+\frac{1}{\pi}\cos{[2(k_F x-\phi)]},
\end{equation}
we have $\hat{M}=e^{-W/2\pi\int {\rm d}x \cos{(2\phi)}}$, where we keep the leading term and neglect higher-order terms like $\cos{[2(nk_F x+\phi)]}$ with $n\neq0$.
Here, notice that $\phi(x) = \phi(x,\tau = 0)$.

The density-density correlation for the post-selected state is $\left<n(x)n(0)\right>={\rm Tr}[\hat{M}\rho\hat{M}^\dagger n(x)n(0)]/{\rm Tr}[\hat{M}\rho\hat{M}^\dagger]$.
Since $\left<n(x)n(0)\right>$ commutes with the measurement operator, the numerator equals ${\rm Tr}[\rho\hat{M}^2 n(x)n(0)]$. 
Using path integral representation, the numerator can be expressed as 
\begin{equation}
    {\rm Tr}[\rho\hat{M}^2 n(x)n(0)]=\int \mathcal{D}\phi e^{-S} n(x)n(0),
\end{equation}
with the action $S$
\begin{equation}
\label{effective action with measurement}
    S=\int {\rm d}x{\rm d}\tau\left[\frac{1}{2\pi K}[(\partial_\tau \phi)^2+(\partial_x \phi)^2]-\delta(\tau)v\cos{2\phi}\right],
\end{equation}
and $v=W/\pi$. 
For other case where the operators do not commute with the measurement $\hat{M}$, we need to be careful about the order of operators.

With the effective action~\eqref{Luttinger liquid model} we can integrate out the time direction and keep the ``boundary" at $\tau=0$~\cite{jian_2019}. 
Here we consider a more general case with an additional mass term $\frac{1}{2\pi K}\int {\rm d}x{\rm d}\tau \Omega^2\phi^2$. 
To integrate out the time direction, we first solve the equation of motion $\partial_\tau^2\phi+ \partial_x^2\phi-\Omega^2\phi=0$. 
In the momentum space, $\phi(p,\tau) = \int dx \phi(p,\tau) e^{i p x} $, we have $\partial_\tau^2 \phi(p,\tau)= (p^2+\Omega^2)\phi(p,\tau)$, which has the general solution
\begin{equation}
    \phi(p,\tau)=\phi(p,0)(\cosh{\omega_p\tau}-\coth{\omega_p\beta}\sinh{\omega_p\tau})+\phi_(p,\beta)\frac{\sinh{\omega_p\tau}}{\sinh{\omega_p\beta}},
\end{equation}
where $\omega_p^2=p^2+\Omega^2$. 
Then plugging it in to the action we have
\begin{equation}
    S = \frac{1}{\pi K}\int {\rm d}p \frac{\omega_p}{2} \left(\phi_c^2(p)\tanh{\frac{\beta\omega_p}{2}}+\phi_q^2(p)\coth{\frac{\beta\omega_p}{2}}\right),
\end{equation}
where $\phi_c(p)=[\phi(p,\beta)+\phi(p,0)]/ \sqrt{2}$ and $\phi_q(p)=[\phi(p,\beta)-\phi (p,0)]/ \sqrt{2}$.  
In the case of gapless fermions at zero temperature, $\Omega=0$ and $\beta=\infty$, we have $\omega_p=|p|$ and $S=\frac{1}{\pi K}\int {\rm d}p \frac{|p|}{2} \left(\phi_c^2(p)+\phi_q^2(p)\right)=\frac{1}{\pi K}\int {\rm d}p |p|\phi^2(p)$, where we apply the periodic boundary condition. 
With measurement, the final effective theory is
\begin{equation}
\label{Luttinger liquid model integrate time}
    s[\phi]=s_0[\phi]-v\int {\rm d}x \cos{[2\phi]},\quad s_0[\phi]=\frac{1}{\pi K}\int \frac{{\rm d}q}{2\pi}|q||\phi(q)|^2.
\end{equation}
Also a remark is that the sign of $v$ is unimportant, since a simple translation relates the two cases.

In Ref.~\cite{garratt2022measurements}, authors choose the convention that the field $\phi$ is related to the fermion density operator, and the dual field $\theta$, defined by $ \left[ \partial_x\phi,\theta \right]={\rm i}\pi\delta(x-x')$, is related to the phase. 
Then the corresponding actions are
\begin{subequations}
\label{Garratt's action}
\begin{equation}
    S_G[\phi]=\frac{1}{2\pi K}\int {\rm d}x{\rm d}\tau\left[ (\nabla\phi)^2 + \Dot{\phi}^2\right],\quad
    S_G[\theta]=\frac{K}{2\pi}\int {\rm d}x{\rm d}\tau\left[ (\nabla\theta)^2 + \Dot{\theta}^2\right].
\end{equation}
\end{subequations}
After integrating out time direction, the actions are 
\begin{subequations}
\label{Garratt's 1d action}
\begin{equation}
    s_G[\phi]=\frac{1}{\pi K}\int\frac{{\rm d}q}{2\pi} |q| |\phi(q)|^2,\quad
    s_G[\theta]=\frac{K}{\pi}\int\frac{{\rm d}q}{2\pi} |q| |\theta(q)|^2.
\end{equation}
\end{subequations}

\subsection{Entanglement entropy transition induced by measurement}

\subsubsection{Transformation of Different Fields with Strong Interaction}

The interactions in dual theories have a strong-weak duality. 
Therefore, we can use the dual theory at strong measurement strength or when the measurement is relevant, i.e., $v \gg 1$.
In the following, we discuss the dual transformation in more detail~\cite{altland2010condensed}. 
The initial action of the field $\phi$ is 
\begin{equation}
\label{1d action with measurement}
    s[\phi]=\frac{1}{\pi K}\int\frac{{\rm d}q}{2\pi} |q| |\phi(q)|^2-v\int {\rm d}x \cos{[2m\phi]},
\end{equation}
where $m\in\mathbb{Z_+}$ is an integer.
For $v\gg 1$, we know that the configuration of $\phi$ must be consisted of domain walls. 
Therefore, we define $h = \partial_x \phi=\sum_{i=1}^n e_i f(x-x_i)$ with the constraint $\int_{-\infty}^{+\infty}{\rm d}x f(x-x_i)=\frac{2\pi}{2m}$. 
And $e_i = \pm 1$ denotes the (anti) domain wall. 
So, the Fourier transform of $f(x)$ gives $f(0)=\frac{2\pi}{2m}$, and a large $v$ means that $f(x)$ can be approximate by $\delta$-function, i.e., $f(x)=\int\frac{{\rm d}q}{2\pi} e^{{\rm i}kx} f(q)\approx\int\frac{{\rm d}q}{2\pi} e^{{\rm i}kx} f(0)=\frac{2\pi}{2m}\delta(x-x_i)$. 
Then $\phi(q)=\frac{1}{{\rm i}q}h(q)=\frac{f(0)}{{\rm i}q} \sum_{i=1}^n e_i e^{-{\rm i}q x_i}$.

Plugging the relation in action \eqref{1d action with measurement}, we arrive at
\begin{equation}
    s[\Bar{\phi}]=\frac{1}{\pi K}\int\frac{{\rm d}q}{2\pi} |q| \frac{|f(0)|^2}{|q|^2}\sum_{ij}^n e_i e_j e^{-{\rm i}q(x_i-x_j)}+nS_{{\rm d.w.}},
\end{equation}
Using Hubbard–Stratonovich (HS) transformation, we will get 
\begin{equation}
\label{HS trasformation action}
    Z=\sum_n \sum_{\{e_i=\pm1\}}e^{-nS_{{\rm d.w.}}} \int \mathcal{D}\theta e^{-\frac{4m^2K}{16\pi}\int \frac{{\rm d}q}{2\pi} |q| |\theta(q)|^2} \frac{1}{n!}\prod_{i=1}^n \int {\rm d}x_i e^{{\rm i}\sum_{i=1}^n e_i\int\frac{{\rm d}q}{2\pi} e^{-{\rm i}qx_i}\theta(q)},
\end{equation}
where $\theta(-x_i)=\int\frac{{\rm d}q}{2\pi} e^{-{\rm i}qx_i}\theta(q)$. Defining $\gamma=2e^{-S_{{\rm d.w.}}}$ and summing all possible $\{e_i\}$, we will get the final result
\begin{equation}
\begin{split}
\label{domain wall action}
    Z&=\int \mathcal{D}\theta \ e^{-\frac{4m^2K}{16\pi}\int \frac{{\rm d}q}{2\pi} |q| |\theta(q)|^2} \sum_n \gamma^n  \frac{1}{n!}\prod_{i=1}^n \left(\int {\rm d}x_i \cos{\theta(-x_i)}\right)
    =\int \mathcal{D}\theta \ e^{-\frac{4m^2K}{16\pi}\int \frac{{\rm d}q}{2\pi} |q| |\theta(q)|^2 +\gamma\int {\rm d}x\cos{\theta(x)}}.
\end{split}
\end{equation}
If we only consider $m=1$ and define $g=e^{-S_{{\rm d.w.}}}$ where $S_{{\rm d.w.}}$ is attributed to the action of a single domain wall, we get the consistent results with Ref.~\cite{garratt2022measurements}.

Here, we briefly discuss the action of domain wall configuration. 
In Ref.~\cite{garratt2022measurements}, to introduce a UV-cutoff, the authors add another term $\frac{1}{2}\int {\rm d}x (\nabla \phi)^2$. 
For large $v$ limit, rescaling $x=v^{-1/2}x'$ and solving equation of motion for domain wall configuration, they give
\begin{equation}
    \phi_{\rm d.w.}(x)=\frac{\pi}{2}+\arctan[\sinh(2v^{1/2}x)].
\end{equation}
For the contribution of domain wall configuration to \eqref{1d action with measurement}, there are two parts. 
One comes from the last term in \eqref{1d action with measurement} and the additional term, which gives $\Delta s=4v^{1/2}$. 
Another part is attributed to the first term in \eqref{1d action with measurement}. 
With the rescaling $x=v^{-1/2}x'$, and $q=v^{1/2}q'$, the first term becomes
\begin{equation}
\label{eq:action of instanton}
    \frac{1}{\pi K}\int\frac{{\rm d}q}{2\pi} |q| |\phi_{\rm d.w.}(v^{-1/2}q)|^2
    =a v, \quad 
    a = \frac{1}{\pi K}\int\frac{{\rm d}q'}{2\pi} |q'| |\phi_{\rm d.w.}(q')|^2,
\end{equation}
where $\phi_{\rm d.w.}(q')$ is the Fourier transformation of $\phi_{\rm d.w.}(x')=\frac{\pi}{2}+\arctan[\sinh(2x')]$. Therefore, $S_{{\rm d.w.}}=a v+4v^{1/2}$.

We remark on the commutation relation. 
When we apply HS transformation in \eqref{HS trasformation action}, the commutation relation is $\left[e_i, \theta(x_j) \right]={\rm i}\delta_{ij}$, which means $\left[\partial_x\phi(x), \theta(x_j) \right]=\left[\sum_{i=1}^n e_i f(x-x_i), \theta(x_j) \right]={\rm i}\frac{2\pi}{2m}\delta(x-x_j)$. 
Define $\Tilde{\theta}=m\theta$, $\left[\partial_x\phi(x),\Tilde{\theta}(x') \right] = {\rm i} \delta(x-x')$, $\Tilde{\theta}$ is the dual field of $\phi$.   
Using $\Tilde{\theta}$ to rewrite the action, we will get the action
\begin{equation}
\label{strong measurement action}
    s[\Tilde{\theta}]=\frac{K}{4\pi}\int \frac{{\rm d}q}{2\pi} |q| |\Tilde{\theta}(q)|^2 -\gamma\int {\rm d}x\cos{\frac{\Tilde{\theta}(x)}{m}}.
\end{equation}
While in the following, we mainly consider $m=1$, we remark that the results will not rely on $m$ up to the zero order with $\gamma=0$.

\subsubsection{Correlation Functions}
\label{sec:Correlation Functions in Zero Order}

In the following, we consider three different correlation functions, including $\left<e^{{\rm i}[\theta(x)-\theta(0)]}\right>$, $\left<e^{{\rm i}[\phi(x)-\phi(0)]}\right>$ and $\left< {\mathcal{T}}_{nk} {\mathcal{T}}^{-1}_{nk}\right> =\left<e^{-{\rm i}\frac{2k}{\sqrt{K}n}[\phi(x)-\phi(0)]}\right>$.

We first consider zero order results. 
For the phase correlation function, in the free case ($v=0$) with the action \eqref{1d action with measurement}, Ref.~\cite{garratt2022measurements} shows that
\begin{equation}
\label{phase correlation function for free case}
    \left<e^{{\rm i}[\theta(x_0)-\theta(0)]}\right>=e^{-\frac{\pi}{4K}\int \frac{{\rm d}q}{2\pi} |q||T_{0,x_0}(q)|^2}\sim x_0^{-\frac{1}{2K}}.
\end{equation}
Here, with abuse of notation, we define
\bea
T_{0,x_0}(x) = \begin{cases} 1 , \quad & 0<x<x_0 \\
0, \quad & x<0, x>x_0
\end{cases}.
\eea
It should be clear from the context that this is different from the twist operator $T_A$. 

In the strong measurement limit ($v\gg1$), we take the action \eqref{strong measurement action} with $\gamma=0$. In Ref.~\cite{garratt2022measurements} the authors have shown that the result should be $\left<e^{{\rm i}[\theta(x_0)-\theta(0)]}\right>\sim x_0^{-\frac{1}{K}}$.

We derive the correlation function above in another way which takes care of the order of operators, and perform an explicit calculation. 
We first consider $s[\phi+\pi T_{0,x_0}]$. 
With similar approximations, we have
\begin{equation}
\label{measure}
    \phi+\pi T_{0,x_0}=\frac{1}{{\rm i}q}f(q)\sum_{i=1}^n(e_i e^{{\rm i}q x_i}+e_0 e^{{\rm i}q x_0}+e_{n+1}e^{{\rm i}q x_{n+1}}),
\end{equation}
where $e_0=1, e_{n+1}=-1, x_0=0, x_{n+1}=x_0$. 
Plugging it in the action \eqref{1d action with measurement} with $v\rightarrow\infty, \gamma=0$, we have 
\begin{equation}
\begin{split}
    s_0&=\frac{\pi}{K}\int\frac{{\rm d}q}{2\pi} \frac{1}{|q|} \sum_{i,j=1}^n e_i e_j e^{-{\rm i}q(x_i-x_j)}\\&+\frac{\pi}{K}\int\frac{{\rm d}q}{2\pi} \frac{1}{|q|}\left[ \sum_{j=1}^n 1\cdot e_j e^{-{\rm i}q(0-x_j)}
    +\sum_{j=1}^n (-1)\cdot e_j e^{-{\rm i}q(x_0-x_j)}+(-1)e^{-{\rm i}q(0-x_0)}+h.c.+2 \right].
\end{split}
\end{equation}
It means that the function $T_{0,x_0}(x)$ behaves like two domain walls at $x=0, x_0$ with opposite signs. 
Then the partition function will be
\begin{equation}
\begin{split}
\label{phase correlation details}
    Z=&\sum_n \sum_{\{e_i=\pm1\}}e^{-nS_{{\rm d.w.}}} \int \mathcal{D}\theta e^{-\frac{K}{4\pi}\int \frac{{\rm d}q}{2\pi} |q| |\theta(q)|^2} e^{\frac{\pi}{2K}\int\frac{{\rm d}q}{2\pi}\frac{1}{|q|}(2\cos{qx_0}-2)}\\
    &\frac{1}{n!}\prod_{i=1}^n \int {\rm d}x_i e^{{\rm i}\sum_{i=1}^n e_i\int\frac{{\rm d}q}{2\pi} e^{-{\rm i}qx_i}\theta(q)-\frac{\pi}{2K}\int\frac{{\rm d}q}{2\pi}\frac{1}{|q|}\sum_{i=1}^n e_i(e^{-{\rm i}q(0-x_i)}-e^{-{\rm i}q(x_0-x_i)}+h.c.)}\\
    =&\int \mathcal{D}\theta e^{-\frac{K}{4\pi}\int \frac{{\rm d}q}{2\pi} |q| |\theta(q)|^2+\gamma \int {\rm d}x \cos{[\theta(x)+\Delta_{x_0}(x)]}} e^{\frac{\pi}{2K}\int\frac{{\rm d}q}{2\pi}\frac{1}{|q|}(2\cos{qx_0}-2)},
\end{split}
\end{equation}
where $\Delta_{x_0}(x)={\rm i}\frac{\pi}{2K}\int\frac{{\rm d}q}{2\pi}\frac{1}{|q|}(e^{{\rm i}qx}-e^{{\rm i}q(x-x_0)}+h.c.)$. 
Then directly we get
\begin{equation}
\label{phase correlation result}
    \left<e^{{\rm i}[\theta(x_0)-\theta(0)]}\right>=\frac{\int \mathcal{D}\phi e^{-\frac{1}{2}(s[\phi]+s[\phi+\pi T_{0,x_0}])}}{\int \mathcal{D}\phi e^{-s[\phi]}}=e^{\frac{\pi}{2K}\int\frac{{\rm d}q}{2\pi}\frac{1}{|q|}(2\cos{qx_0}-2)}\sim x_0^{-\frac{1}{K}},
\end{equation}
which is consistent with Ref.~\cite{garratt2022measurements}.

Now we consider the correlation function of vertex operators and twist operators. 
For the free case ($v=0$) with the action \eqref{1d action with measurement}, we have 
\begin{equation}
    \left<e^{{\rm i}[\phi(x_0)-\phi(0)]}\right>=e^{G(x_0)}\sim x_0^{-\frac{K}{2}},
\end{equation}
where $G(x_0)=-\frac{\pi K}{2\pi}\ln{x_0}$. 
Similarly, we can calculate the correlation function of twist operators, which will give entanglement entropy. 
For the free case ($v=0$) with the action \eqref{1d action with measurement}, we have 
\begin{equation}
\label{entanglement correlation function for free case}
    \left<e^{-{\rm i}\frac{2k}{\sqrt{K}n}[\phi(x_0)-\phi(0)]}\right>=e^{\frac{4k^2}{Kn^2}G(x_0)}\sim x_0^{-\frac{2k^2}{n^2}},
\end{equation}
which is consistent with Ref.~\cite{casini2005entanglement} and gives entanglement entropy $S=\frac{1}{3}\log{L}$.

In the following, we derive the results of the correlation functions of vertex operators and twist operators for strong measurement strength by using the similar method of \eqref{phase correlation details}. 
For the correlation function of vertex operators, we have
\begin{equation}
    \left<e^{{\rm i}[\phi(x_0)-\phi(0)]}\right>=\left<e^{-\int \frac{{\rm d}q}{2\pi} q\phi(q)T_{0,x_0}(-q)}\right>=\left< {\rm exp}\left(-\int \frac{{\rm d}q}{2\pi} \frac{1}{{\rm i}q}\sum_{i=1}^n e_i e^{-{\rm i}qx_i}\pi qT_{0,x_0}(-q)\right) \right>,
\end{equation}
where the expectation value is according to the action \eqref{strong measurement action}. 
Then the numerator is 
\begin{equation}
\label{vertex correlation details}
\begin{split}
    Z=&\sum_n \sum_{\{e_i=\pm1\}}e^{-nS_{{\rm d.w.}}} \int \mathcal{D}\theta e^{-\frac{K}{4\pi}\int \frac{{\rm d}q}{2\pi} |q| |\theta(q)|^2}
    \frac{1}{n!}\prod_{i=1}^n \int {\rm d}x_i e^{{\rm i}\sum_{i=1}^n e_i\int\frac{{\rm d}q}{2\pi} e^{-{\rm i}qx_i}\theta(q)+{\rm i}\int\frac{{\rm d}q}{2\pi}\pi T_{0,x_0}(-q)\sum_{i=1}^n e_i e^{-{\rm i}qx_i}}\\
    =&\int \mathcal{D}\theta \ e^{-\frac{K}{4\pi}\int \frac{{\rm d}q}{2\pi} |q| |\theta(q)|^2+\gamma \int {\rm d}x \cos{[\theta(x)+\Delta_{x_0}(x)]}},
\end{split}
\end{equation}
where $\Delta_{x_0}(x)=\int\frac{{\rm d}q}{2\pi}\pi T_{0,x_0}(-q) e^{-{\rm i}qx_i}=\pi T_{0,x_0}(x)$. Therefore, up to the zeroth order $\gamma=0$, the denominator equals the numerator, which means $\left<e^{{\rm i}[\phi(x_0)-\phi(0)]}\right>=1$. 
There is another way to double-check the result above. 
We can rewrite the exponent of numerator as follows.
\begin{equation}
\begin{split}
    -s'=&-\frac{1}{\pi K}\int\frac{{\rm d}q}{2\pi} |q| |\phi(q)|^2+v\int {\rm d}x \cos{[2\phi]}-\int \frac{{\rm d}q}{2\pi} q\phi(q)T_{0,x_0}(-q)\\
    =&-\frac{1}{\pi K}\int\frac{{\rm d}q}{2\pi} |q| (\phi(q)+\frac{\pi K q}{2|q|}T_{0,x_0}(q))(\phi(-q)+\frac{\pi K q}{2|q|}T_{0,x_0}(-q)) +v\int {\rm d}x \cos{[2\phi]}
    +\frac{1}{\pi K}\int \frac{{\rm d}q}{2\pi} |q| \frac{\pi^2 K^2}{4}|T_{0,x_0}(q)|^2,
\end{split}
\end{equation}
where $(-\int \frac{{\rm d}q}{2\pi} q\phi(q)T_{0,x_0}(-q))^*=-\int \frac{{\rm d}q}{2\pi} q\phi(-q)T_{0,x_0}(q)=\int \frac{{\rm d}q}{2\pi} q\phi(q)T_{0,x_0}(-q)$. 
Then there are two terms contributing to the final result. 
One is $-\frac{1}{\pi K}\int\frac{{\rm d}q}{2\pi} |q| |\phi(q)+\frac{\pi K q}{2|q|}T_{0,x_0}(q)|^2$, which is similar to \eqref{phase correlation result} and gives $e^{\frac{\pi}{K}\int\frac{{\rm d}q}{2\pi}\frac{1}{|q|}(2\cos{qx_0}-2)\cdot\left(\frac{Kq}{2|q|}\right)^2}\sim x_0^{-\frac{K}{2}}$. 
Another term is $\frac{1}{\pi K}\int \frac{{\rm d}q}{2\pi} |q| \frac{\pi^2 K^2}{4}|T_{0,x_0}(q)|^2$ which is similar to \eqref{phase correlation function for free case} and gives $\sim x_0^{\frac{K}{2}}$. 
Therefore, two terms' contributions will cancel out and lead to $\left<e^{{\rm i}[\phi(x_0)-\phi(0)]}\right>=1$. 

For twist operators, the correlation function is similar to vertex operators case. 
The only difference in \eqref{vertex correlation details} is that there is a prefactor in front of $\Delta_{x_0}(x)$, which will not change the zeroth order results and gives a trivial correlation function $\left<{\mathcal{T}}_{nk} {\mathcal{T}}^{-1}_{nk}\right>=1$.

Here, we briefly discuss the strong measurement results of several correlation functions. 
The nontrivial result of phase correlation function is because of the interaction of two domain walls in $T_{0,x_0}(x)$. But for vertex operator, the additional term is linear with $\phi$, which doesn't contribute the quadratic terms like phase correlation. So the result is trivial.

Finally, we consider the correlation functions of density operators. 
For the free case ($v=0$) we have 
\begin{equation}
    \left<\nabla\phi(x_0)\nabla\phi(0)\right>=-\nabla^2G(x_0)=\frac{K}{2}x_0^{-2}.
\end{equation}
For the strong measurement case, Ref.~\cite{garratt2022measurements} shows that $\left<\nabla\phi(x_0)\nabla\phi(0)\right>\sim \gamma^2 x_0^{-\frac{2}{K}}$. It is worth mentioning that this density correlation function will rely on different $m$.

According to the discussion above, for the strong measurement case, we always get vanishing results at the zeroth order.
So now we consider the first order of $\gamma$. 
To simplify the problem, we only consider $m=1$ case (For general $m$ the results will be similar.) 
In the following, we will only use the method of \eqref{phase correlation details} and \eqref{vertex correlation details}.

\subsubsection{Phase Correlation Function}

Firstly, we consider phase correlation function. 
To consider the domain wall contribution, we include another term to \eqref{1d action with measurement},
\begin{equation}
\label{1d action with measurement and self-energy}
\begin{split}
    s[\phi]=&\frac{1}{\pi K}\int\frac{{\rm d}q}{2\pi} |q| |\phi(q)|^2-v\int {\rm d}x \cos{[2\phi]}+\frac{1}{2}\int {\rm d}x(\nabla\phi)^2\\
    =&\frac{1}{\pi K}\int\frac{{\rm d}q}{2\pi} |q| |\phi(q)|^2-v^{\frac{1}{2}}\left[\int {\rm d}x' \cos{[2\phi]}-\frac{1}{2}\int {\rm d}x'(\nabla'\phi)^2\right],
\end{split}
\end{equation}
where $x'=v^{\frac{1}{2}}x$. Similar to \eqref{phase correlation result} we have
\begin{equation}
\label{first order phase correlation}
\begin{split}
    &\frac{1}{2}(s[\phi]+s[\phi+\pi T_{0,x_0}])=\frac{1}{\pi K}\int\frac{{\rm d}q}{2\pi} |q| |\phi(q)|^2+\frac{1}{K}\int\frac{{\rm d}q}{2\pi} |q| \phi(q)T_{0,x_0}(-q)+\frac{\pi}{2 K}\int\frac{{\rm d}q}{2\pi} |q| |T_{0,x_0}(q)|^2\\
    &-\frac{1}{2}v^{\frac{1}{2}}\left[\int {\rm d}x' \cos{[2(\phi+\pi T_{0,x_0})]}-\frac{1}{2}\int {\rm d}x'(\nabla'(\phi+\pi T_{0,x_0}))^2+\int {\rm d}x' \cos{[2\phi]}-\frac{1}{2}\int {\rm d}x'(\nabla'\phi)^2\right].
\end{split}
\end{equation}
Therefore, there are two additional terms in the second line \eqref{first order phase correlation}. 
$T_{0,x_0}$ is unimportant for $\cos$ term, because we can safely ignore the step function. 
But we need to consider the contribution of it to quadratic terms. 
For the strong measurement strength, we need to change the configuration of $\phi$ to minimize two $\cos$ terms, which means for first order we take $\phi$ as two domain walls (beside two domain walls of $T_{0,x_0}$).
Now we have 
\begin{equation}
\begin{split}
    \frac{1}{2}(s[\phi]+s[\phi+\pi T_{0,x_0}])=&\frac{1}{2\pi K}\int\frac{{\rm d}q}{2\pi} |q| |\phi(q)|^2+\frac{1}{2\pi K}\int\frac{{\rm d}q}{2\pi} |q| |\phi(q)+\pi T_{0,x_0}(q)|^2\\
    &-v^{\frac{1}{2}}\left[\int {\rm d}x' \cos{[2\phi]}-\frac{1}{2}\int {\rm d}x'(\nabla'\phi)^2\right]+\frac{\pi}{2}\int {\rm d}x\nabla\phi\nabla T_{0,x_0}+\frac{\Tilde{C}}{2},
\end{split}
\end{equation}
where $\Tilde{C}=\frac{1}{2}\int {\rm d}x(\pi\nabla T_{0,x_0})^2$. 
Then we consider \eqref{phase correlation details} with the additional terms
\begin{equation}
\label{phase correlation details up to first order}
\begin{split}
    Z=&e^{-\frac{\Tilde{C}}{2}}\sum_n \sum_{\{e_i=\pm1\}}e^{-nS_{{\rm d.w.}}} \int \mathcal{D}\theta e^{-\frac{K}{4\pi}\int \frac{{\rm d}q}{2\pi} |q| |\theta(q)|^2} e^{\frac{\pi}{2K}\int\frac{{\rm d}q}{2\pi}\frac{1}{|q|}(2\cos{qx_0}-2)}
    \frac{1}{n!}\prod_{i=1}^n \int {\rm d}x_i \\
    &e^{{\rm i}\sum_{i=1}^n e_i\int\frac{{\rm d}q}{2\pi} e^{{\rm i}qx_i}\theta(q)-\frac{\pi}{2K}\int\frac{{\rm d}q}{2\pi}\frac{1}{|q|}\sum_{i=1}^n e_i(e^{-{\rm i}q(0-x_i)}-e^{-{\rm i}q(x_0-x_i)}+h.c.)-\frac{\pi^2}{2}\int\frac{{\rm d}q}{2\pi} q^2 T_{0,x_0}(-q)\cdot\frac{1}{{\rm i}q}\sum_{i=1}^n e_i e^{-{\rm i}q x_i}}\\
    =&e^{-\frac{\Tilde{C}}{2}}\int \mathcal{D}\theta e^{-\frac{K}{4\pi}\int \frac{{\rm d}q}{2\pi} |q| |\theta(q)|^2+\gamma \int {\rm d}x \cos{[\theta(x)+\Delta_{x_0}(x)+\Tilde{\Delta}_{x_0}(x)]}} e^{\frac{\pi}{2K}\int\frac{{\rm d}q}{2\pi}\frac{1}{|q|}(2\cos{qx_0}-2)},
\end{split}
\end{equation}
where 
\begin{equation}
\begin{split}
    \Delta_{x_0}(x_i)=&{\rm i}\frac{\pi}{2K}\int\frac{{\rm d}q}{2\pi}\frac{1}{|q|}(e^{-{\rm i}q(0-x_i)}-e^{-{\rm i}q(x_0-x_i)}+h.c.)
    ={\rm i}\frac{\pi}{K}\int\frac{{\rm d}q}{2\pi}\frac{1}{|q|}\left[\cos{qx_i}-\cos{q(x_0-x_i)} \right],
\end{split}
\end{equation}
\begin{equation}
\begin{split}
\label{definition of Delta_tilde}
    \Tilde{\Delta}_{x_0}(x_i)=&\frac{\pi^2}{2}\int\frac{{\rm d}q}{2\pi} q T_{0,x_0}(-q)e^{-{\rm i}q x_i}={\rm i}\frac{\pi^2}{2} \partial_{x_i}T_{0,x_0}(x_i).
\end{split}
\end{equation}
Then we double-check the results above and give some comments. (i) The zero order term satisfies
\begin{equation}
    e^{\frac{\pi}{2K}\int\frac{{\rm d}q}{2\pi}\frac{1}{|q|}(2\cos{qx_0}-2)}=e^{-\frac{\pi}{2K}\int \frac{{\rm d}q}{2\pi} |q||T_{0,x_0}(q)|^2}\sim x_0^{-\frac{2}{K}}.
\end{equation}
(ii) The additional $\Delta_{x_0}(x)$ is from the crossing term
\begin{equation}
\label{phase correlation Delta function}
\begin{split}
    &-\frac{1}{2\pi K}\int\frac{{\rm d}q}{2\pi} |q| \left[\phi(q)\pi T_{0,x_0}(-q)+\phi(-q)\pi T_{0,x_0}(q)\right]=-\frac{1}{K}\int\frac{{\rm d}q}{2\pi}|q| T_{0,x_0}(-q)\cdot\frac{1}{{\rm i}q}\sum_{i=1}^n e_i e^{-{\rm i}q x_i}f(q)\\
    =&{\rm i}\sum_{i=1}^n e_i \frac{\pi}{K}\int\frac{{\rm d}q}{2\pi} \frac{|q|}{q} e^{-{\rm i}qx_i}T_{0,x_0}(-q)={\rm i}\sum_{i=1}^n e_i \frac{{\rm i}\pi}{K}\int\frac{{\rm d}q}{2\pi} \frac{1}{|q|} e^{-{\rm i}qx_i}(1-e^{{\rm i}qx})={\rm i}\sum_{i=1}^n e_i\Delta_{x_0}(x_i),
\end{split}
\end{equation}
where $f(q)=\pi$. Simplifying $\Delta_{x_0}(x)$ we have 
\begin{equation}
    \Delta_{x_0}(x)={\rm i}\frac{\pi}{K}\int\frac{{\rm d}q}{2\pi}\frac{1}{|q|} \left[2\sin^2{\frac{q(x_i-x_0)}{2}}-2\sin^2{\frac{qx_i}{2}} \right]={\rm i}\frac{1}{K}\ln{\left|\frac{x_i-x_0}{x_i}\right|}.
\end{equation}
(iii) For $\Tilde{\Delta}_{x_0}(x_i)$ we have 
\begin{equation}
    \Tilde{\Delta}_{x_0}(x_i)=\frac{\pi^2}{2}\int\frac{{\rm d}q}{2\pi} q \frac{1}{-{\rm i}q}(e^{{\rm i}q\cdot0}-e^{{\rm i}q x_0})e^{-{\rm i}q x_i}={\rm i}\frac{\pi^2}{2} \left[\delta(x_i)-\delta(x_i-x_0)\right],
\end{equation}
which is consistent with \eqref{definition of Delta_tilde}. 

We calculate the phase correlation function as follows,
\begin{equation}
    \left<e^{{\rm i}[\theta(x_0)-\theta(0)]}\right>=e^{-\frac{\Tilde{C}}{2}}e^{\frac{\pi}{2K}\int\frac{{\rm d}q}{2\pi}\frac{1}{|q|}(2\cos{qx_0}-2)} \frac{\int \mathcal{D}\theta e^{-\frac{K}{4\pi}\int \frac{{\rm d}q}{2\pi} |q| |\theta(q)|^2+\gamma \int {\rm d}x \cos{[\theta(x)+\Delta_{x_0}(x)+\Tilde{\Delta}_{x_0}(x)]}}}{\int \mathcal{D}\theta e^{-\frac{K}{4\pi}\int \frac{{\rm d}q}{2\pi} |q| |\theta(q)|^2+\gamma \int {\rm d}x \cos{[\theta(x)]}}},
\end{equation}
where the first two parts above give the zero order result $e^{-\frac{\Tilde{C}}{2}}x_0^{-\frac{1}{K}}$. 
In the following, we evaluate the path integral. 
Expanding the $\cos$ term gives 
\begin{equation}
\label{expanding of phase correlation}
\begin{split}
    {\rm num}=&\int \mathcal{D}\theta e^{-\frac{K}{4\pi}\int \frac{{\rm d}q}{2\pi} |q| |\theta(q)|^2+\gamma \int {\rm d}x \cos{[\theta(x)+\Delta_{x_0}(x)+\Tilde{\Delta}_{x_0}(x)]}}\\
    =&\int \mathcal{D}\theta e^{-\frac{K}{4\pi}\int \frac{{\rm d}q}{2\pi} |q| |\theta(q)|^2}\sum_{n=0}^\infty \frac{1}{n!}\left(\gamma \int {\rm d}x \cos{[\theta(x)+\Delta_{x_0}(x)+\Tilde{\Delta}_{x_0}(x)]}\right)^n.
\end{split}
\end{equation}
For the denominator we can just set $\Delta_{x_0}(x)=\Tilde{\Delta}_{x_0}(x)=0$. 
As we know, the expectation value of ``vertex" operators is nontrivial only when the total charge is zero, which means the first order term is $n=2$. 
With the form of ``vertex" operator we have $\gamma^2$ term
\begin{equation}
\begin{split}
    &\frac{1}{2}\gamma^2\int {\rm d}x_1 {\rm d}x_2\left<\cos{[\theta(x_1)+\Delta_{x_0}(x_1)+\Tilde{\Delta}_{x_0}(x_1)]}\cos{[\theta(x_2)+\Delta_{x_0}(x_2)+\Tilde{\Delta}_{x_0}(x_2)]}\right>\\
    =&\frac{1}{2}\gamma^2\times\frac{1}{4}\int {\rm d}x_1 {\rm d}x_2\left( e^{{\rm i}[\Delta_{x_0}(x_1)+\Tilde{\Delta}_{x_0}(x_1)-\Delta_{x_0}(x_2)-\Tilde{\Delta}_{x_0}(x_2)]}\left<e^{{\rm i}\theta(x_1)}e^{-{\rm i}\theta(x_2)}\right>+h.c.\right),
\end{split}
\end{equation}
where
\begin{equation}
    \left<e^{{\rm i}\theta(x_1)}e^{-{\rm i}\theta(x_2)}\right>=e^{\frac{1}{2}G(x_1-x_2)+\frac{1}{2}G(x_2-x_1)}=e^{-\frac{1}{2\pi}\frac{4\pi}{K}\ln{|x_1-x_2|}}.
\end{equation}
So the $\gamma^2$ term is 
\begin{equation}
\label{phase correlation form}
    \frac{1}{2}\gamma^2\times\frac{1}{4}\int {\rm d}x_1 {\rm d}x_2\left( e^{{\rm i}[\Delta_{x_0}(x_1)+\Tilde{\Delta}_{x_0}(x_1)-\Delta_{x_0}(x_2)-\Tilde{\Delta}_{x_0}(x_2)]-\frac{2}{K}\ln{|x_1-x_2|}}+h.c.\right)=\frac{1}{2}\gamma^2I_1.
\end{equation}
Similarly, $\gamma^2$ term in denominator is
\begin{equation}
\label{denominator}
    \frac{1}{2}\gamma^2\times\frac{1}{4}\int {\rm d}x_1 {\rm d}x_2\left( e^{-\frac{2}{K}\ln{|x_1-x_2|}}+h.c.\right)=\frac{1}{2}\gamma^2I_2.
\end{equation}
Therefore, we get the phase correlation function
\begin{equation}
\label{simplified phase correlation}
    \left<e^{{\rm i}[\theta(x_0)-\theta(0)]}\right>\approx e^{-\frac{\Tilde{C}}{2}}x_0^{-\frac{1}{K}}(1+\frac{\gamma^2}{2}I_1)(1+\frac{\gamma^2}{2}I_2)^{-1}\approx e^{-\frac{\Tilde{C}}{2}}x_0^{-\frac{1}{K}}(1+\frac{\gamma^2}{2}(I_1-I_2)).
\end{equation}
Now we evaluate the integral $I_1-I_2$. 
Plugging $\Delta_{x_0}(x)$ and $\Tilde{\Delta}_{x_0}(x)$ in $I_1$ we have
\begin{equation}
\label{integral I1}
\begin{split}
    I_1=&2\times\frac{1}{4}\int {\rm d}x_1 {\rm d}x_2 e^{\left[ -\frac{1}{K}\ln{\left|\frac{(x_1-x_0)x_2}{x_1(x_2-x_0)}\right|}-\frac{\pi^2}{2}(\delta(x_1)-\delta(x_1-x_0)-\delta(x_2)+\delta(x_2-x_0))-\frac{2}{K}\ln{|x_1-x_2|}\right]}\\
    =&\frac{1}{2}\int {\rm d}x_1 {\rm d}x_2 e^{-\frac{\pi^2}{2}(\delta(x_1)-\delta(x_1-x_0)-\delta(x_2)+\delta(x_2-x_0))}\left[\left|\frac{(x_1-x_0)x_2}{x_1(x_2-x_0)}\right|(x_1-x_2)^2\right]^{-\frac{1}{K}}\\
    \approx&\frac{1}{2}\int {\rm d}x_1 {\rm d}x_2 \left[\left|\frac{(x_1-x_0)x_2}{x_1(x_2-x_0)}\right|(x_1-x_2)^2\right]^{-\frac{1}{K}}\\
    =&\frac{1}{2}\int {\rm d}x_1 {\rm d}x_2 e^{-\frac{1}{K}\left(\ln{|x_1-x_0|}+\ln{|x_2-0|}-\ln{|x_1-0|}-\ln{|x_2-x_0|}+2\ln{|x_1-x_2|}\right)}.
\end{split}
\end{equation}
From the above result we find that the contribution of $\Tilde{\Delta}_{x_0}(x)$ is only on some isolated points with the measure that is dimension zero. 
If we apply some UV cutoff that require different domain walls cannot be too close to each other the exponent will always be zero. 
Besides, there is a comment on \eqref{integral I1}. 
From the last line we can understand the meaning of the contribution. Because we consider the sub-leading term, we assume there are two domain walls with opposite signs. 
And there are also two domain walls because of $T_{0,x_0}(x)$. 
Therefore, we actually have two domain walls at $x=0,x_1$ with positive charge and two domain walls at $x=x_0,x_2$ with negative charge. 
There are $C_4^2=6$ pairs, where the interaction between $x=0$ and $x=x_0$ is zero order result in \eqref{simplified phase correlation} and the other five pairs are in \eqref{integral I1}. 
The pairs with opposite (same) charge will have factor $+1(-1)$. 
And the factor $\frac{1}{2}$ is because in \eqref{phase correlation result} numerator the pair $(x_1,x_2)$ contributes two actions and other pairs only contribute $s[\phi+\pi T_{0,x_0}]$.

Finally, we have 
\begin{equation}
\begin{split}
    I_1-I_2=&\frac{1}{2}\int {\rm d}x_1 {\rm d}x_2 \left(\left|\frac{(x_1-x_0)x_2}{x_1(x_2-x_0)}\right|^{-\frac{1}{K}}-1\right) |x_1-x_2|^{-\frac{2}{K}}\\
    =&\frac{1}{2}\int {\rm d}\Tilde{x}_1 {\rm d}\Tilde{x}_2 \left(\left|\frac{(\Tilde{x}_1-1)\Tilde{x}_2}{\Tilde{x}_1(\Tilde{x}_2-1)}\right|^{-\frac{1}{K}}-1\right) |\Tilde{x}_1-\Tilde{x}_2|^{-\frac{2}{K}}\cdot x_0^{2-\frac{2}{K}}=\Delta_{I_1}\cdot x_0^{2-\frac{2}{K}},
\end{split}
\end{equation}
where $\Tilde{x}_i=x_i/x_0$. Therefore, we get the phase correlation function is 
\begin{equation}
    \left<e^{{\rm i}[\theta(x_0)-\theta(0)]}\right>\approx e^{-\frac{\Tilde{C}}{2}}x_0^{-\frac{1}{K}}(1+\frac{\gamma^2}{2}\Delta_{I_1}\cdot x_0^{2-\frac{2}{K}}),
\end{equation}
where we have obtained subleading correction.

\subsubsection{Vertex Operator Correlation and Entanglement Entropy}

Now we consider the correlation function $\left<e^{{\rm i}[\phi(x)-\phi(0)]}\right>$. With \eqref{vertex correlation details} as numerator, we can expand it like \eqref{expanding of phase correlation},
\begin{equation}
    {\rm num}=\int \mathcal{D}\theta e^{-\frac{K}{4\pi}\int \frac{{\rm d}q}{2\pi} |q| |\theta(q)|^2}\sum_{n=0}^\infty\frac{1}{n!}\left(\gamma \int {\rm d}x \cos{[\theta(x)+\Delta_{x_0}(x)]}\right)^n,
\end{equation}
where $\Delta_{x_0}(x)=\pi T_{0,x_0}(x)$. 
The first nontrivial term is $\gamma^2$ term
\begin{equation}
\label{vertex correlation form}
\begin{split}
    &\frac{1}{2}\gamma^2\int {\rm d}x_1 {\rm d}x_2\left<\cos{[\theta(x_1)+\pi T_{0,x_0}(x_1)]}\cos{[\theta(x_2)+\pi T_{0,x_0}(x_2)]}\right>\\
    =&\frac{\gamma^2}{2} \times \frac{1}{4}\int {\rm d}x_1{\rm d}x_2 e^{{\rm i}\pi [T_{0,x_0}(x_1)-T_{0,x_0}(x_2)]} \left<e^{{\rm i}\theta(x_1)}e^{-{\rm i}\theta(x_2)}\right>+h.c.=\frac{1}{2}\gamma^2 I_3,
\end{split}
\end{equation}
where $\left<e^{{\rm i}\theta(x_1)}e^{-{\rm i}\theta(x_2)}\right>=e^{-\frac{2}{K}\ln{|x_1-x_2|}}$. And the denominator is the same as \eqref{denominator}. So, 
\begin{equation}
    \left<e^{{\rm i}[\phi(x_0)-\phi(0)]}\right>\approx (1+\frac{\gamma^2}{2}I_3)(1+\frac{\gamma^2}{2}I_2)^{-1}\approx (1+\frac{\gamma^2}{2}(I_3-I_2)).
\end{equation}
Here is a remark about the result above. 
It seems that \eqref{vertex correlation form} is similar to \eqref{phase correlation form}. 
But actually, their final result is not the same. 
For \eqref{phase correlation form} the function $\Delta_{x_0}(x)$ represent the interaction of two domain walls which contribute a real factor. 
For \eqref{vertex correlation form}, here the function $\Delta_{x_0}(x)$ play the role of phase factor that makes the integrand oscillate. 
The key difference is a sign factor $|q|=q\cdot {\rm sgn}(q)$ in \eqref{phase correlation Delta function}. 
Comparing \eqref{vertex correlation details} and \eqref{phase correlation Delta function}, we can find that although both $\Delta_{x_0}(x)$ is from the linear term to $\phi$. 
But for phase correlation case \eqref{phase correlation Delta function} the linear term is $\int\frac{{\rm d}q}{2\pi} |q|\phi(q)\pi T_{0,x_0}(-q)$ with the factor $|q|$, but for the vertex correlation \eqref{vertex correlation details} the linear term is $\int \frac{{\rm d}q}{2\pi} q\phi(q)T_{0,x_0}(-q)$ with the factor $q$. 
It is the difference ${\rm sgn}(q)$ that makes the difference.

\begin{figure}
    \centering
\subfigure[]{
\includegraphics[width=0.25\linewidth]{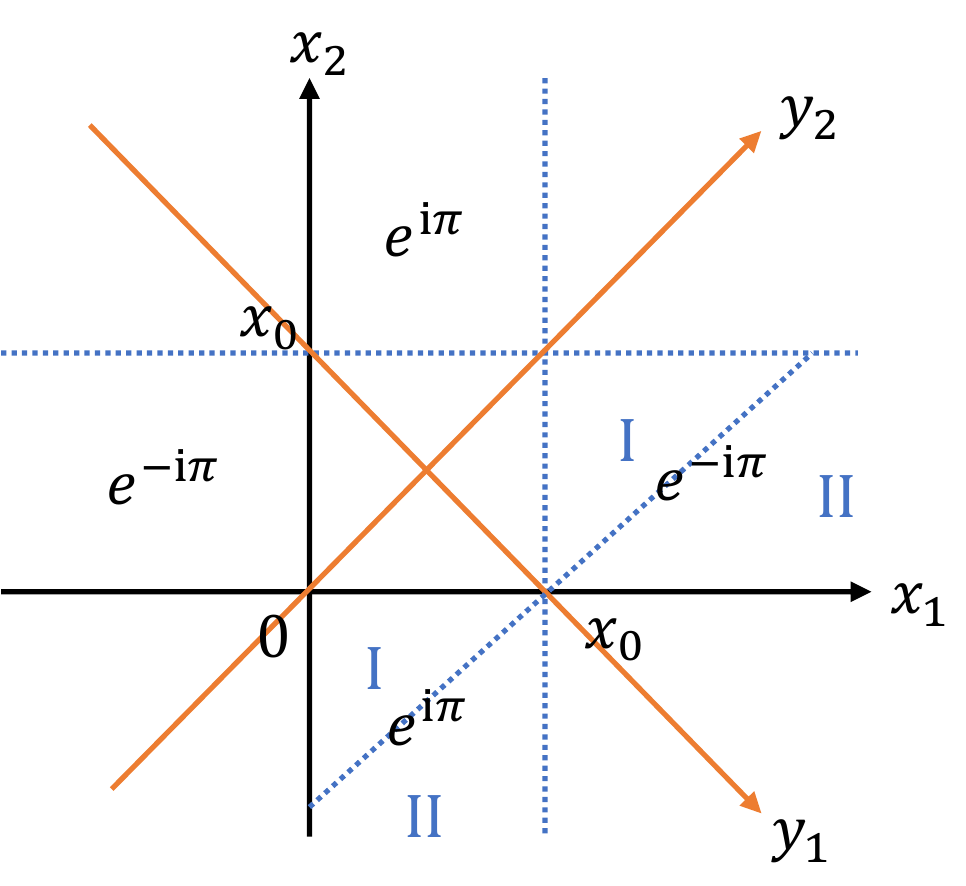}}
\subfigure[]{
\includegraphics[width=0.25\linewidth]{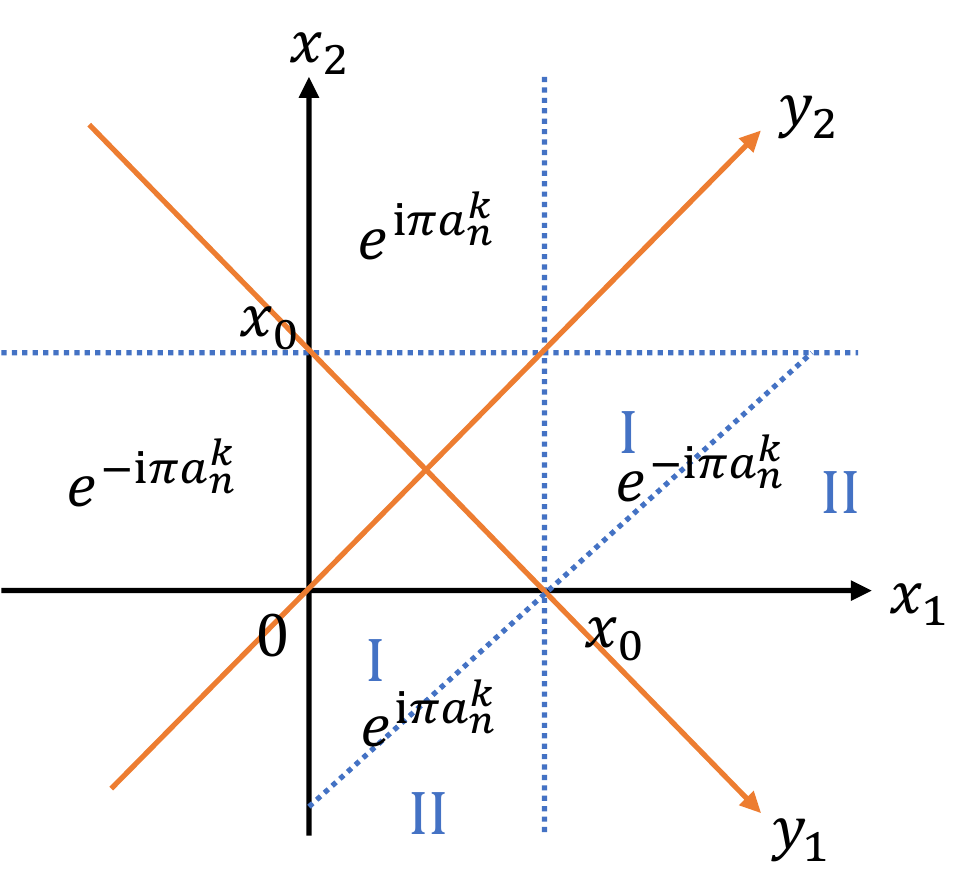}
}
    \caption{(a) Diagram of Integral for integral \eqref{eq:integral dI3}.
    (b) Diagram of Integral for integral \eqref{eq:integral dI4}}
    \label{fig:Diagram of Integral}
\end{figure}

We can calculate the integral $I_3$ and $I_2$.
\begin{equation}
\begin{split}
    I_3=&\frac{1}{4}\int {\rm d}x_1{\rm d}x_2 e^{{\rm i}\pi [T_{0,x_0}(x_1)-T_{0,x_0}(x_2)]-\frac{2}{K}\ln{|x_1-x_2|}}+h.c\\
    =&\frac{1}{2}\int {\rm d}x_1{\rm d}x_2 e^{-\frac{2}{K}\ln{|x_1-x_2|}}e^{{\rm i}\pi\Theta},
\end{split}
\end{equation}
where $\Theta=T_{0,x_0}(x_1)-T_{0,x_0}(x_2)$. Then 
\begin{equation}
\begin{split}
    \Delta_{I_3}=I_3-I_2=2\times\frac{1}{2}\int_{F(\Theta=1)} {\rm d}x_1{\rm d}x_2 e^{-\frac{2}{K}\ln{|x_1-x_2|}}\times(-1),
\end{split}
\end{equation}
where $F(\Theta=1)$ means the region that $(x_1,x_2)$ satisfies $\Theta=1$. In Fig.~\ref{fig:Diagram of Integral} (a), the region $F(\Theta=1)$ is labeled by $e^{\pm {\rm i}\pi}$.

If we change the coordinate $(x_1,x_2)$ to $(y_1,y_2)=(\frac{x_1-x_2}{\sqrt{2}}, \frac{x_1+x_2}{\sqrt{2}})$, it gives 
\begin{equation}
\begin{split}
\label{eq:integral dI3}
    \Delta_{I_3}=-\int_{F(\Theta=1)} {\rm d}y_1{\rm d}y_2 e^{-\frac{2}{K}\ln{|\sqrt{2}y_1|}}=-2\int_{\rm I+II} {\rm d}y_1{\rm d}y_2 e^{-\frac{2}{K}\ln{|\sqrt{2}y_1|}},
\end{split}
\end{equation}
where the region I and II is also shown in Fig.~\ref{fig:Diagram of Integral} (a). 
So, we have 
\begin{equation}
\begin{split}
    \Delta_{I_3,{\rm I}}=-2\int_0^{x_0/\sqrt{2}} {\rm d}y_1 \cdot4y_1 e^{-\frac{2}{K}\ln{\sqrt{2}y_1}}=-8\int_0^{x_0/\sqrt{2}} {\rm d}y_1 (\sqrt{2}y_1)^{-\frac{2}{K}}\cdot y_1=\frac{-4}{2-\frac{2}{K}}x_0^{2-\frac{2}{K}},
\end{split}
\end{equation}
\begin{equation}
\begin{split}
    \Delta_{I_3,{\rm II}}=-2\int_{-\sqrt{2}x_0}^{\sqrt{2}x_0} {\rm d}y_2 \int_{x_0/\sqrt{2}}^{+\infty} {\rm d}y_1 e^{-\frac{2}{K}\ln{\sqrt{2}y_1}} =-4\sqrt{2}x_0 \int_{x_0/\sqrt{2}}^{+\infty} {\rm d}y_1 (\sqrt{2}y_1)^{-\frac{2}{K}}=\frac{4}{1-\frac{2}{K}}x_0^{2-\frac{2}{K}},
\end{split}
\end{equation}
where we assume $K<1$. Therefore, we have
\begin{equation}
    \Delta_{I_3}=\Delta_{I_3,{\rm I}}+\Delta_{I_3,{\rm II}}=\left(\frac{4}{1-\frac{2}{K}}-\frac{4}{2-\frac{2}{K}}\right)x_0^{2-\frac{2}{K}},
\end{equation}
which means the correlation function is 
\begin{equation}
    \left<e^{{\rm i}[\phi(x_0)-\phi(0)]}\right>\approx 1+\frac{\gamma^2}{2}\cdot\left(\frac{4}{1-\frac{2}{K}}-\frac{4}{2-\frac{2}{K}}\right)x_0^{2-\frac{2}{K}}.
\end{equation}

For entanglement entropy, we consider the correlation function $\left<e^{-{\rm i}a_n^k[\phi(x_0)-\phi(0)]}\right>$ with $a_n^k=\frac{2k}{\sqrt{K}n}$. Therefore, we just need to rescale the function $T_{0,x_0}$ with $a_n^k$. The numerator is 
\begin{equation}
    {\rm num}=\int \mathcal{D}\theta e^{-\frac{K}{4\pi}\int \frac{{\rm d}q}{2\pi} |q| |\theta(q)|^2}\sum_{n=0}^\infty\frac{1}{n!}\left(\gamma \int {\rm d}x \cos{[\theta(x)+a_n^k \pi T_{0,x_0}(x)]}\right)^n.
\end{equation}
If we define 
\begin{equation}
    I_4=\frac{1}{4}\int {\rm d}x_1{\rm d}x_2 e^{{\rm i}\pi a_n^k[T_{0,x_0}(x_1)-T_{0,x_0}(x_2)]-\frac{2}{K}\ln{|x_1-x_2|}}+h.c.,
\end{equation}
the correlation function is 
\begin{equation}
    \left<e^{-{\rm i}\frac{2k}{\sqrt{K}n}[\phi(x_0)-\phi(0)]}\right>\approx (1+\frac{\gamma^2}{2}I_4)(1+\frac{\gamma^2}{2}I_2)^{-1}\approx (1+\frac{\gamma^2}{2}(I_4-I_2)).
\end{equation}
Similar to the results above,
\begin{equation}
\label{eq:integral dI4}
\begin{split}
    \Delta_{I_4}=&\frac{1}{4}\int {\rm d}x_1{\rm d}x_2 e^{-\frac{2}{K}\ln{|x_1-x_2|}}\left(e^{{\rm i}\pi a_n^k [T_{0,x_0}(x_1)-T_{0,x_0}(x_2)]}-1\right)+h.c\\
    =&\frac{1}{2}\int_{\rm I+II} {\rm d}x_1{\rm d}x_2 e^{-\frac{2}{K}\ln{|x_1-x_2|}}\left(e^{{\rm i}\pi a_n^k [T_{0,x_0}(x_1)-T_{0,x_0}(x_2)]}-1\right)+h.c\\
    =&[\cos{(\pi a_n^k)}-1]\int_{\rm I+II} {\rm d}x_1{\rm d}x_2 e^{-\frac{2}{K}\ln{|x_1-x_2|}}=[\cos{(\pi a_n^k)}-1]\int_{\rm I+II} {\rm d}y_1{\rm d}y_2 e^{-\frac{2}{K}\ln{\sqrt{2}y_1}}.
\end{split}
\end{equation}
Using the result above with Fig.~\ref{fig:Diagram of Integral} (b), we have
\begin{equation}
    \Delta_{I_4}=[\cos{(\pi a_n^k)}-1]\left(\frac{-2}{1-\frac{2}{K}}-\frac{-2}{2-\frac{2}{K}}\right)x_0^{2-\frac{2}{K}}.
\end{equation}
Therefore, we have correlation function 
\begin{equation}
    \left<e^{-{\rm i}\frac{2k}{\sqrt{K}n}[\phi(x_0)-\phi(0)]}\right>\approx 1+\frac{\gamma^2}{2}[\cos{(\pi a_n^k)}-1]\left(\frac{-2}{1-\frac{2}{K}}-\frac{-2}{2-\frac{2}{K}}\right)x_0^{2-\frac{2}{K}}.
\end{equation}
Finally, according to Section~\ref{sec:Free Field Twist Operators}, we have
\begin{equation}
    \ln{Z_k}=\ln{\left[1+\frac{\gamma^2}{2}[\cos{(\pi a_n^k)}-1]\left(\frac{-2}{1-\frac{2}{K}}-\frac{-2}{2-\frac{2}{K}}\right)x_0^{2-\frac{2}{K}}\right]}\approx[\cos{(\pi a_n^k)}-1]\frac{\gamma^2}{2}\left(\frac{-2}{1-\frac{2}{K}}-\frac{-2}{2-\frac{2}{K}}\right)x_0^{2-\frac{2}{K}},
\end{equation}
so the entanglement entropy is 
\begin{equation}
\begin{split}
\label{eq:final result of EE with power law}
    S=\lim_{n\rightarrow1}\frac{1}{1-n}\sum_k \ln{Z_k}=-\lim_{n\rightarrow1}\frac{\partial}{\partial n}\sum_k [\cos{(\pi a_n^k)}-1]\frac{\gamma^2}{2}\left(\frac{-2}{1-\frac{2}{K}}-\frac{-2}{2-\frac{2}{K}}\right)x_0^{2-\frac{2}{K}}=\gamma^2 f(K)x_0^{2-\frac{2}{K}},
\end{split}
\end{equation}
where $f(K)=\left(\frac{1}{1-\frac{2}{K}}-\frac{1}{2-\frac{2}{K}}\right)\lim_{n\rightarrow1}\frac{\partial}{\partial n}\sum_k [\cos{(\pi a_n^k)}-1]=\left(\frac{1}{1-\frac{2}{K}}-\frac{1}{2-\frac{2}{K}}\right)[\frac{\pi}{\sqrt{K}}\cot{\frac{\pi}{\sqrt{K}}}-1]$.

\subsection{Entanglement entropy at the critical point}

For the case we consider here, the measurement effectively equals to a local defect line along the space direction. 
Closely related to this, there existed research~\cite{brehm2015entanglement} on the $1d$ quantum chain $[-L,L]$ with a point defect at $x=0$ along the time direction, where the entanglement entropy between the two parts separated by the point defect was considered.
With $2D$ conformal field theory, the problem leads to a defect line along the imaginary axis ${\rm Re}w=0$ (the temporal direction) and the branch cut along the positive real axis ${\rm Im}w>0$ (the spatial direction).
Applying replica trick and conformal transformation $z=\log w$, for $n$ replicas, we will get a $\log{\left(L/\varepsilon\right)}\times 2\pi n$ stripe with a periodic boundary condition along the imaginary axis.
Here, $\varepsilon$ is a short-distance cutoff~\cite{brehm2015entanglement}.
The defect line after conformal transformation lies on ${\rm Im}z=(2i-1)\frac{\pi}{2}, i=1,2,...,2n$.
Then the entanglement entropy can be obtained by evaluating the partition function of such a manifold after the transformation.

In our case, we can use a similar method above. 
The only difference is that the defect line is also along the real axis ${\rm Im}w=0$. 
Then, after the conformal transformation, the defect line will lie on ${\rm Im}z=(i-1)\pi, i=1,2,...,2n$.
However, because the stripe has a periodic boundary condition along the imaginary axis of $z$, the translation of defect lines by $\pi/2$ along the imaginary axis will not change the partition function. 
Therefore, we can transform our problem into a quantum system with a point defect at $x=0$.

A $1d$ transverse field Ising chain with a defect leads to an effective central charge ~\cite{eisler_solution_2010} 
\begin{equation} \label{eq:effective_central_charge_Ising}
    \tilde c_{\rm eff}=-\frac{3}{\pi^2}\left\{\left[(1+s)\log(1+s)+(1- s)\log(1-s)\right]\log(s)+(1+s){\rm Li}_2(-s)+(1-s){\rm Li}_2(s) \right\}.
\end{equation}
This effective central charge is defined by the prefactor of the logarithmic entanglement entropy between two regions separated by the defect.
Moreover, Ref.~\cite{yang2023entanglement} performed a space-time rotation to relate the measurement and the defect.
In the following we will show that our model can be mapped to two copies of transverse field Ising model, so we expect that the effective central charge in our case is ${c}_{\rm eff}=2 \tilde c_{\rm eff}$, where variable $s$ is given by $s=1/\cosh{2W}$, and $W$ is the strength of measurement in the lattice model.

In the following, we will show an exact map from our XXZ lattice model to two copies of the transverse Ising model with corresponding measurements.
In Ref.~\cite{yang2023entanglement}, after spin rotation around the $y$ axis, the transverse field Ising model $H_{\rm tf}$ and the measurement operator $U_{\rm tf}$ read
\begin{equation}
    H_{\rm tf}=-\sum_i (\sigma_i^x\sigma_{i+1}^x-\sigma_i^z), \quad U_{\rm tf}=e^{W\sum_i \sigma_i^z}.
\end{equation}
With Jordan-Wigner transformation 
\begin{equation}
\label{JW transformation}
    \sigma^{+}_l=c_l^\dagger e^{{\rm i}\pi \sum_{j<l}n_j}, \quad \sigma^-_l=e^{-{\rm i}\pi \sum_{j<l}n_j} c_l, \quad \sigma^z_l=2c_l^\dagger c_l-1, 
\end{equation}
and Majorana representation of complex fermion ${\gamma}_l^1=c_l^\dagger+c_l, {\gamma}_l^2=(c_l-c_l^\dagger)/{\rm i}$, the Hamiltonian $H_{\rm tf}$ and measurement $U_{\rm tf}$ will become
\begin{equation}
\label{Majorana rep of Htf}
    H_{\rm tf}=\sum_i ({\rm i}\gamma_i^2\gamma_{i+1}^1+{\rm i}\gamma_{i+1}^1\gamma_{i+1}^2), \quad U_{\rm tf}=e^{W\sum_i {\rm i}\gamma_i^1\gamma_{i}^2},
\end{equation}
where $\sigma_i^z$ term induces density operator, i.e., a coupling between two Majorana fermions on the same site, and $\sigma_i^x\sigma_{i+1}^x$ induces the coupling ${\rm i}\gamma_i^2\gamma_{i+1}^1$.

\begin{figure}
    \centering
\includegraphics[width=0.4\linewidth]{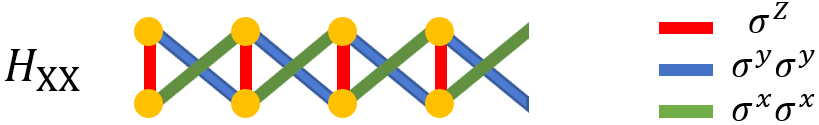} 
    \caption{The diagram of Hamiltonian $H_{\rm XX}$ in terms of Majorana operators. The yellow dots represent two Majorana fermions at each site.}
    \label{fig:Majorana_representation_of_Hamiltonian_and_measurement}
\end{figure}

For our XXZ model at $\Delta=0$ (compared with DMRG method, we change the total sign and a constant of Hamiltonian with rotation around $z$ axis for all even sites), the Hamiltonian and measurement are 
\begin{equation}
    H_{\rm XX}=-\sum_i (\sigma_i^x \sigma_{i+1}^x+\sigma_i^y \sigma_{i+1}^y), \quad U_{\rm XX}=e^{W\sum_i (-1)^i \sigma_i^z}.
\end{equation}
With \eqref{JW transformation} and Majorana fermion representation, up to a total constant we have
\begin{equation}
\label{Majorana rep of HXX}
\begin{split}
    H_{\rm XX}=&\sum_i ({\rm i}\gamma_i^2\gamma_{i+1}^1-{\rm i}\gamma_{i}^1\gamma_{i+1}^2)=\sum_i ({\rm i}\gamma_{2i}^2\gamma_{2i+1}^1-{\rm i}\gamma_{2i}^1\gamma_{2i+1}^2+{\rm i}\gamma_{2i+1}^2\gamma_{2i+2}^1-{\rm i}\gamma_{2i+1}^1\gamma_{2i+2}^2)\\
    =&\sum_i \left[({\rm i}\gamma_{2i}^2\gamma_{2i+1}^1-{\rm i}\gamma_{2i+1}^1\gamma_{2i+2}^2)-({\rm i}\gamma_{2i}^1\gamma_{2i+1}^2-{\rm i}\gamma_{2i+1}^2\gamma_{2i+2}^1)\right]=H_{\rm even}+H_{\rm odd},
\end{split}
\end{equation}
and $U_{\rm XX}=e^{W\sum_i ({\rm i}\gamma_{2i}^1\gamma_{2i}^2 -{\rm i}\gamma_{2i+1}^1 \gamma_{2i+1}^2)}$. 
The different sign in $H_{\rm XX}$ regarding the transverse Ising field can be retained by time reversal transformation or spin rotation along $z$ axis.
This can also be realized for $H_{\rm XX}$ if we rotate the spin at $4i+1$ and $4i+2$ along $z$ axis which flips the sign of $\sigma^x, \sigma^y$. It will induce a phase $(-1)^i$ for each pair $\left<i,i+1\right>$ and set the relative sign in \eqref{Majorana rep of HXX} the same as \eqref{Majorana rep of Htf}.
However, although both $H_{\rm XX}$ and $U_{\rm XX}$ are decoupled into two parts, the measurement $U_{\rm XX}$ will couple two parts in $H_{\rm XX}$. 
It is shown with Majorana representation in Fig.~\ref{fig:Majorana_representation_of_Hamiltonian_and_measurement}.

In the following, we generalize the discussion to include more general measurements.
From \eqref{Majorana rep of HXX}, we consider different types of measurement $U^{2}_{\rm XX}=e^{W_2\sum_i \sigma_i^x\sigma_{i+1}^x}$ or $U^{1}_{\rm XX}=e^{W_1\sum_i (\sigma_{2i}^x\sigma_{2i+1}^x+ \sigma_{2i}^y \sigma_{2i+1}^y)}$.
With Majorana fermion representation, they are
\begin{equation}
\begin{split}
    U^{2}_{\rm XX}=&e^{W_2\sum_i (-{\rm i} \gamma_i^2\gamma_{i+1}^1)}=e^{W_1\sum_i (-{\rm i} \gamma_{2i}^2\gamma_{2i+1}^1-{\rm i} \gamma_{2i+1}^2\gamma_{2i+2}^1)}=U^{2}_{\rm even}U^{2}_{\rm odd},\\
    U^{1}_{\rm XX}=&e^{W_1\sum_i (-{\rm i} \gamma_{2i}^2\gamma_{2i+1}^1+{\rm i} \gamma_{2i}^1\gamma_{2i+1}^2)}=U^{1}_{\rm even}U^{1}_{\rm odd}.
\end{split}
\end{equation}
Actually for these two kinds of measurement we can directly find two decoupled Ising chains from Majorana representation.

Now we consider a complex fermion representation, which will give a uniform understanding of all three measurements considered above.
With complex fermion in Jordan-Wigner transformation \eqref{JW transformation}, we have 
\begin{equation}
    H_{\rm XX}=-2\sum_i (c_i^\dagger c_{i+1}+c_{i+1}^\dagger c_i)=-4\sum_k \cos{k}c_k^\dagger c_k.
\end{equation}
Then for the low-energy theory, we just consider $k$ near $\pm \frac{\pi}{2}$.
Denoting $c_{L,q}=c_{k=-\frac{\pi}{2}+q}, c_{R,q}=c_{k=\frac{\pi}{2}+q}$ we have 
\begin{equation}
    H_{\rm XX}=4\sum_q q(c_{R,q}^\dagger c_{R,q}-c_{L,q}^\dagger c_{L,q})=4\int_x \ {\rm d}x c_R^\dagger (-{\rm i}\partial_x) c_R+c_L^\dagger ({\rm i}\partial_x) c_L=4\int_x \ {\rm d}x c^\dagger (-{\rm i}\partial_x \sigma_z) c,
\end{equation}
where $c=(c_R \ c_L)^T$.
For the measurement $U_{\rm XX}=e^{W\sum_i (-1)^i \sigma_i^z}= e^{h_{\rm XX}}$, in terms of the complex fermion, up to a constant it is 
\begin{equation}
    h_{\rm XX}=2W\sum_i (-1)^i c_i^\dagger c_i=2W\sum_k c_k^\dagger c_{k+\pi}.
\end{equation}
At half filling $h_{\rm XX}=2W\sum_q (c_{q,L}^\dagger c_{q,R}+c_{q,R}^\dagger c_{q,L})= \int_x {\rm d}x \ c^\dagger (2W\sigma_x) c$.

To include the other two measurements, we need the Nambu space basis $\psi_N=(c \ c^\dagger)^T$ with Pauli matrix $\mu_i$, then $H_{\rm XX}=2\int_x  {\rm d}x \ \psi_N^\dagger (-{\rm i}\partial_x \sigma_z\mu_0) \psi_N$ and $h_{\rm XX}= \int_x {\rm d}x \ \psi_N^\dagger (W\sigma_x\mu_z) \psi_N$. For the measurement $U_{\rm XX}^2=e^{h_{\rm XX}^2}$, it will be
\begin{equation}
    h_{\rm XX}^2=W_2\sum_i (c_i^\dagger-c_i)(c_{i+1}^\dagger+ c_{i+1})=W_2\sum_i (c_i^\dagger c_{i+1}^\dagger-c_i c_{i+1}-c_i c_{i+1}^\dagger+c_i^\dagger c_{i+1}),
\end{equation}
where the last two terms will only renormalize the velocity in $H_{\rm XX}$, and the first two terms will open a gap.
We only consider the first two terms. 
In terms of complex fermion representation, it is 
\begin{equation}
    h_{\rm XX}^2=W_2\sum_k (e^{{\rm i}k} c_k^\dagger c_{-k}^\dagger+e^{-{\rm i}k} c_{-k} c_{k})=W_2\sum_k ({\rm i}\sin{k} c_k^\dagger c_{-k}^\dagger-{\rm i}\sin{k} c_{-k} c_{k}).
\end{equation}
At low-energies theory, we only need to consider $k$ near $\pm \frac{\pi}{2}$. Then $h_{\rm XX}^2$ in Nambu space is 
\begin{equation}
    h_{\rm XX}^2=2W_2\sum_q ({\rm i} c_{R,q}^\dagger c_{L,-q}^\dagger-{\rm i} c_{L,-q} c_{R,q})=2W_2\int_x {\rm d}x \ ({\rm i} c_{R}^\dagger c_{L}^\dagger-{\rm i}c_{L} c_{R} )=\int_x {\rm d}x \ \psi_{N}^\dagger (-W_2\sigma_y\mu_x) \psi_{N}.
\end{equation}
Similarly, the measurement $U^{1}_{\rm XX}=e^{h_{\rm XX}^1}$, in complex fermion representation, is given by
\begin{equation}
\begin{split}
    h_{\rm XX}^1=&W_1\sum_i (\sigma_{2i}^x\sigma_{2i+1}^x+ \sigma_{2i}^y \sigma_{2i+1}^y)=2W_1\sum_i (c_{2i}^\dagger c_{2i+1}+c_{2i+1}^\dagger c_{2i})\\
    =&W_1\sum_i (c_{i}^\dagger c_{i+1}+c_{i+1}^\dagger c_{i})+W_1\sum_i (-1)^i(c_{i}^\dagger c_{i+1}+c_{i+1}^\dagger c_{i}),
\end{split}
\end{equation}
where again the first two terms will only renormalize the velocity in $H_{\rm XX}$ and the second two terms will open the gap.
Then we only consider the last two terms. For the low-energy theory in which $k$ is near $\pm \frac{\pi}{2}$ it leads to
\begin{equation}
\begin{split}
    h_{\rm XX}^1=&W_1\sum_k (-e^{{\rm i}k}c_{k}^\dagger c_{k+\pi}-e^{-{\rm i}k}c_{k+\pi}^\dagger c_{k})=W_1\sum_q (-{\rm i}c_{R,q}^\dagger c_{L,q}+{\rm i}c_{L,q}^\dagger c_{R,q}+{\rm i}c_{L,q}^\dagger c_{R,q}-{\rm i}c_{R,q}^\dagger c_{L,q})\\
    =&2W_1 \int_x {\rm d}x \ (-{\rm i}c_R^\dagger c_L +{\rm i}c_L^\dagger c_R)=2W_1 \int_x {\rm d}x \ c^\dagger (\sigma_y) c=\int_x {\rm d}x \ \psi_N^\dagger (W_1\sigma_y \mu_0) \psi_N.
\end{split}
\end{equation}
To summarize, we now have a Hamiltonian $H_{\rm XX}$ and three kinds of measurements in the following,
\begin{equation}
\begin{split}
    H_{\rm XX}=&2\int_x  {\rm d}x \ \psi_N^\dagger (-{\rm i}\partial_x \sigma_z\mu_0) \psi_N,\\
    h_{\rm XX}= \int_x {\rm d}x \ \psi_N^\dagger (W\sigma_x\mu_z) \psi_N, \quad h_{\rm XX}^1=&\int_x {\rm d}x \ \psi_N^\dagger (W_1\sigma_y \mu_0) \psi_N, \quad h_{\rm XX}^2=\int_x {\rm d}x \ \psi_{N}^\dagger (-W_2\sigma_y\mu_x) \psi_{N}.
\end{split}
\end{equation}
Before, we have mentioned that measurements $h_{\rm XX}^1$ and $h_{\rm XX}^2$ will directly induce to decoupled transverse Ising chains with corresponding measurements.
For $h_{\rm XX}$, it can be transformed to other two kinds of measurement with unitary transformation, which means it can also induce two decoupled Ising chains.
Actually, we cannot find a microscopic unitary transformation such that the model is decoupled because the symmetry between three kinds of measurements is emergent symmetry only for low energy continuous model.
Finally, All the three kinds of measurements will open a gap and lead to the same behavior of entanglement which is exactly two copies of transverse Ising model with measurements in Ref.~\cite{yang2023entanglement}.
Therefore, they will give the same effective central charge ${c}_{\rm eff} =2 \Tilde c_\text{eff}$.
To be consistent with the measurement of Luttinger liquid, there is an additional factor $2$ in measurement strength $W$.
Therefore, different with Ref.~\cite{yang2023entanglement}, here $s=1/\cosh{2W}$ in \eqref{eq:effective_central_charge_Ising}.

\subsection{Numerical calculation of the entanglement entropy}

\subsubsection{Free fermion calculation with Julia}

Here we consider the Luttinger liquid model with $K=1$, which corresponds to a free fermion system with periodic boundary condition
\begin{equation}
\label{hopping model}
    H=\sum_i(c_i^\dagger c_{i+1}-c_i c_{i+1}^\dagger),
\end{equation}
where $c_i^\dag$ ($c_i$) denotes the fermion creation (annihilation) operator at site $i$. 
To calculate entanglement entropy after measurement~\cite{surace2022fermionic}, we first diagonalize the Hamiltonian \eqref{hopping model} and calculate its correlation matrix $\Gamma$
\begin{equation}
\Gamma:=\left(
\begin{aligned}
    &\Gamma^{c^\dagger c} &\Gamma^{c^\dagger c^\dagger}\\
    &\Gamma^{c c} &\Gamma^{c c^\dagger}
\end{aligned}
\right),
\end{equation}
where $\Gamma^{c^\dagger c}_{ij}=\left<c^\dagger_i c_j\right>$ and $\Gamma^{c^\dagger c^\dagger}_{ij}=\left<c^\dagger_i c^\dagger_j\right>$. 
Then we apply imaginary time evolution with measurement Hamiltonian $H_m$, which has the same effect as measurement operator $\hat{M}$. 
For different system sizes $L$ and strength of measurement $W$ that is represented by the imaginary time $\tau$ of evolution, we calculate the entanglement entropy of half of the system. 
Then we obtain the relation of entanglement entropy and $L, W$.

\begin{figure}
    \centering
\subfigure[]{
\includegraphics[width=0.3\linewidth]{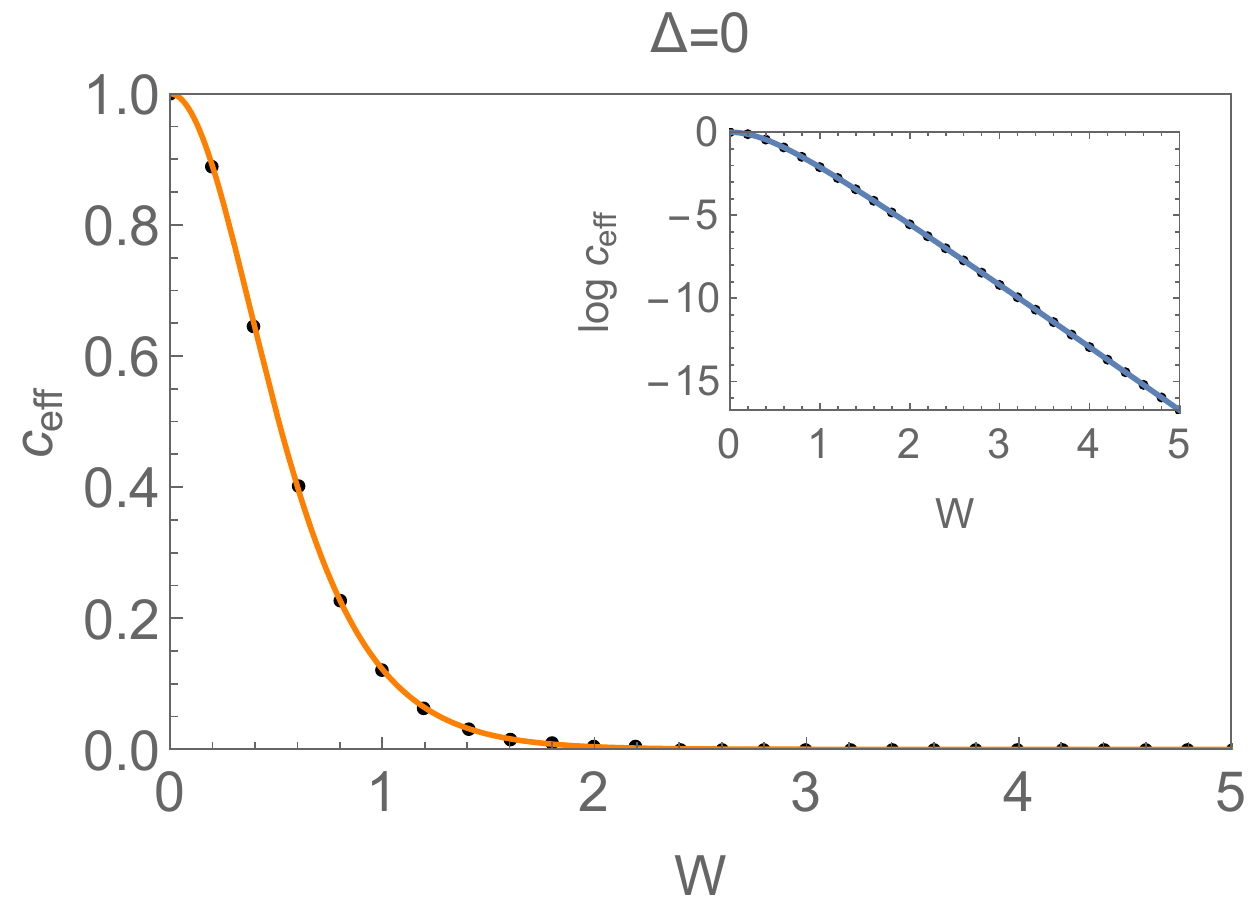}} \quad \quad
\subfigure[]{
\includegraphics[width=0.34\linewidth]{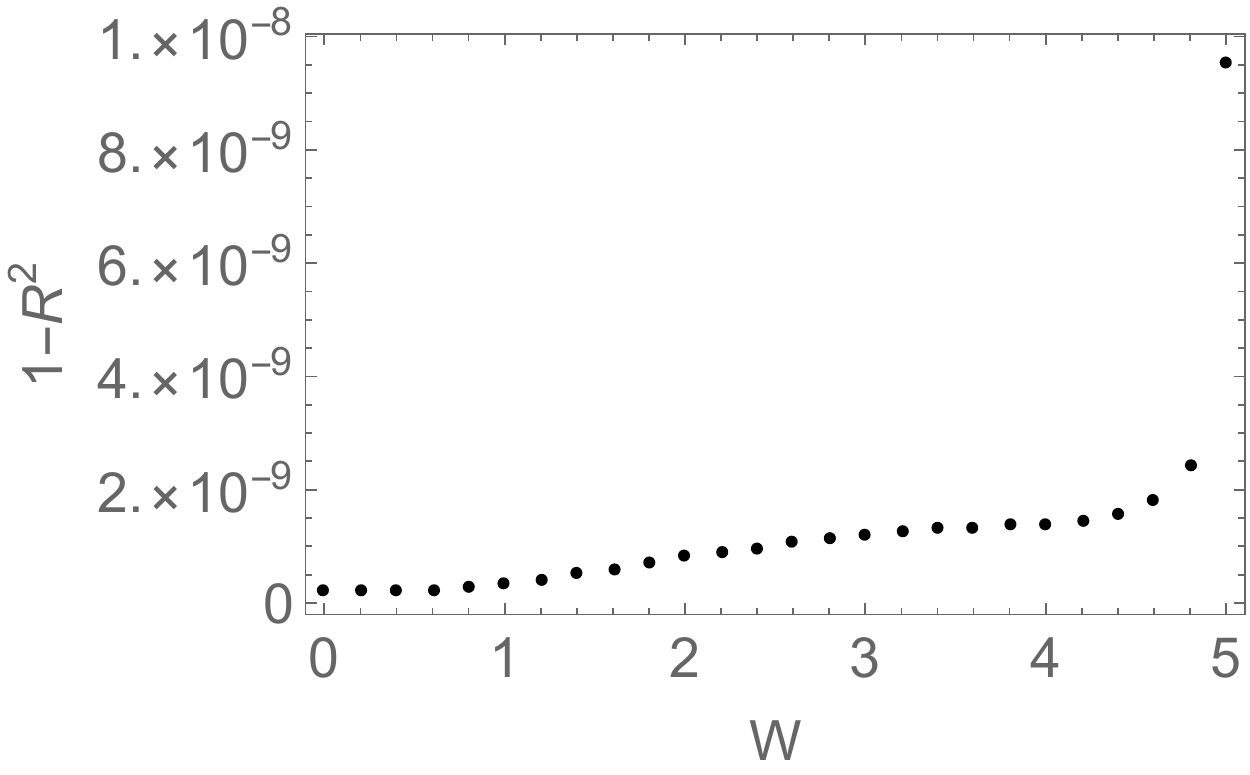}
}
    \caption{
    (a) The effective central charge as a function of the measurement strength $W$ for measurement $H_m$ with free fermion calculation.
    The black dots are numerical results and colorful lines are analytical predictions.
    Inner figure plots $\log{c_{\rm eff}}$ which shows perfect matching for very small $c_{\rm eff}$.
    (b) Fitting coefficient $1-R^2$ at different measurement strength.}
    \label{fig: Free fermion Gaussian state calculation}
\end{figure}

We consider the system size $L \in [2,200]$ and the measurement $W \in [0,5]$. 
For different measurement strength we fit the relation between entanglement entropy and system size $L$ with $a+b\log{L}$, the prefactor $b$ is related to the effective central charge, i.e. $ b = \frac{c_\text{eff}}{3}$. 
In Fig.~\ref{fig: Free fermion Gaussian state calculation} (a), we plot the relation between effective central charge and measurement strength $W$.  
The dot points are the numerical results we get from Julia, and the orange line is the analytical result $2\Tilde{c}_{\rm eff}$ where $\Tilde{c}_{\rm eff}$ is \eqref{eq:effective_central_charge_Ising}. 
Here we take measurement Hamiltonian $H_m={\rm Diag}(0,1,0,1,...,0,-1,0,-1,...)$ and imaginary time evolution $e^{W H_m}$. 
It is equivalent to the measurement $ {\rm Diag}(-1/2,1/2,-1/2,1/2,...,1/2,-1/2,1/2,-1/2,...)$ considered in the main text~(3) up to an identity operator.
At $W=0$, we have central charge $c=1$ with prefactor $b=1/3$. 
For $W\neq0$ numerical results and analytical results are consistent with each other very well. 
Fig.~\ref{fig: Free fermion Gaussian state calculation} (b) shows the coefficient of determination $R^2$ when we interpolate the entanglement entropy $S$ with $\log{L}$. 
$1-R^2\approx 10^{-9}$ means the fitting is accurate.

\subsubsection{Density matrix renormalization group calculation of the XXZ model}

\begin{figure}
    \centering
\subfigure[]{
\includegraphics[width=0.3\linewidth]{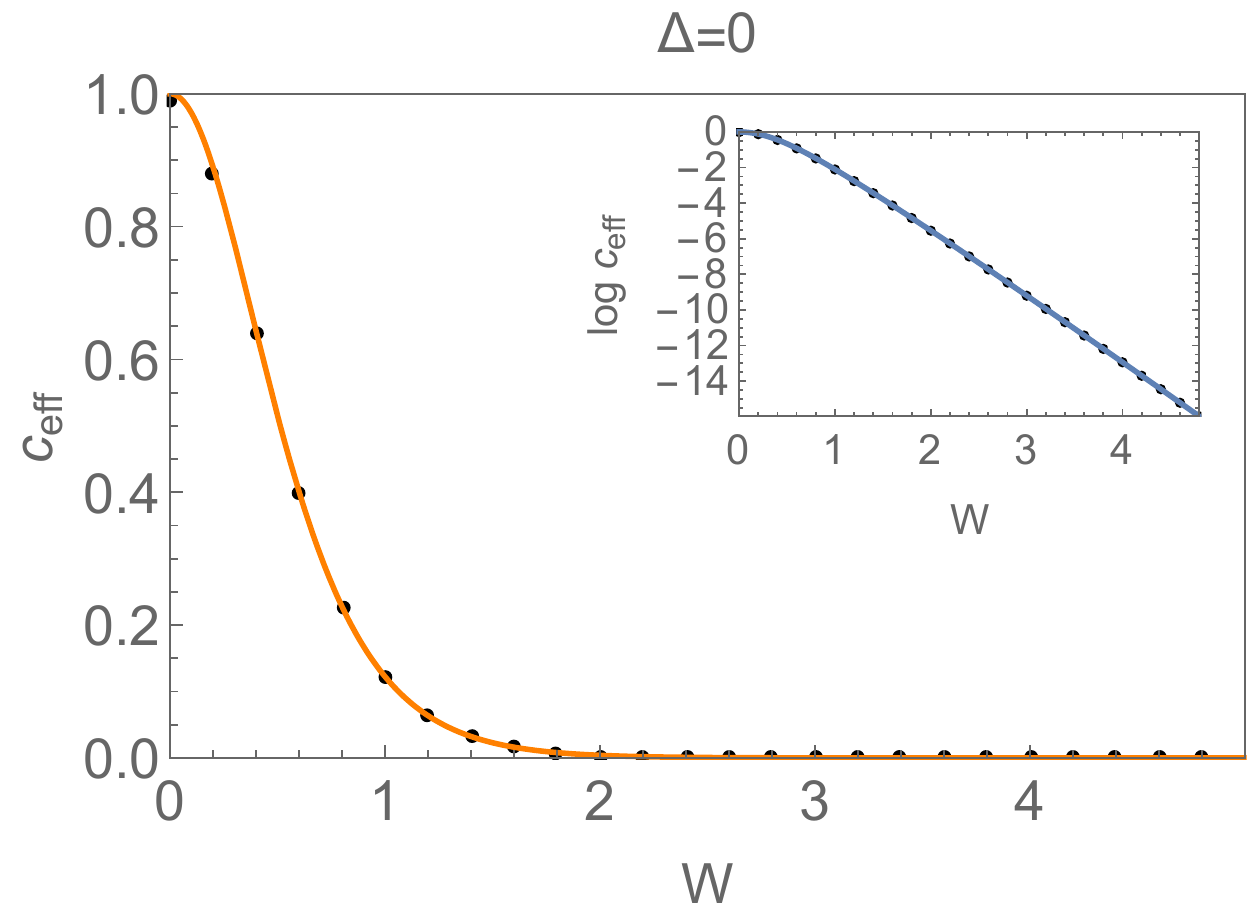}} \quad \quad
\subfigure[]{
\includegraphics[width=0.34\linewidth]{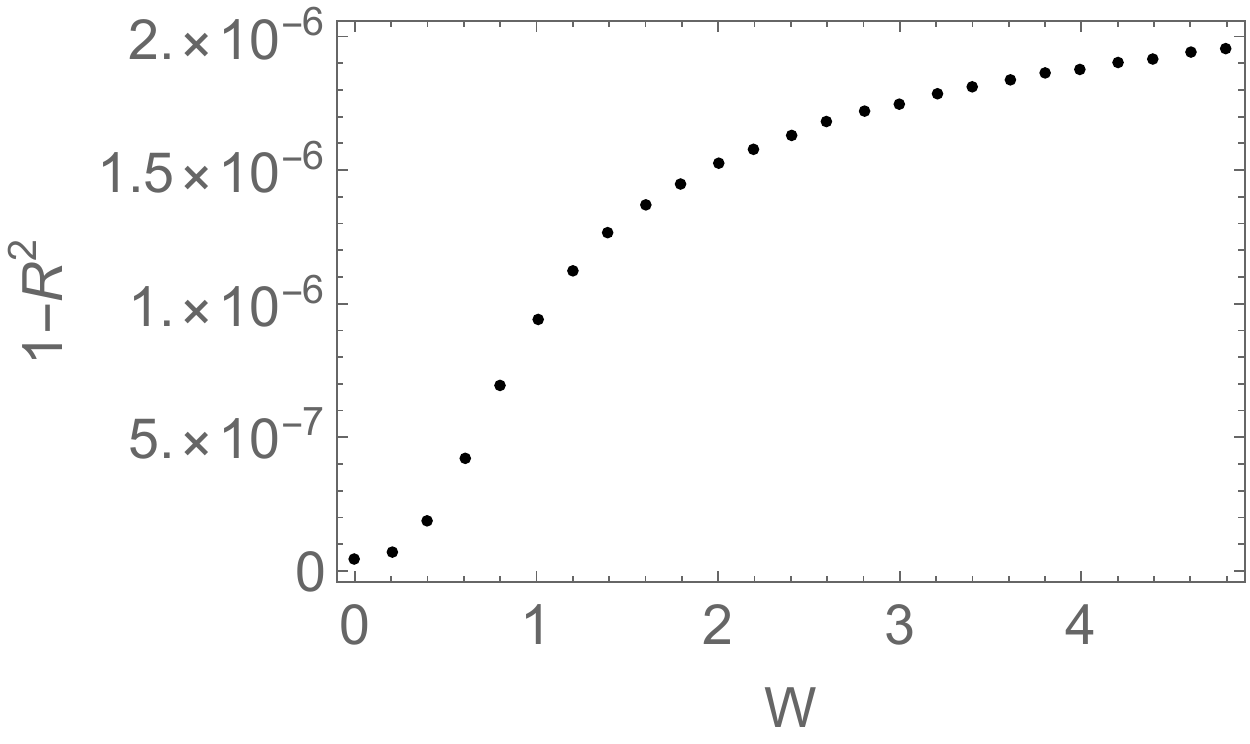}}
\subfigure[]{
\includegraphics[width=0.3\linewidth]{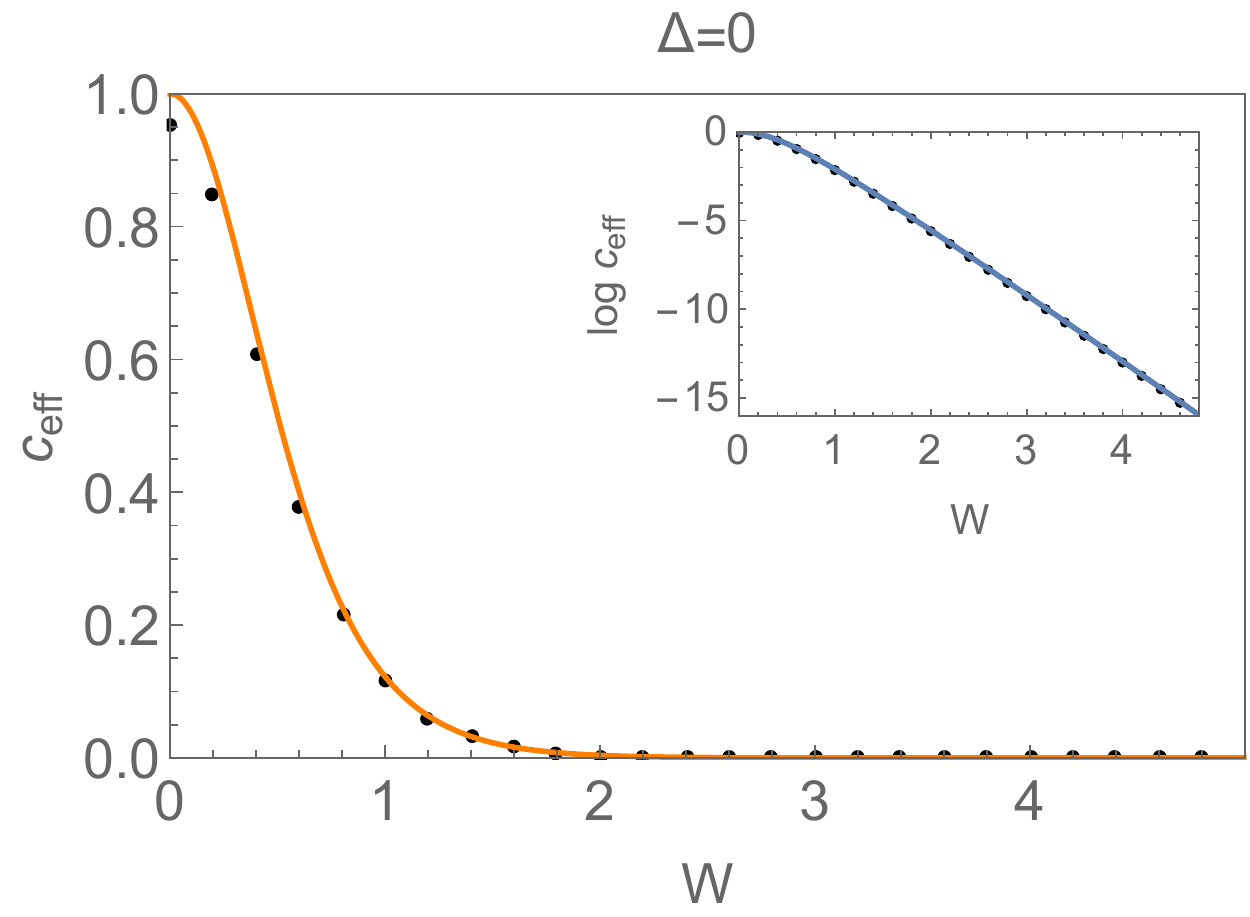}} \quad \quad \quad
\subfigure[]{
\includegraphics[width=0.3\linewidth]{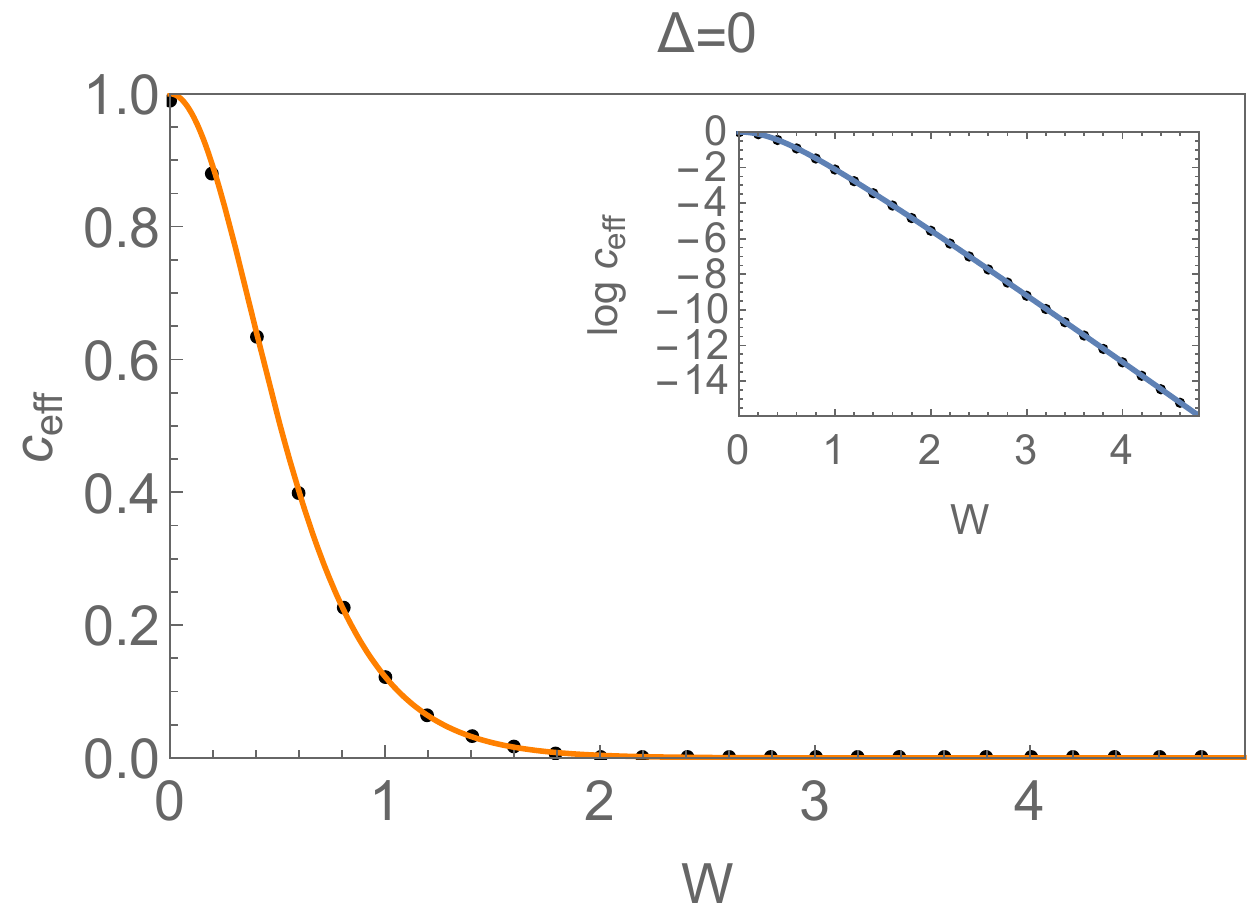}
}
    \caption{
    The results for $\Delta =0$ from DMRG calculation of XXZ model. 
    The black dots are numerical results and the colored curves are analytical predictions.
    Inner figure plots $\log{c_{\rm eff}}$ which shows perfect matching for very small $c_{\rm eff}$.
    (a) The effective central charge as a function of the measurement strength $W$ for measurement $H_m'$.
    (b) Fitting coefficient $1-R^2$ at different measurement strength.
    (c) The effective central charge as a function of the measurement strength $W$ for measurement $H_m^1$.
    (d) The effective central charge as a function of the measurement strength $W$ for measurement $H_m^2$.}
    \label{fig: DMRG for K=1}
\end{figure}

Using Jordan-Wigner transformation, the Hamiltonian~(1) becomes (recall that we set $t=1$)
\bea
    H = \sum_i (S^x_i S^x_{i+1} + S^y_i S^y_{i+1} + \Delta S^z_i S^z_{i+1}) ,
\eea
with $S_\alpha^j$, $\alpha = x, y, z$ spin 1/2 operators at site $i$. 
We first calculate the ground state wave function with Matrix Product State (MPS), then apply the measurement by considering the imaginary time evolution of the ground state wave function. 
Here we take $H_m$ as evolution Hamiltonian. We consider the system size $L\in[10,200]$ and the strength of measurement $W\in[0,5]$. 

\begin{figure}
    \centering
\subfigure[]{
\includegraphics[width=0.33\linewidth]{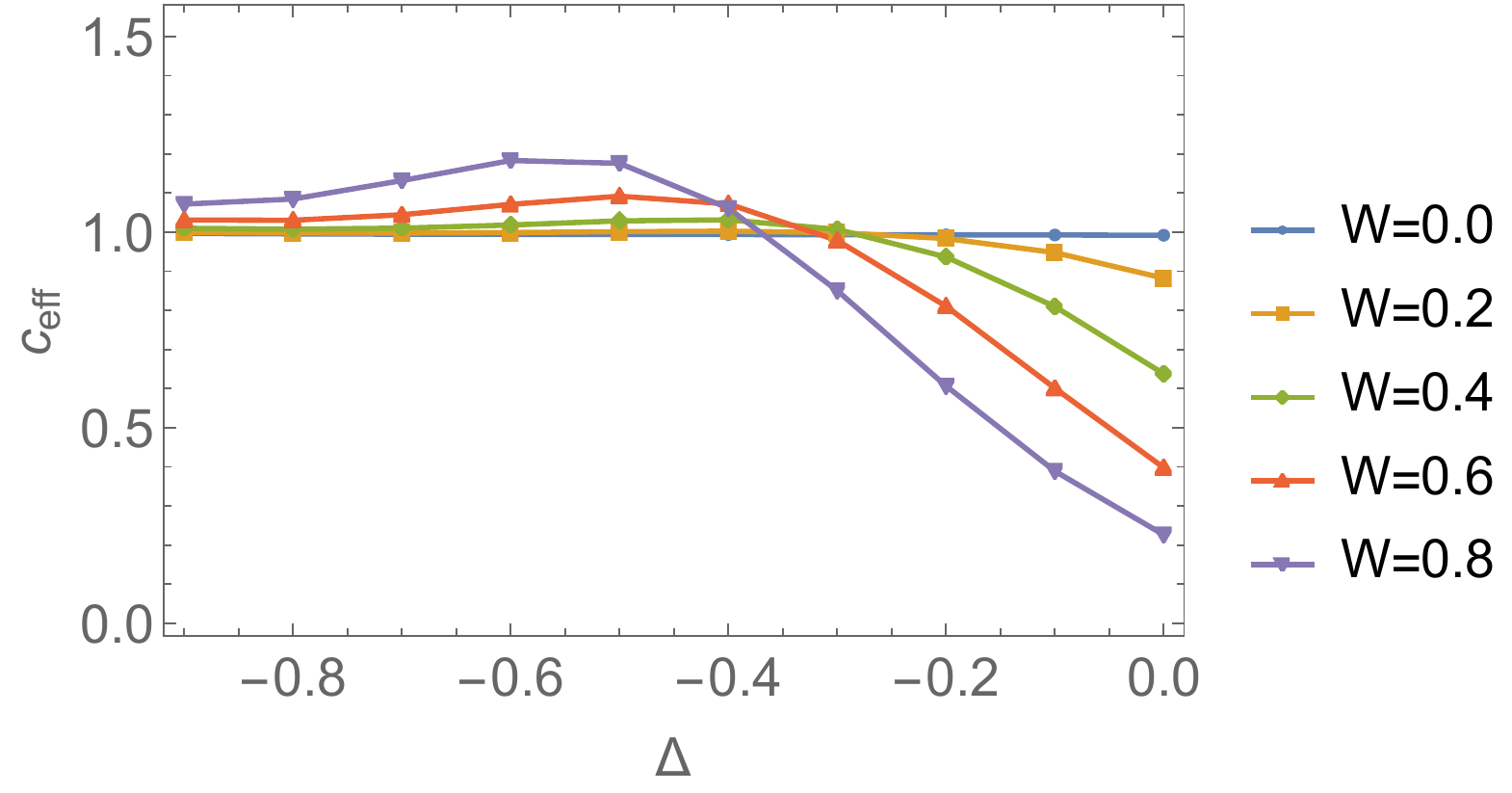}} \quad \quad
\subfigure[]{
\includegraphics[width=0.34\linewidth]{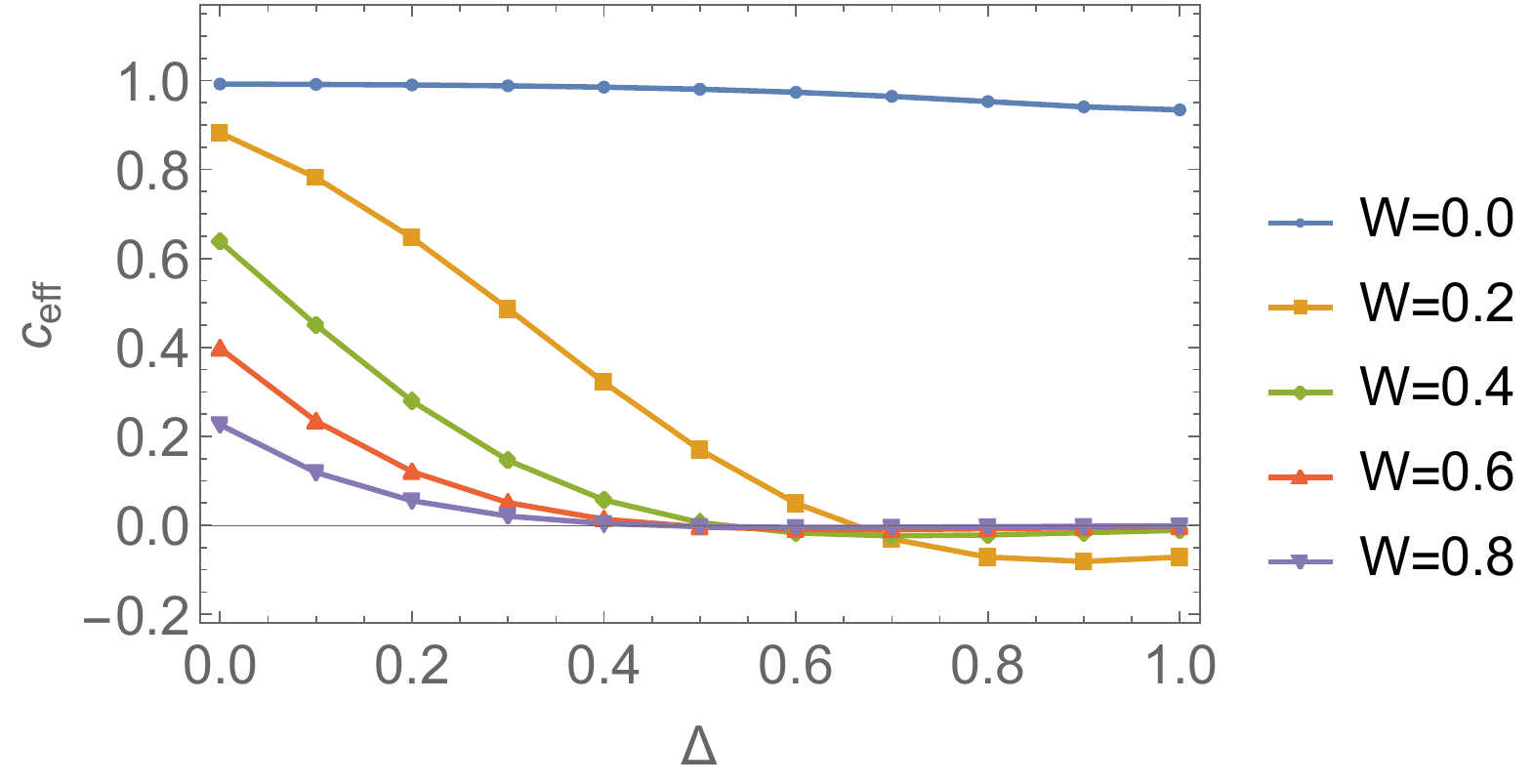}}
    \caption{Effective central charge $c_\text{eff}=3b$ as a function of $\Delta$ for different measurement strength: (a) $\Delta < 0$, (b) $\Delta > 0$.
    }
    \label{fig:DMRG_effective_central_charge}
\end{figure}

At $K=1 (\Delta=0)$ we can plot similar results as free fermion in Fig.~\ref{fig: DMRG for K=1}. 
The results are consistent with the free fermion case. 
Here we define $H_m '=W\sum_i (-1)^i S_{i}^z$, then the effective central charge is \eqref{eq:effective_central_charge_Ising} with $s=\frac{1}{\cosh{2W}}$.
If we take the measurement Hamiltonian $H_m=W\sum_i S_{2i+1}^z$.  
The only difference is that the variable $s$ is $s=\frac{1}{\cosh{W}}$. 
Besides, we also consider other two kinds of measurements $H_m^2=\frac{W_2}{2}\sum_i \sigma_i^x\sigma_{i+1}^x, H^{1}_{m}=\frac{W_1}{2}\sum_i (\sigma_{2i}^x\sigma_{2i+1}^x+ \sigma_{2i}^y \sigma_{2i+1}^y)$ and plot the similar results in Fig.~\ref{fig: DMRG for K=1} (c) and (d), with $W_1 = W_2 = W$.

\begin{figure}
    \centering
\subfigure[]{
\includegraphics[width=0.3\linewidth]{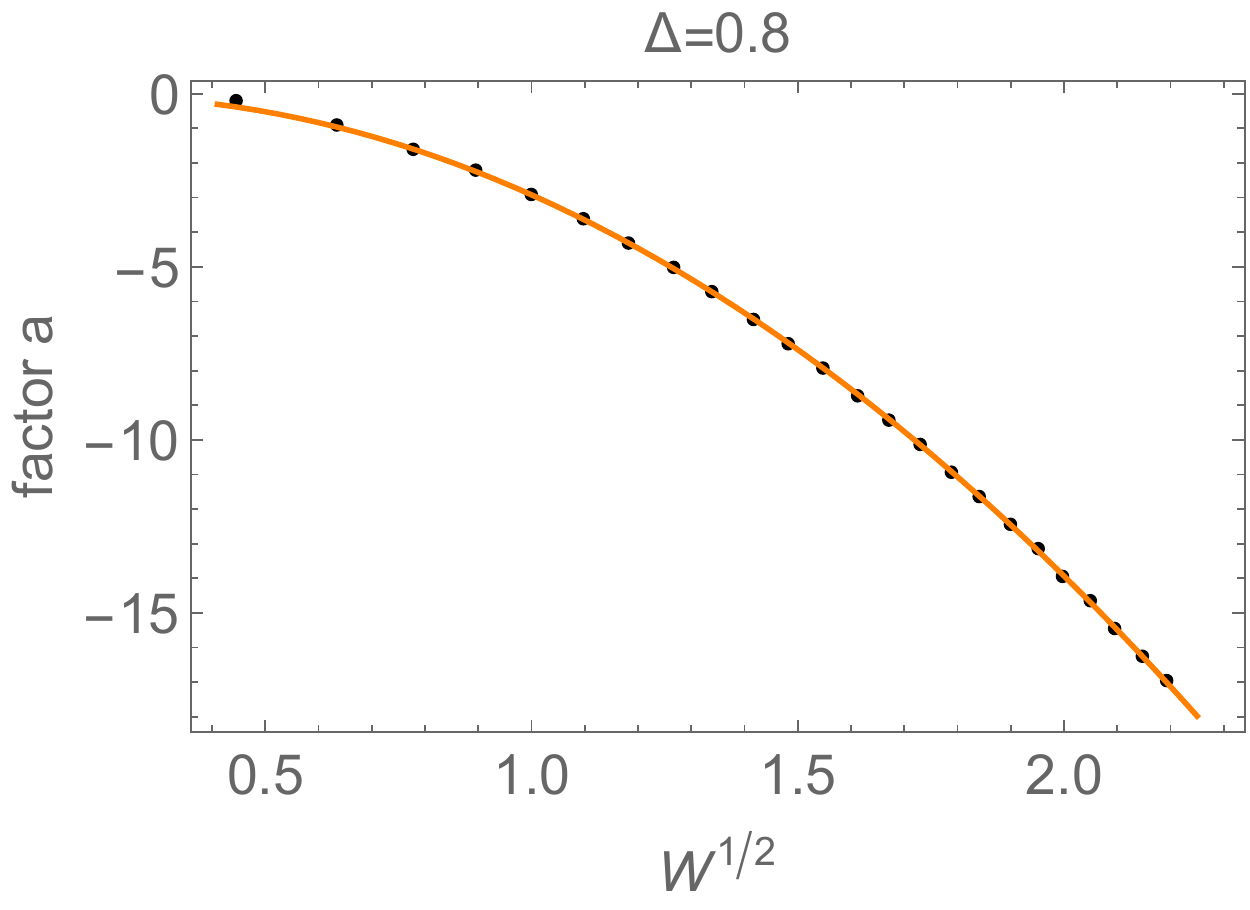}} \quad \quad
\subfigure[]{
\includegraphics[width=0.3\linewidth]{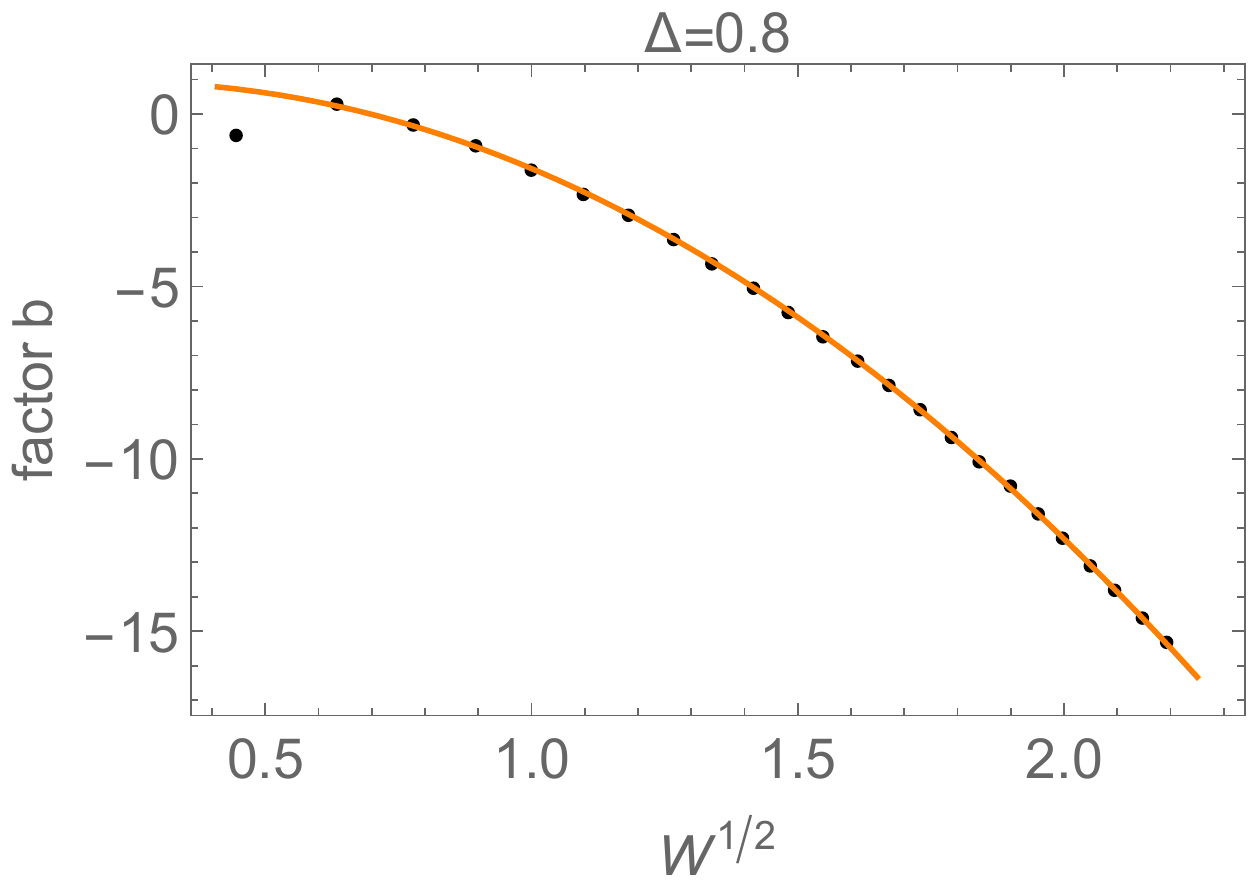}}
    \caption{Prefactor fitting with quadratic polynomials: (a) prefactor $a$ with fitting results $\log{a}\sim-4.14W+1.40\sqrt{W}-0.19$, (b) prefactor $b$ with fitting results $\log{b}\sim-4.23W+1.95\sqrt{W}+0.70$.
    }
    \label{fig:prefactor fitting}
\end{figure}

For $K>1 (\Delta<0)$ we consider two cases and fit the entanglement entropy with $S=a+b\log{L}$ and plot the effective central charge $c_\text{eff}=3b$ with respect to $\Delta$.
This is shown in Fig.~\ref{fig:DMRG_effective_central_charge} (a). 
For $W=0$ and $-0.9 <\Delta <0$, the central charge is $c_{\text{eff}}=1$.
For $\Delta \in(-0.8,-0.2)$ and $W \leq 0.6$, the prefactor $b\approx1/3$ which means the central charge remains approximately $c_\text{eff} \approx 1$. 
Because the effective central charge exactly at $\Delta =0$ is a function of $W$, the deviation of the central charge from $c_{\text{eff}}=1$ near $\Delta = 0$ is due to the finite size effect.
These results verify that the measurement is irrelevant for $K>1$.

For $K<1 (\Delta>0)$, we similarly plot the fitting effective central charge $c_{\rm eff}=3b$ with respect to different $\Delta$. 
In Fig.~\ref{fig:DMRG_effective_central_charge} (b), at $W=0$ we always have $b=1/3$, showing the central charge $c_{\rm eff}=1$, while
for $W > 0$, we see that the effective central charge decreases to zero, indicating that the $\log{L}$ behavior of the entanglement entropy breaks down. 
To characterize the entanglement behavior, we explore the algebraic correction, and identify the power, as shown in Fig.~2 in the main text. 
Besides, we also fit the prefactors $a$ and $b$ in the power law fitting $S=a+b/L^c$ with respect to measurement strength $W$. 
With \eqref{eq:action of instanton} and \eqref{eq:final result of EE with power law}, we know $a$ and $b$ exponentially decay with exponents which are a quadratic polynomial of $\sqrt{W}$. 
In Fig.~\ref{fig:prefactor fitting}, we plot $\log{a}$ and $\log{b}$ with respect to $W^{1/2}$ and fit them with quadratic polynomials, which show perfect fitting results.
Although the prefactor of $\sqrt{W}$ is negative and different from analytical results, here it is unimportant because we have leading term proportional to $W$ which makes the theory stable and the sub-leading term is more artificial.
Moreover, two quadratic polynomials have coefficients of the quadratic term $(\sqrt{W})^2$.

\subsubsection{Experimental realization with a general filling factor}

For a concrete protocol to realize the measurement operator, we consider the implementation of ancilla in Ref.~\cite{garratt2022measurements}, and propose that tuning the filling factor can increase the probability of success exponentially.
A more detailed description of how to couple and measure the ancilla and how to post-select to get (2) is in the following.
After the ground state of the 1D fermion chain (1) is prepared, we couple each site of the chain to an ancillary qubit in the state $\left|\uparrow\right>$.
Then we apply a real-time evolution using the Hamiltonian $H_j= P_j\otimes\hat{\sigma}^x$,where $P_j$ is the projection of particle density $\hat n_j$ or hole density $1-\hat n_j$ at site $j$ in the fermion chain and $\hat{\sigma}^x$ is the Pauli operator acting on the ancillary qubit.
After time evolution $t=u_j$, we arrive at
\begin{align}
     U_j(\left|\psi_{\rm g.s.}\right>\otimes\left|\uparrow\right>)=& e^{iu_j P_j\otimes\hat{\sigma}^x}(\left|\psi_{\rm g.s.}\right>\otimes\left|\uparrow\right>) \\
     =&(1+(\cos{u_j}-1)P_j+i\sin{u_j}P_j \otimes\hat{\sigma}^x)(\left|\psi_{\rm g.s.}\right>\otimes\left|\uparrow\right>)\\
     =&(1+(\cos{u_j}-1)P_j)\left|\psi_{\rm g.s.}\right>\otimes\left|\uparrow\right>+i\sin{u_j}P_j\left|\psi_{\rm g.s.}\right>\otimes\left|\downarrow\right>,
 \end{align}
where $\left|\psi_{\rm g.s.}\right>$ is the ground state of the fermion chain.
Applying the measurement $\hat\sigma^z$ of ancillary qubit, we will have two possibilities.
Then we only post-select the measurement outcome $\left|\uparrow\right>$. 
It is equivalent to apply operator $1+(\cos{u_j}-1)P_j$ at site $j$ to the ground state $\left|\psi_{\rm g.s.}\right>$.
Defining $e^{-W_j P_j}=1+(\cos{u_j}-1) P_j$ we have $W_j=-\log{|\cos{u_j}|}$ with $0 \leq u_j <\pi/2$ and $0\leq W_j < \infty$.
To realize the desired measurement operator in (2) in the main text, one considers the coupling Hamiltonian $H_j =  P_j \otimes \hat \sigma^x$ at site $j$, where $P_{2j-1} = 1-\hat n_{2j-1}$, $P_{2j} = \hat n_{2j}$.
After a proper time evolution, $W = -\log{|\cos{u_j}|}$ and post-selections on $\left|\uparrow \right>$ for all sites, it is not hard to see that (2) can be realized.

The probability of postselection on the state $\left|\uparrow \right>$ is then given by 
\bea
    p_j = \left< \psi_{\rm g.s.} \right| ( 1 - (1- e^{-2W_j}) P_j) \left| \psi_{\rm g.s.} \right>.
\eea
Here $P_j$ can be either particle or hole density operator. 
However, because the particle number is conserved, and the ground state is also an eigenstate of total particle number, we can apply an operator $e^{\tilde W \sum_i \hat n_j }$ without changing the effect of measurement.
For example, in order to realize the measurement operator $e^{-W \sum_j (-1)^j \hat n_j}$, we can instead consider $ e^{-W \sum_j (-1)^j \hat n_j}  e^{W \sum_j \hat n_j} =e^{ -W \sum_j \hat n_{2j}} $.
This allows us to consider the projective operator to be $P_j = \hat n_j$, and bring the probability of success as 
\bea
    p_j = \left< \psi_{\rm g.s.} \right| ( 1 - (1- e^{-2W_j}) \hat n_j) \left| \psi_{\rm g.s.} \right> = 1 - (1- e^{-2W_j}) n,
\eea
where $n = \left< \psi_{\rm g.s.} \right| \hat n_j  \left| \psi_{\rm g.s.} \right>$, if we assume the ground state has translation symmetry. 
This is used to give the lower bound for the success probability in the main text.

Now we discuss a concrete example to show that tuning the filling factor can increase the success probability exponentially without changing the universality class. 
Consider a filling factor $n=1/4$, the measurement should have a 4-site periodicity. 
To simplify the notation, we denote the measurement strength by a four-tuple, $\{W_{4i}, W_{4i+1}, W_{4i+2}, W_{4i+3} \}$, which specifies the strength of four sites in one period. 
A compatible measurement strength that can induce the scattering process between the left and right movers is $W \{1,0,-1,0 \}$.
This is nothing but $W_j = W \cos(2 k_F j) = W \cos(\frac\pi2 j)$, with $j=0,1,2,3$.

\begin{figure}
    \centering
    \includegraphics[width=0.4\textwidth]{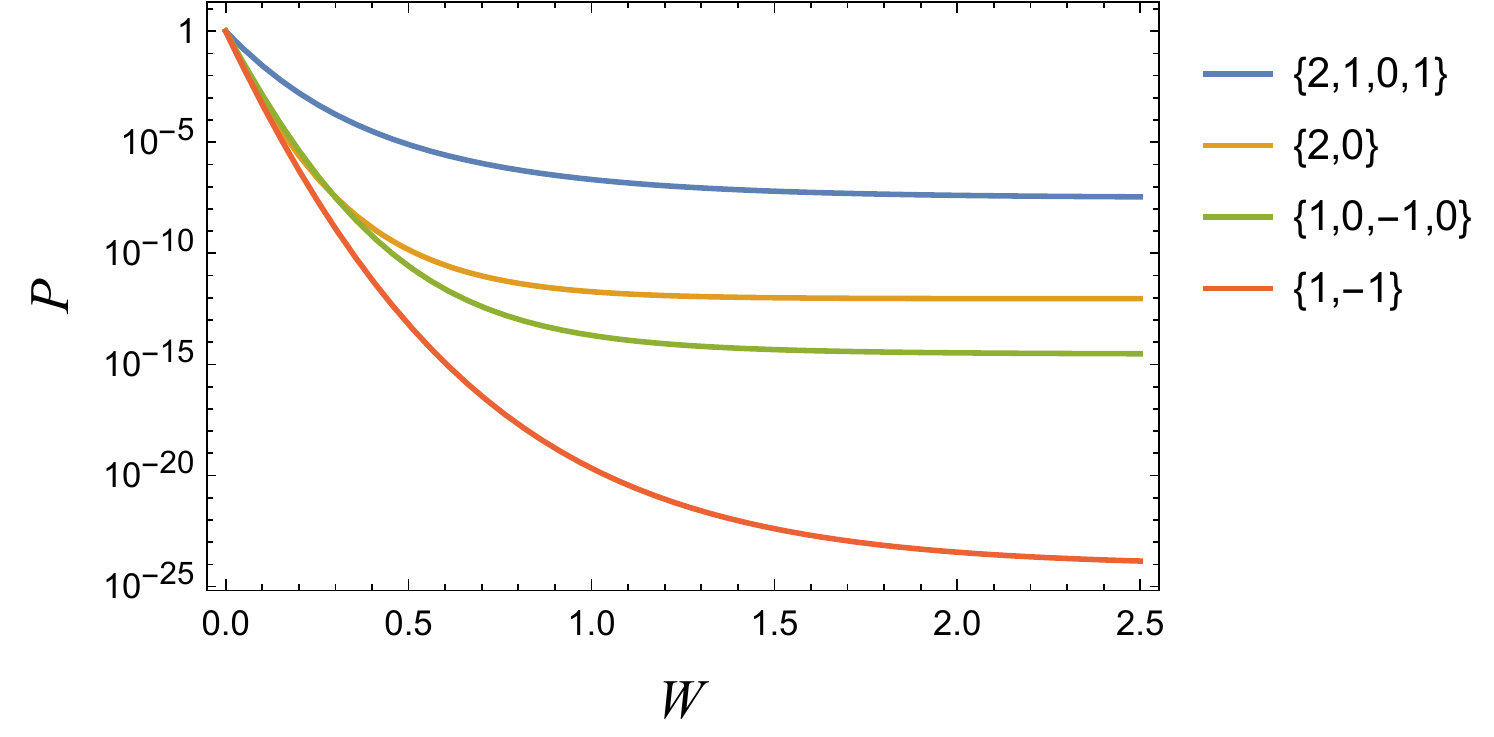}
    \caption{The probability of success of each run of experiments. The total length of the chain is set to be $L_\text{tot} = 80$. The curves with different colors correspond to different measurement protocols shown in the legends.}
    \label{fig:probability}
\end{figure}

Due to the particle conservation, there are different ways to implement the measurement operator.
(1) We can measure the particle number at site $4i$ and the hole number at site $4i+2$. 
This directly implements $W \{1,0,-1,0 \}$, and the success probability is 
\bea
    P^{(1)} = \left[(1 - (1- e^{-2W}) \cdot 1/4)(1 - (1- e^{-2W}) \cdot 3/4) \right]^{L_\text{tot}/4}.
\eea
(2) We can apply an operator $e^{W \sum_j \hat n_j}$ to shift the measurement to be $W\{2,1,0,1\}$. 
The success probability is
\bea
    P^{(2)} = \left[(1 - (1- e^{-4W}) \cdot 1/4)(1 - (1- e^{-2W}) \cdot 1/4)^2 \right]^{L_\text{tot}/4}.
\eea
It is not hard to see that $P^{(2)} \ge P^{(1)}$.
Also, we should compare it with the probability of half-filling, i.e.,
\bea 
P^{(2)} \ge (1 - (1- e^{-4W})\cdot 1/2) ^{L_\text{tot}/2}, 
\eea
where the right-hand side is the success probability of the measurement $W\{2,0\}$ (which is shifted from $W\{1,-1\}$).
In Fig.~\ref{fig:probability}, we compare the probability of success in a chain with length $L_\text{tot} = 80$ for all the measurement protocols for filling factors $n=1/2$ and $1/4$ at different measurement strength.
The measurement protocol $W \{2,1,0,1\}$ for the filling factor $n=1/4$ has an exponentially better probability of success.
In particular, the probability is of order $P \sim 10^{-6}$ near $W \approx 0.7$, which is feasible in NISQ device.
In summary, in addition to the lower bound, $P > (1-n)^{L_\text{tot}}$, given in the main text, our model provides a vast possibility to tune the filling factor and the measurement strength to increase the probability of success.
On the contrary, for the MIPT in the random quantum gates, the probability of success is exponentially suppressed as $p^{-L_\text{tot}^2}$, where $p < 1$ is almost a constant and the exponent $L_\text{tot}^2$ is because the dynamical measurements occur at the time direction.

Now that we have seen how to increase the success probability by tuning the filling factor, we should check that the universality class is independent of the filling factor, and is also robust with imperfect measurement strength.
To this end, we present numerical results of measurement $W\{2,2,0,2\}$ at a $n=1/4$ filling factor in the following. 
The reason for choosing this measurement strength is two-fold: first to show the same universality class for a different filling factor, and second to show that the critical theory is robust again imperfection of measurement strength.
The deviation of measurement strength from $W\{2,1,0,1\}$ results in a higher frequency component, i.e., it is given by $ \cos(\pi/2 \cdot j ) -1/2 \cos(\pi \cdot j ) + 3/2$ with an additional frequency $\pi$.

\begin{figure}
    \centering
    \subfigure[]{\includegraphics[width=0.3\linewidth]{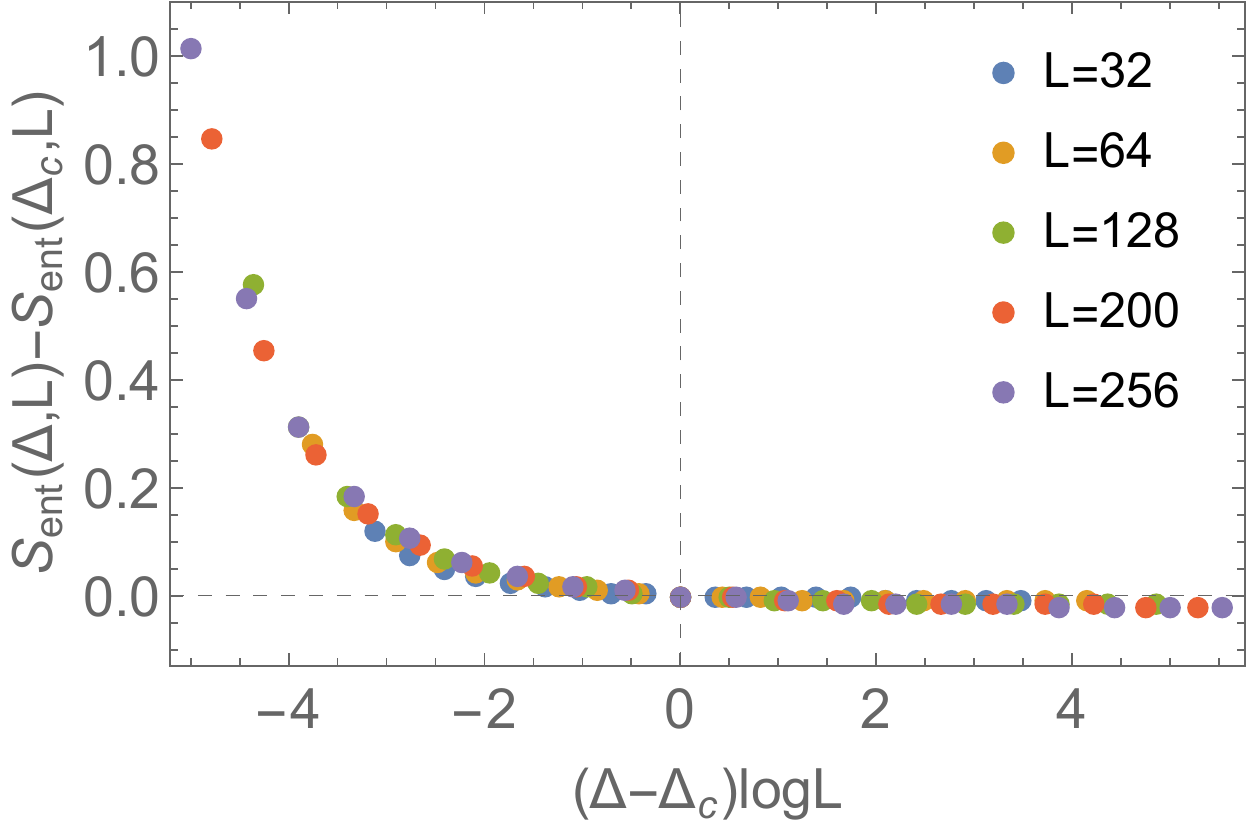}} \quad \quad
    \subfigure[]{\includegraphics[width=0.32\linewidth]{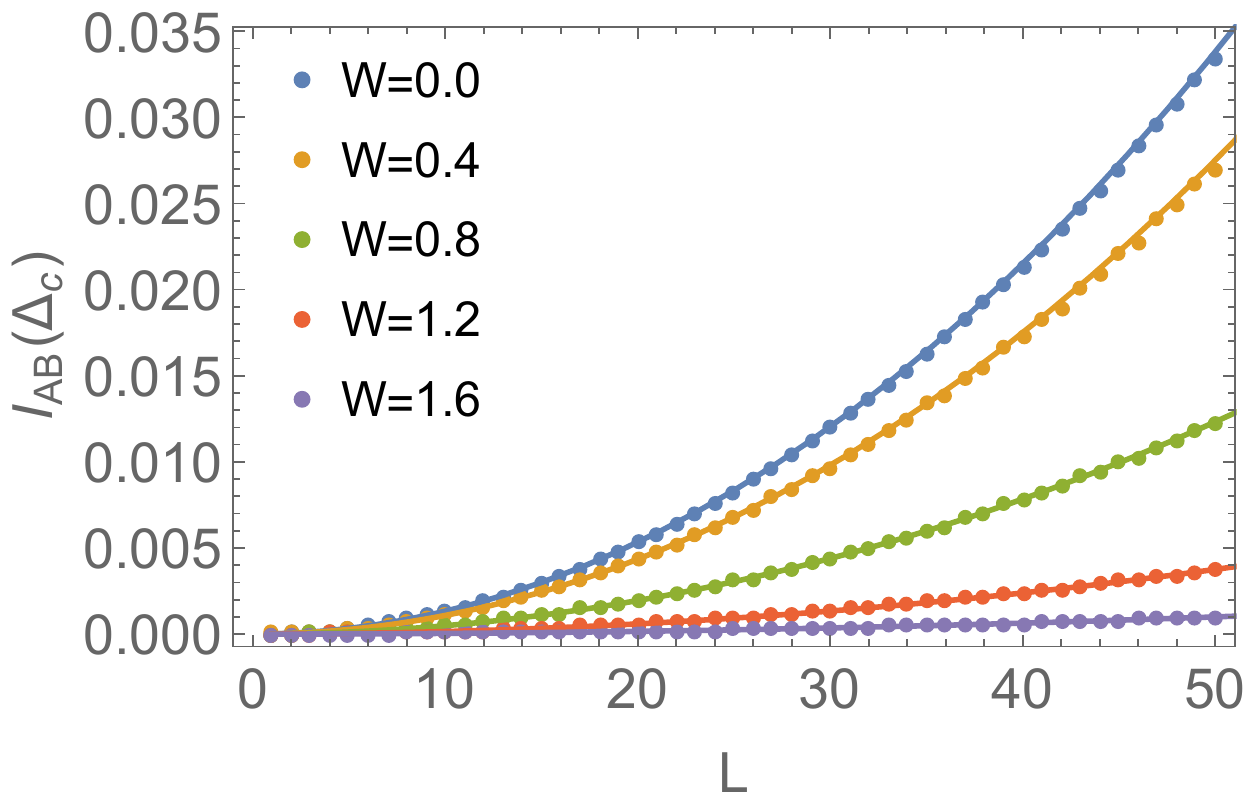}}
    \subfigure[]{\includegraphics[width=0.5\linewidth]{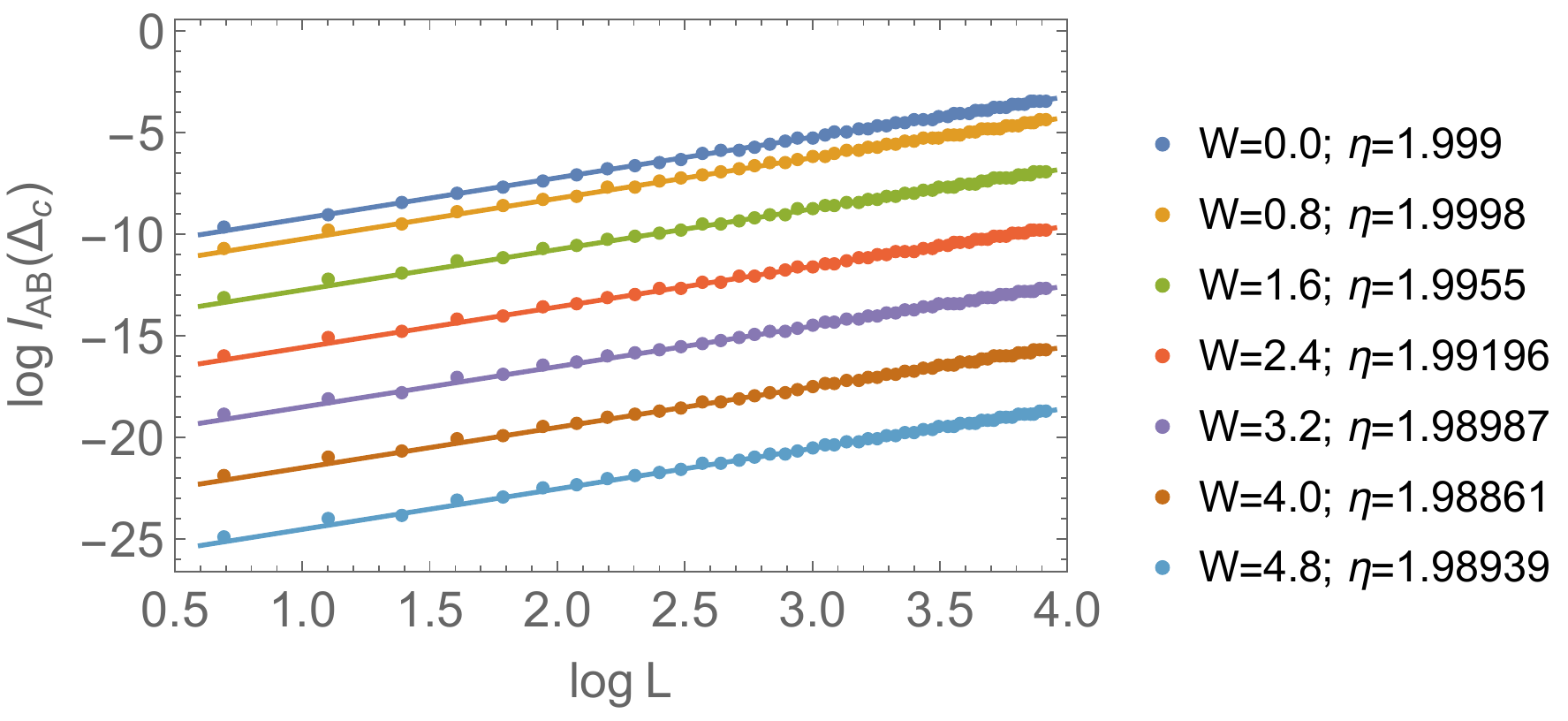}}
    \caption{(a) Half-chain entanglement entropy as a function of $\Delta$ for different sizes $L$. 
    The measurement strength is $W=2.0$.
    (b) Mutual information $I_{\rm AB}$ as a function of L at critical point for different measurement strength $W$. 
    The colorful dots are numerical results and the lines are analytical predictions.
    (c) log-log plot of mutual information $I_{\rm AB}$ at critical point for different measurement strength $W$ with fitting exponent $\eta$. 
    The colorful dots are numerical results and the lines are linear fitting results.
    }
    \label{fig:transition for 1/4}
\end{figure}

Similar to the main text, we calculate two exponents $\nu$ and $\eta$ near the critical point $\Delta_c=0$.
For $\nu$, we plot the data collapse of half-chain entanglement entropy as a function of $\Delta$ for different sizes $L$ in Fig.~\ref{fig:transition for 1/4} (a).
All data collapse onto a smooth function when the argument is chosen to be $(\Delta-\Delta_c) \log L$. 
For $\eta$, we plot mutual information $I_{\rm AB}$ as a function of length $L$ for different measurement strength $W$ in Fig.~\ref{fig:transition for 1/4} (b).
The dots are numerical results and the colored curves are analytical predictions, $I_{\rm AB}=-c'_{\rm eff}/3\log{[\cos^2{(\pi L/L_{\rm tot})}]}$. 
Here the prefactor $c'_{\rm eff}$ is different from $c_{\rm eff}$ in the main text.
The reason is that it gets renormalized due to the higher frequency component in the measurement strength. 
Nevertheless, $c'_{\rm eff}$ that is used to compare with the numerical results is not obtained from a fitting of the mutual information but from an independent numerical calculation of the half-chain entanglement entropy at the critical point (which is not shown here).
To further confirm the results above, we also show log-log plot in Fig.~\ref{fig:transition for 1/4} (c). 
The dots are numerical results and the colored lines are linear fitting results, in which the slope is the critical exponent $\eta$.
The fitting exponent also shows that $\eta\approx2$.

\subsection{Realization of MIPT in variational quantum algorithms}

\begin{figure}
    \centering
    \includegraphics[width=0.5\textwidth]{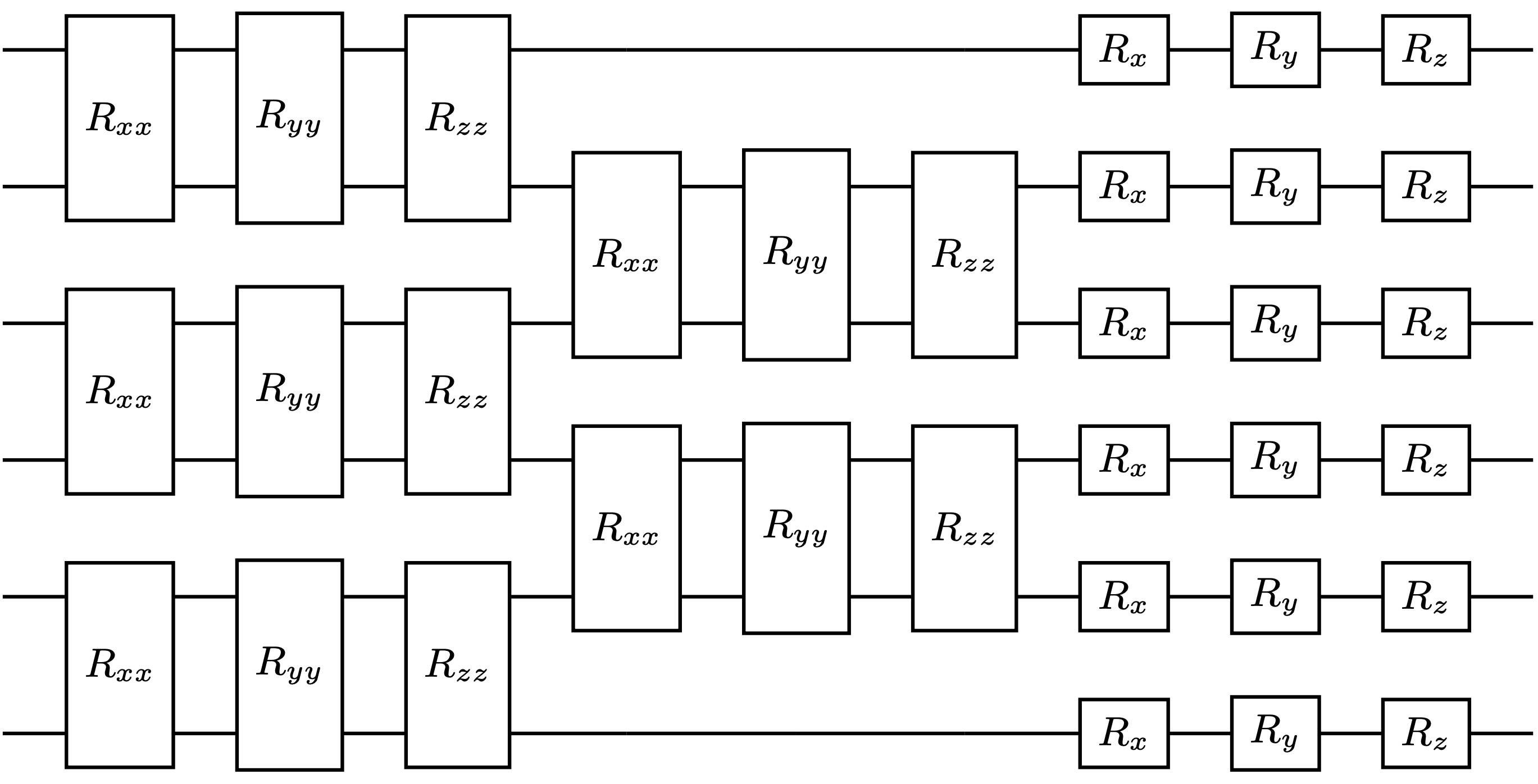}
    \caption{An illustration of a layer of the parametrized quantum circuit ansatz. Here $L_\text{tot} = 6$, $l=1$, and we use open boundary condition.}
    \label{fig:circuit}
\end{figure}

To realize our protocol in experiment, here we use variational quantum algorithm (VQA)~\cite{mcclean2016theory} to simulate our model.
Because the measurement with post-selection is equivalent to imaginary time evolution, we can use parametrized quantum circuits (PQC) to represent the imaginary time evolved target state and implement variational principle to train the parameters~\cite{yuan2019theory}.
We choose an ansatz with $l$ layers of parametrized quantum circuit, and the initial state is the ground state of the XXZ model~\footnote{We can also use the VQA first to get the ground state of the XXZ model, and further train the PQC to simulate the imaginary time evolution.}.
Each layer contains six small layers of two-qubit gates arranged in an alternating pattern and three small layers of one-qubit gates.
The concrete PQC is given by
\bea
    U(\vec \theta)  = \prod_{i=1}^l \left[ \prod_{\sigma = {x,y,z}} R_{\sigma}^{i} \prod_{r=\text{odd,even}} \left(\prod_{\sigma = {x,y,z}} R_{\sigma \sigma}^{i,r} \right) \right],
\eea
where $R_{\sigma \sigma}^{i,\text{odd}} = \prod_j \exp \left( i \theta_{i,2j-1}^{\sigma\sigma} \sigma_{2j-1} \sigma_{2j} \right)$, $R_{\sigma \sigma}^{i,\text{even}} = \prod_j \exp \left( i \theta_{i,2j}^{\sigma\sigma} \sigma_{2j} \sigma_{2j+1} \right)$, and $R_\sigma^i = \prod_j \exp(i \theta^{\sigma}_{i,j} \sigma_j/2)$.
Here $\theta^{\sigma\sigma}_{i,j}$ ($\theta^{\sigma}_{i,j}$) denotes the parametrization for the two-qubit (single-qubit) gate.
An illustration of $L_\text{tot} = 6$ and $l=1$ is shown in Fig.~\ref{fig:circuit}.
The ansatz for the imaginary time evolution is $|\psi(\vec \theta) \rangle = U(\vec \theta) \left|\psi_{\rm g.s.} \right>$, where $\left|\psi_{\rm g.s.} \right>$ denotes the ground state of the XXZ chain.

We consider $L_{\rm tot}=14$ with $l=14$.
Under the imaginary time evolution, the parameters $\vec \theta(\tau)$ will satisfy the following differential equation~\cite{yuan2019theory}
\bea \label{eq:imag_time}
    \sum_b A_{ab} \dot\theta_b  = C_{a},
\eea
where  
\bea
    A_{ab} = \Re\left( \frac{ \partial \langle \psi (\vec \theta)|}{\partial \theta_a} \frac{\partial | \psi(\vec \theta) \rangle}{\partial \theta_b} \right), \quad C_a = - \Re \left( \frac{ \partial \langle \psi (\vec \theta)|}{\partial \theta_a} H | \psi(\vec \theta) \rangle \right).
\eea
where $a$ denotes the general index for the parametrization, including both layers and sites, and $\Re\left(\cdot\right)$ means taking real part of $\left(\cdot\right)$.  
If we start from the ground state with all initial parameterization $\vec\theta(0) = 0$, then numerically we will see that $C_a\approx0$ for all $a$, which means parameters in \eqref{eq:imag_time} will not change under differential equation.
To fix this problem, we can consider a gate $\exp(i \theta^{z}_{1,1} \sigma^z_1/2)$ with $\theta^{z}_{1,1}=0.5$ acting on the first site, and redefine the initial parameter $\theta = 0$ except $\theta^{z}_{1,1}=-0.5$.
Then the initial $|\psi(\vec \theta(0)) \rangle=\left|\psi_{\rm g.s.} \right>$ and corresponding $C\neq 0$~\footnote{If we do not start from the ground state, but use the PQC to get the ground state before the imaginary time evolution, we just need to continue train the parameters after the state is converged.}.
For our purpose, the Hamiltonian for the imaginary time evolution reads
\bea
    H = \sum_j (-1)^j \hat n_j,
\eea
and the evolution time is $W$.

\begin{figure}
    \centering
    \includegraphics[width=0.45\linewidth]{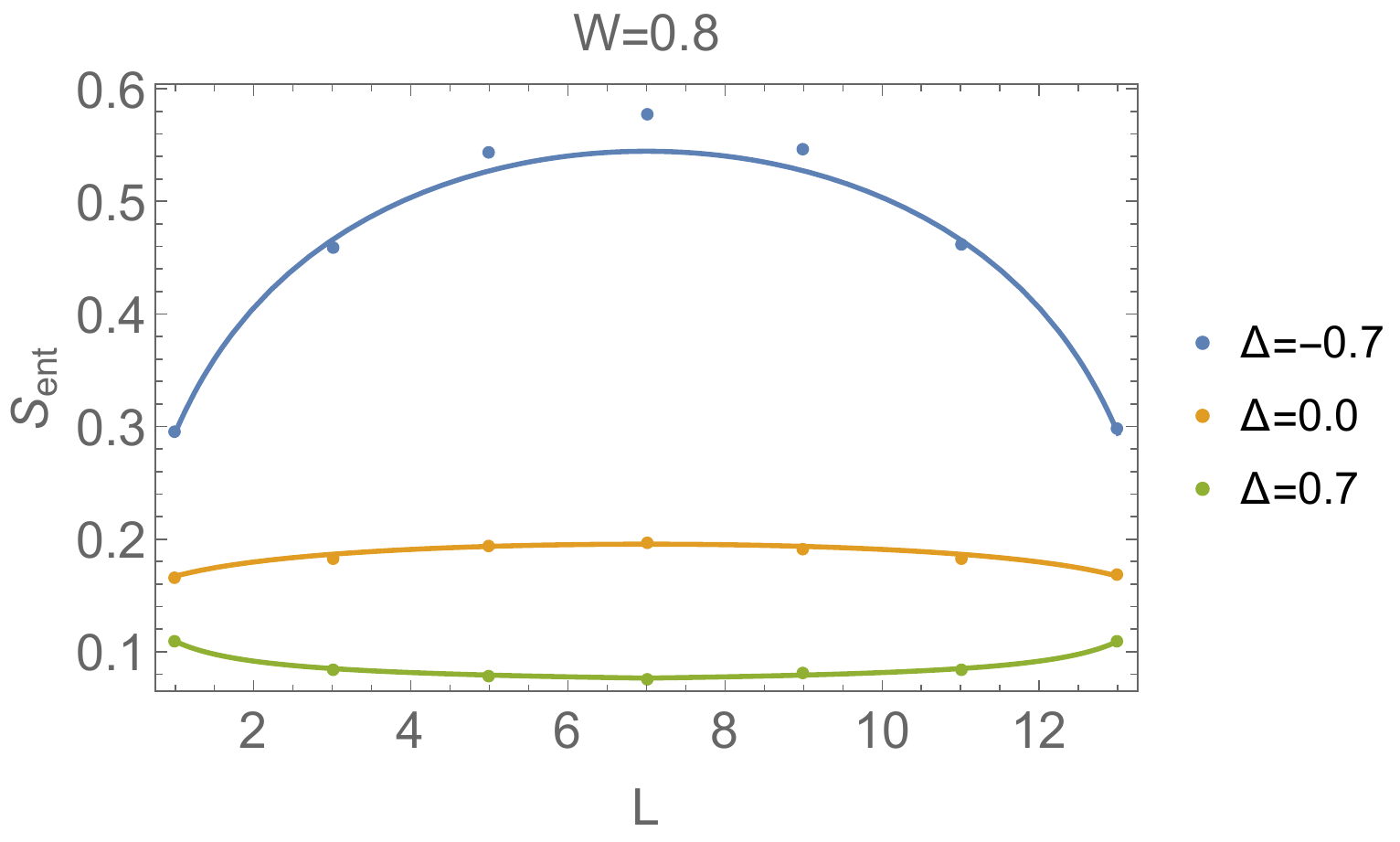}
    \caption{Entanglement entropy with open boundary condition as a function of $L$ for different interaction strength $\Delta$ with measurement strength $W=0.8$. 
    The results are from imaginary time evolution implemented using VQA.
    The colored dots (curves) are numerical (analytical) results for different interaction strength.
    }
    \label{fig:VQA}
\end{figure}

We use the TensorCircuit package~\cite{zhang2023tensorcircuit} to implement the imaginary time evolution~(\ref{eq:imag_time}).
Here, we plot our simulation results of entanglement entropy of the state after the imaginary time evolution in Fig.~\ref{fig:VQA}.
The system has open boundary condition with effective measurement strength $W=0.8$. 
The dots are numerical results. 
The blue curve is the analytical prediction
\bea
S(\Delta=-0.7)=\frac{1}{6} \log{\left[\frac{2L_{\rm tot}}{\pi}\sin{\frac{\pi L}{L_{\rm tot}}} \right]}+c_1
\eea
at $\Delta=-0.7$, where $L $ denotes the subsystem length.
The orange curve is the analytical prediction with effective central charge
\bea
S(\Delta=0.0)=\frac{c_{\rm eff}}{6} \log{\left[\frac{2L_{\rm tot}}{\pi}\sin{\frac{\pi L}{L_{\rm tot}}} \right]}+c_2
\eea
at the critical point. 
$c_{\rm eff}$ is given by Eq. 9 in the main text with $W=0.8$.
The green curve is the fitting result,
\bea
S(\Delta=0.7)=\frac{0.0416}{(L_\text{tot}/2 - |L - L_\text{tot}/2|)^{0.799}}+c_3
\eea
where the entanglement is a power law with fitting power $0.799$. 
Here $L_{\rm tot}=14$ and $c_i$, $i=1,2,3$ are non-universal constants. 
For open boundary condition, there are oscillations in the entanglement entropy due to a parity effect~\cite{calabrese2010parity}, so we only show the data for odd sites to show different phases clearly.
The numerical results show that for $\Delta=-0.7$ the system is in a log-law phase where the measurement is irrelevant, and consequently the state has the same entanglement as the non-measurement case.
For $\Delta=0$ the system is at the critical point with continuous effective central charge given in Eq.~(9).
For $\Delta=0.7$ it is in a pow-law phase with a fitting power $0.799$, which is close to our prediction $2/K-2\approx 0.987$ and the deviation may result from the parity effect~\cite{calabrese2010parity} and the finite size effect.
Although for small system sizes, there is deviation from our prediction, a decent match between the numerical results and the analytical prediction both in different phases and at the phase transition is obvious and promising.

\end{document}